\documentclass[11pt]{article}
\usepackage{algorithm,algpseudocode} 
\usepackage{setspace} 
\usepackage[margin=1in]{geometry} 
\usepackage{rotating} 
\usepackage{amsmath} 
\usepackage{amsfonts, amssymb} 
\usepackage{amsthm}
\usepackage{graphicx} 
\usepackage{epstopdf} 
\usepackage{hyperref} 
\usepackage{bm} 
\usepackage{bbm}
\usepackage{multirow} 
\usepackage{latexsym} 
\usepackage{natbib}
\usepackage{rotating} 
\usepackage{titlesec} 
\usepackage{caption} 
\usepackage{mathtools} 
\usepackage{color} 
\usepackage{enumitem} 
\graphicspath{ {plot/} }
\usepackage{diagbox}
\usepackage{pict2e}
\usepackage{etoolbox}
\usepackage[english]{babel}

\usepackage{tikz}
\usetikzlibrary{arrows, chains, positioning, quotes, shapes.geometric}

\newcommand{\indep}{\rotatebox[origin=c]{90}{$\models$} \, }

\newcommand{\R}{\mathbb{R}}
\newcommand{\T}{^{\intercal}}
\newcommand{\sT}{^{*, \intercal}}
\newcommand{\kT}{^{(-k), \intercal}}
\newcommand{\kk}{^{(-k)}}
\newcommand{\cond}{\,\vert\, }
\newcommand{\Cond}{\,\Big\vert\, }
\newcommand{\con}{\, ; \, }

\DeclareMathOperator*{\argmin}{arg\,min}
\newcommand{\bO}{\mathbf{O}}

\newcommand{\bX}{\mathbf{X}}

\newcommand{\bx}{\mathbf{x}}

\newcommand{\bV}{\mathbf{V}}

\newcommand{\bI}{\mathbf{I}}
\newcommand{\bT}{\bm{\tau}}
\newcommand{\bTH}{\bm{\beta}}
\newcommand{\OV}{{\rm OV}}
\newcommand{\EIF}{{\rm NP}}
\newcommand{\SSLS}{{\rm SP}}

\newcommand{\CR}{{\rm CR}}

\newcommand{\bZ}{\mathbf{Z}}

\newcommand{\bxi}{\bm{\xi}}
\newcommand{\ETA}{^{(\eta)}}

\newcommand{\bIF}{\textbf{IF}}
\newcommand{\EFF}{(\text{eff})}

\newcommand{\EXP}{{\rm E}}
\newcommand{\AVAR}{{\rm aVar}}
\newcommand{\VAR}{{\rm Var}}

\newcommand{\ind}{\mathbbm{1}}

\DeclarePairedDelimiter\norm{\lVert}{\rVert}

\newcommand{\reals}{\mathbb{R}}

\DeclareMathOperator*{\median}{median}

\newcommand{\model}{\mathcal{M}}
\newcommand{\NP}{{\rm NP}}
\newcommand{\SP}{{\rm SP}}
\newcommand{\PH}{ {{\rm PLM.Hom}} }
\newcommand{\SE}{ {{\rm Str.Exp}} }

\hypersetup{
colorlinks=true,
linkcolor=blue,
filecolor=blue,
urlcolor=blue,
citecolor=blue
}

\newtheorem{theorem}{Theorem}[section]
\newtheorem{coro}[theorem]{Corollary}

\theoremstyle{definition}						
\newtheorem{lemma}{Lemma}[section]				

\theoremstyle{definition}						

\theoremstyle{definition}
\newtheorem{assumption}{Assumption}[section]

\theoremstyle{definition}								

\theoremstyle{definition}								

\spacing{1.25}

\numberwithin{table}{section}

\numberwithin{figure}{section}

\begin{document}

\setlength{\abovedisplayskip}{6pt}
\setlength{\belowdisplayskip}{6pt}

\title{\vspace*{-1cm}A Groupwise Approach for Inferring Heterogeneous Treatment Effects in Causal Inference}
\author{Chan Park$^{1}$ and Hyunseung Kang$^{2}$\\[0.25cm]
{\small 1: Department of Statistics and Data Science, University of Pennsylvania}\\
{\small 2: Department of Statistics, University of Wisconsin--Madison}}
\date{ }
\maketitle

\begin{abstract}
Recently, there has been great interest in estimating the conditional average treatment effect using flexible machine learning methods. However, in practice, investigators often have working hypotheses about effect heterogeneity across pre-defined subgroups of study units, which we call the groupwise approach. The paper compares two modern ways to estimate groupwise treatment effects, a nonparametric approach and a semiparametric approach, with the goal of better informing practice. Specifically, we compare (a) the underlying assumptions, (b) efficiency and adaption to the underlying data generating models, and (c) a way to combine the two approaches. We also discuss how to test a key assumption concerning the semiparametric estimator and to obtain cluster-robust standard errors if study units in the same subgroups are correlated. We demonstrate our findings by conducting simulation studies and reanalyzing the Early Childhood Longitudinal Study.\\

\end{abstract}

\section{Introduction}								\label{sec:1}
\subsection{Motivation: A Groupwise Approach for Inferring Effect Heterogeneity} \label{sec:1-1} 

Recently, there has been great interest in estimating heterogeneous treatment effects using machine learning methods  \citep{Su2009, Hill2011, AtheyImbens2016, Shalit2017, victor2018, Dorie2019, Hahn2020, Kennedy2020, NieWager2020}. A common focus in these works is estimating the conditional average treatment effect (CATE) given a specific value of $p$ covariates $\bX_i \in \reals^p$, i.e., $\tau^*(\bX_i) = \EXP \big\{ Y_i^{(1)} - Y_{i}^{(0)} \cond \bX_i \big\} $ where $Y_{i}^{(a)}$ is the potential outcome of study unit $i$ if he/she were to receive a binary treatment value $a \in \{0 ,1\}$. However, in practice, investigators often hypothesize and discuss effect heterogeneity based on pre-defined, meaningful subgroups of study units. For example, in our empirical example from \citet{ECLSK1}, the authors studied the differential effects of early childhood care on children’s academic achievements among urban and rural communities; see Section \ref{sec:data2} for details. We call this approach to studying effect heterogeneity the \emph{groupwise approach} and the target estimand $\bT^* = (\tau_1^*,\ldots,\tau_G^*)\T$ is formally represented as 
\begin{align}			\label{eq1-001}
\tau_g^* = \EXP \big\{ Y_{i}^{(1)} - Y_{i}^{(0)} \cond M(\bX_i) = g \big\}
= \EXP \big\{ \tau^*(\bX_i) \cond M(\bX_i) = g \big\}  \ , \
g=1,\ldots,G \ .
\end{align}
The function $M: \bX_i \to \{1,\ldots,G\}$ is a fixed, well-defined (i.e., $\Pr\{M(\bX_i) = g\}$ is away from $0$ and $1$) function that partitions the $p$-dimensional covariates $\bX_i$ into $G$ non-overlapping subgroups and $\tau_g^*$ is the average treatment effect within the $g$th subgroup. The  main theme of the paper is to carefully examine recent, modern approaches of estimating $\tau_g^*$ based on different models of the observed data and to use the insights from our investigation to better inform practice.

\subsection{Nonparametric Versus Semiparametric Approaches to Estimate $\tau_g^*$}							\label{sec:1-2}

Estimation and inference of $\tau_g^*$ have been discussed in many prior works \citep{Imai2013, Victor2017,Kunzel2018, GRF, Kennedy2020, NieWager2020, ImaiLi2022}. These works can be roughly divided into two types, a nonparametric approach and a semiparametric approach. A nonparametric approach usually starts by estimating $\tau^*(\bX_i)$ with nonparamteric, machine learning methods, say by the generalized random forest (GRF) \citep{GRF}, the $X$-learner \citep{Kunzel2019}, the $R$-learner \citep{NieWager2020}, or the $DR$-learner \citep{Kennedy2020}, and averaging $\tau^*(\bX_i)$ over $\bX_i$ with $M(\bX_i) = g$. This approach typically makes no parametric assumptions about the functional form of $\tau^*(\bX_i)$, the outcome regression, or the propensity score \citep{rosenbaum1983}. In contrast, a semiparametric approach 
usually makes a semiparametric modeling assumption about $\tau^*(\bX_i)$ where the parametric component of the semiparametric model often equals the target parameter of interest $\tau_g^*$; see Section \ref{sec:SSLS-GPLM} for details. We remark that there are works that are in-between or outside of the two approaches \citep{NieWager2020,Victor2017,ImaiLi2022}. In particular, \citet{Victor2017} and \citet{ImaiLi2022} focused on estimating a version of the groupwise effects in a randomized experiment with a known propensity score. Specifically, their target estimand is a conditional groupwise effect where the subgroups are random and depend on particular sample-splitting realizations. In contrast, we primarily focus on an observational study with an unknown propensity score. Also, our target estimand, the groupwise effect $\tau_g^*$, and the subgroups are fixed regardless of sample-splitting realizations.

The paper compares and contrasts the semiparametric approach and the nonparametric approach of estimating $\tau_g^*$. Some notable results in the paper include (a) a sufficient and necessary condition for the semiparametric estimator to consistently estimate the groupwise effect, (b) efficiency and data-adaptive properties of the estimators from the two approaches, (c) a new, combined estimator that can be more efficient than both estimators, (d) derivation of cluster-robust standard errors of these estimators, and (e) a simple, multiple testing procedure to control for familywise error rate when each component of $\tau_g^*$ are tested simultaneously. For practitioners, we summarize our findings in Figure \ref{Figure-Models}.

\section{Different Approaches of Estimating Groupwise Effects}	\label{sec:2}
\subsection{Setup} \label{sec:review_as} 
For each study unit $i=1,\ldots,N$, we observe $\bO_i = (Y_i,A_i,\bX_i)$ where $Y_i \in \reals$ is the outcome, $A_i \in \{0,1\}$ is the treatment indicator with 1 indicating treatment and 0 indicating control, and $\bX_i \in \reals^p$ are pre-treatment covariates. Let $Y_{i}^{(1)}$ and $Y_{i}^{(0)}$ be the potential outcomes under treatment and control, respectively. Let $\bT^*=(\tau_1^*,\ldots,\tau_G^*)\T$ be the vector of the groupwise treatment effects, which are formally defined in equation \eqref{eq1-001}.

We use the following notations for sets, norms, and convergence. For a subset $\mathcal{S} \subseteq \{1,\ldots,N\}$, we denote its complement $\mathcal{S}^C = \{1,\ldots,N\} \setminus \mathcal{S}$. We denote both the 2-norm of a vector and the spectral norm of a matrix as $\| \cdot \|_2$. For a vector $\bm{v}$, let $\bm{v}^{\otimes 2}$ be the outer product of $\bm{v}$, i.e., $\bm{v}^{\otimes 2} = \bm{v}\bm{v}\T$. Let $L_2(P)$-norm for a random variable $ \mathbf{W}$ and its function $h(\mathbf{W})$ be denoted as $\| \mathbf{W} \|_{P,r} = \big\{ \int \norm{\mathbf{w}}_2^r \, dP(\mathbf{w}) \big\}^{1/r}$ and $ \| h \|_{P,r}  = \| h(\mathbf{W})\|_{P,r} = \big\{ \int \norm{h(\mathbf{w})}_2^r \, dP(\mathbf{w}) \big\}^{1/r}$, respectively, where $P(\mathbf{w})$ is the distribution of $\mathbf{W}$. For a sequence $a_N$, let $\mathbf{W}_N = O_P(a_N)$ and $\mathbf{W}_N = o_P(a_N)$ be the usual big-O and little-O notations, respectively. Let $\mathbf{W}_N \stackrel{D}{\rightarrow} \mathbf{W}$ mean that a random variable $\mathbf{W}_N$ converges to $\mathbf{W}$ in distribution as $N \rightarrow \infty$.

We make the standard causal assumptions for observational data; see \citet{ImbensRubin2015} and \citet{HR2020} for textbook discussions. 

\begin{assumption}			\label{assp:CausalAssumption}
Suppose the following conditions hold. 
\begin{enumerate}[topsep=0.05cm,itemsep=0.0cm,label=(\alph*),leftmargin=0.75cm,labelsep=0.25cm]
\item[\hypertarget{(A1)}{(A1)}]  \emph{Stable Unit Treatment Value Assumption (SUTVA)}: $Y_{i} = Y_{i}^{(A_i)}$ almost surely. 
\item[\hypertarget{(A2)}{(A2)}] \emph{Conditional Ignorability/Unconfoundedness}: $\{ Y_i^{(1)}, Y_i^{(0)} \} \indep A_i \cond \bX_i$ for all $\bX_i$.
\item[\hypertarget{(A3)}{(A3)}]  \emph{Overlap/Positivity}:  For all $\bx$ and some positive constant $c_e$, the propensity score $e^* (\bx) = \Pr\big( A_i = 1\cond \bX_i = \bx \big)$ satisfies $c_e \leq e^*(\bx) \leq 1- c_e$.
\item[\hypertarget{(A4)}{(A4)}] \emph{Well-defined Subgroups}: For $g=1,\ldots,G$ and for some positive constant $c_M$, we have $c_M \leq \Pr\big\{M(\bX_i)=g \big\} \leq 1- c_M$.
\end{enumerate}
\end{assumption}
\noindent Under Assumption \ref{assp:CausalAssumption}, the CATE and the target estimand $\tau_g^*$ are identifiable from the observed data as follows:
\begin{align} \label{eq:target}
&
\tau^*(\bx)
=
\mu^*(1,\bx) - \mu^*(0,\bx)
\ ,
\nonumber
\\
&
\tau_g^* 
=
\frac{\int \{ \mu^*(1,\bx) - \mu^*(0,\bx) \} \ind \{ M(\bx) = g \} \, d P_\bX (\bx)}{\int \ind \{ M(\bx) = g \} \, d P_\bX (\bx)}
\ , \
g=1,\ldots,G 
\ .
\end{align}
Here, the function $\mu^*(A_i,\bX_i) = \EXP \big( Y_i \cond A_i, \bX_i )$ is the outcome regression. The rest of the paper discusses the nonparametric and semiparametric approaches of estimating $\tau_g^*$ from equation \eqref{eq:target}.

\subsection{A Semiparametric Estimator $\widehat{\tau}_{\SSLS, g}$}				\label{sec:SSLS-GPLM}
\label{sec:knownM}
This section reviews a popular, semiparametric approach to estimate $\tau_g^*$ based on the following partially linear outcome model: 
\begin{align}				\label{model-PLM} \tag{PLM}
Y_i &= \mu^*(0,\bX_i) + A_i  \sum_{g=1}^{G} \beta_g^* \ind \big\{ M(\bX_i) = g \big\} + \epsilon_i \ , \ \EXP \big( \epsilon_i \cond A_i, \bX_i \big) = 0, 
\\
&= \mu^*(0,\bX_i) + A_i  \bI(\bX_i) \T \bTH^* + \epsilon_i. \nonumber
\end{align}
Here, $ \bI(\bX_i) = \big( \ind \{ M(\bX_i) = 1 \},\ldots, \ind\{ M(\bX_i) = G \} \big) \T$ and $\bTH^* = (\beta_1^*,\ldots,\beta_G^*) \T$. 
In words, \eqref{model-PLM} states that study unit $i$'s observed outcome shifts by a constant amount $\beta_g^*$ if he/she were treated i.e., $A_i = 1$, and belonged to group $g$. If \eqref{model-PLM} is the correct model for the observed data and the causal identifying assumptions in Assumption \ref{assp:CausalAssumption} hold, the causal parameters based on potential outcomes are related to the model parameters based on the observed data as $\tau^*(\bx) = \sum_{g=1}^G \ind \{ M(\bx) = g \}  \beta_g^*$ for all $\bx$ and $\tau_g^* = \beta_g^*$ for $g=1,\ldots,G$.

\citet{robinson1988} and more recently \citet{victor2018} provided a general approach to estimate semiparametric models such as \eqref{model-PLM} by using the following procedure. First, we remove the nonparametric component $\mu^*(0,\bX_i)$ in \eqref{model-PLM} by subtracting $ \nu^*(\bX_i) = \EXP \big( Y_i \cond \bX_i \big)$ from \eqref{model-PLM}. By the definition of conditional expectations, we arrive at
\begin{align*}					
Y_i - \nu^*(\bX_i)
= \big\{ A_i - e^*(\bX_i) \big\}   \bI(\bX_i) \T  \bTH^* + \epsilon_i. 
\end{align*} 
Second, if we define $Z_i^* = Y_i - \nu^*(\bX_i)$ and $\bV_i^* = \big\{ A_i - e^*(\bX_i) \big\}   \bI(\bX_i)\T$, the above model becomes a classic linear regression model with $Z_i^*$ as a response variable and $\bV_i^*$ as a $G$-dimensional regressor. In particular, if the two functions $\nu^*$ and $e^*$ that define the variables $Z_i^*$ and $\bV_i^*$ are known, we can use ordinary least squares (OLS) to arrive at consistent and asymptotically normal (CAN) estimators of $\bTH^*$. When $\nu^*$ and $e^*$ are unknown, we can use cross-fitting \citep{victor2018} where we (i) split the data into two folds, (ii) estimate the two unknown functions $\nu^*$ and $e^*$ with potentially flexible machine learning methods, (iii) run OLS, and (iv) repeat (i)-(iii); see Algorithm \ref{al1} for details.

\begin{algorithm}[!htb]
\begin{algorithmic}
\Require Original data $\bO_i$, $i=1,\ldots,N$.
\State Randomly split data into $\mathcal{I}_1$, $\mathcal{I}_2 \subseteq \{1,\ldots,N\}$ where $|\mathcal{I}_1| = |\mathcal{I}_2| = N/2$ and $\mathcal{I}_{1} \cap \mathcal{I}_{2} = \emptyset$.
\For{$k=1,2$}
\State Estimate $\nu^*$ and $e^*$ with subsample $\mathcal{I}_k^C$. Denote them as $\widehat{\nu}^{(-k)}$ and $\widehat{e}^{(-k)}$.
\State Evaluate $\widehat{\nu}^{(-k)}(\bX_i)$ and $\widehat{e}^{(-k)}(\bX_i)$ for $i \in \mathcal{I}_k$.
\EndFor
\State Run OLS by using $\{ A_i - \widehat{e}^{(-k)}(\bX_i) \} \bI(\bX_i)$ as the regressors and  $Y_i - \widehat{\nu}^{(-k)}(\bX_i)$ as the dependent variable.
\State \Return A semiparametric estimator of $\bT^*$, denoted by $\widehat{\bT}_{\SP}=(\widehat{\tau}_{\SP,1},\ldots,\widehat{\tau}_{\SP,G})\T$, and the corresponding variance estimator $(\widehat{\sigma}_{\SP,1}^2,\ldots,\widehat{\sigma}_{\SP,G}^2)$:
\begin{align}
\widehat{\tau}_{\SSLS, g} 
&= 
\frac{\sum_{k=1}^2 \sum_{i \in \mathcal{I}_k}  \big\{ Y_i - \widehat{\nu}^{(-k)} \big( \bX_i \big) \big\} \big\{ A_i - \widehat{e}^{(-k)} (\bX_i) \big\} \ind\{ M(\bX_i) = g \} }
{ \sum_{k=1}^2  \sum_{i \in \mathcal{I}_k} \big\{ A_i - \widehat{e}^{(-k)} (\bX_i) \big\}^2 \ind\{ M(\bX_i) = g \} } \ ,
\nonumber
\\
\widehat{\sigma}_{\SSLS,g}^2 &= \frac{ \frac{1}{N} \sum_{k=1}^{2} \sum_{i \in \mathcal{I}_{k}}  \{ \widehat{\epsilon}_i^{(-k)}\} ^2 \big\{ A_i - \widehat{e}^{(-k)}( \bX_i) \big\}^2 \ind \big\{ M(\bX_i) = g \big\} }{\big[ \frac{1}{N} \sum_{k=1}^{2} \sum_{i \in \mathcal{I}_{k}} \big\{ A_i - \widehat{e}^{(-k)}( \bX_i) \big\}^2 \ind \big\{ M(\bX_i) = g \big\}  \big]^2} \ , \
\label{eq-consistentVarSP}
\\
\widehat{\epsilon}_i^{(-k)}
& =
\big\{ Y_i - \widehat{\nu}^{(-k)} \big( \bX_i \big) \big\} -  \big\{ A_i - \widehat{e}^{(-k)}( \bX_i) \big\} \bI(\bX_i) \T \widehat{\bT}_\SSLS
\nonumber
\ .    
\end{align}
\end{algorithmic}
\caption{A Semiparametric Estimator for Groupwise Treatment Effects}
\label{al1}
\end{algorithm}

We make some brief remarks about the semiparametric estimator $\widehat{\bT}_{\SSLS}$. 
First, $\widehat{\bT}_{\SSLS}$ is a special case of the $R$-learner \citep{NieWager2020} where the $R$-learner estimates the entire CATE function $\tau^*(\bX_i)$ across all $\bX_i$ in a data-adaptive manner without specifying subgroups. In contrast, our task is to provide a group-level summary of the CATE function where the subgroups are specified a priori by investigators. Second, under \eqref{model-PLM}, $\tau^*(\bX_i)$ in equation (4) of \citet{NieWager2020} is represented by $G$ finite-dimensional parameters. Third, we can split the sample into more than two folds to stabilize the cross-fitted estimators. 

\subsection{A Nonparametric Estimator $\widehat{\tau}_{\EIF,g}$} \label{sec:DR}
This section reviews a popular, nonparametric approach to estimate groupwise effects based on the following, nonparametric linear model of the outcome: 
\begin{equation} \label{model-nonpara}
Y_i = \mu^*(0,\bX_i) + A_i  \gamma^*(\bX_i) + \epsilon_i, \ \EXP (\epsilon_i \cond A_i, \bX_i) = 0. 
\tag{NPM}
\end{equation}
Here, $\gamma^*$ is a nonparametric function. In words, \eqref{model-nonpara} states that study unit $i$'s observed outcome shifts by an amount $\gamma^*(\bX_i)$ if study unit $i$ is treated. Compared to \eqref{model-PLM}, two people in the same subgroup $g$ may have their outcomes shifted by a different amount depending on their respective covariates. Also, unlike \eqref{model-PLM}, \eqref{model-nonpara} does not make any parametric or semiparametric assumptions about the outcome; the outcome is determined by two nonparametric functions  $\mu^*$ and $\gamma^*$. Finally, if the causal identifying assumptions in Assumption \ref{assp:CausalAssumption} hold, the causal parameters based on potential outcomes are related to the model parameters based on the observed data as follows: $\tau^*(\bx) = \gamma^*(\bx)$ and  $\tau_g^* = \EXP \{ \gamma^*(\bX_i) \cond M(\bX_i) = g\}$ for $g=1,\ldots,G$. 

We can adapt the results in \citet{Robins1994}, \citet{Hahn1998}, \citet{Scharfstein1999}, and \citet{vvLaan2003} to obtain an estimator of $\tau_g^*$ under \eqref{model-nonpara}. Specifically, our nonparametric estimator is based on the efficient influence function for the average treatment effect, denoted by $\varphi^* (\bO_i)$, which is formally stated below:
\begin{align*}	
&
\varphi^* (\bO_i) 
=
\bigg\{ \frac{A_i}{e^*(\bX_i)} - \frac{1-A_i}{1-e^*(\bX_i)}  \bigg\} \{ Y_i - \mu^*(A_i,\bX_i) \}
+ \mu^*(1,\bX_i) - \mu^*(0,\bX_i)  \ .  
\end{align*}
To estimate groupwise effects, we can run a linear regression with $\varphi^*(\bO_i)$ as a response variable and the dummy variables representing each group $g$ (without the intercept term) as regressors. Also, we can use plug-in estimators from cross-fitting, similar to the semiparametric approach, to estimate the two unknown functions $\mu^*$ and $e^*$ in $\varphi^*(\bO_i)$; see Algorithm \ref{al2} for details.

\begin{algorithm}[!htb]
\begin{algorithmic}
\Require Original data $\bO_i$, $i=1,\ldots,N$.
\State Randomly split data into $\mathcal{I}_1$, $\mathcal{I}_2 \subseteq \{1,\ldots,N\}$ where $|\mathcal{I}_1| = |\mathcal{I}_2| = N/2$ and $\mathcal{I}_{1} \cap \mathcal{I}_{2} = \emptyset$.
\For{$k=1,2$}
\State Estimate $\mu^*(1,\bX_i), \mu^*(0,\bX_i)$, and $e^*(\bX_i)$ with subsample $\mathcal{I}_k^C$. 
\State Denote them as  $\widehat{\mu}^{(-k)}(1,\bX_i), \widehat{\mu}^{(-k)}(0,\bX_i)$,  and $\widehat{e}^{(-k)}(\bX_i)$.
\State Evaluate the efficient influence function using the estimated nuisance functions for $i \in \mathcal{I}_k$, i.e.,
\begin{align*}
&
\widehat{ \varphi }^{(-k)} (\bO_i) 	
=
\bigg\{ \frac{A_i}{\widehat{e}^{(-k)}(\bX_i)} - \frac{1-A_i}{1-\widehat{e}^{(-k)}(\bX_i)} \bigg\} \{ Y_i - \widehat{ \mu }^{(-k)} (A_i,\bX_i) \}
\nonumber
\\
&
\hspace*{2cm}
+ \widehat{ \mu }^{(-k)}(1,\bX_i) - \widehat{ \mu }^{(-k)}(0,\bX_i)
\end{align*}
\EndFor
\State  Run OLS by using  $\bI(\bX_i)$ as the regressors and  $\widehat{\varphi}^{(-k)}(\bO_i)$ as the dependent variable.
\State \Return A nonparametric estimator of $\bT^*$, denoted by $\widehat{\bT}_{\NP}=(\widehat{\tau}_{\NP,1},\ldots,\widehat{\tau}_{\NP,G})\T$, and the corresponding variance estimator $(\widehat{\sigma}_{\NP,1}^2,\ldots,\widehat{\sigma}_{\NP,G}^2)$:
\begin{align}
& \widehat{\tau}_{\EIF, g}
=
\frac{ \sum_{k=1}^2 \sum_{i \in \mathcal{I}_k} \widehat{\varphi}^{(-k)} (\bO_i) \ind \big\{ M(\bX_i) = g \big\} }{\sum_{i=1}^N \ind \big\{ M(\bX_i) = g \big\} } \ , 
\nonumber
\\
&		\widehat{\sigma}_{\EIF,g}^2 
= 
\frac{ \frac{1}{N} \sum_{k=1}^2 \sum_{i \in \mathcal{I}_k} \big\{ \widehat{\varphi}^{(-k)} (\bO_i) - \bI(\bX_i)\T \widehat{\bT}_\EIF \big\}^2 \ind \{ M(\bX_i) = g \}  }{ \big[ \frac{1}{N} \sum_{i=1}^N \ind \{ M(\bX_i) = g \} \big]^2 } \ .
\label{eq-consistentVarNP}
\end{align}
\end{algorithmic}
\caption{A Nonparametric Approach for Groupwise Treatment Effects}
\label{al2}
\end{algorithm}

\subsection{A Combined Estimator $\widehat{\tau}_{{\rm W},g}$} \label{sec:weighted}
Suppose an investigator wants to combine the estimators from the semiparametric approach and the nonparametric approach. While the motivation for such an estimator may seem odd at first, especially since the nonparametric approach makes fewer assumptions than the semiparametric approach in terms of modeling assumptions, we show in Section \ref{sec:Property-W} how this combined estimator can achieve better performance than either estimator alone in some settings. Formally, for each group $g$, consider the weighted combination of the nonparametric and the semiparametric estimators
where the weight is the value that minimizes the estimated variance of the combined estimator $\widehat{\tau}_{{\rm W},g}$, i.e.,
\begin{align} \label{eq3-linW}
& \widehat{\tau}_{{\rm W},g} = \widehat{w}_g \widehat{\tau}_{\SSLS,g} + (1-\widehat{w}_g) \widehat{\tau}_{\EIF,g} \ , 
\\
\nonumber
& \widehat{w}_g = \argmin_{w \in [0,1]} \widehat{\VAR} \Big\{  w \widehat{\tau}_{\SSLS,g} + (1-w) \widehat{\tau}_{\EIF,g} \Big\}
= 	\bigg(
\frac{\widehat{\sigma}_{\EIF,g}^2 - \widehat{\sigma}_{\SSLS,\EIF,g}}{ \widehat{\sigma}_{\SSLS,g}^2 - 2 \widehat{\sigma}_{\SSLS,\EIF,g} + \widehat{\sigma}_{\EIF,g}^2} 
\bigg)_{[0,1]}
\ . 
\end{align}
Here, the function $(t)_{[0,1]} = t \ind ( 0 < t < 1 ) + \ind (t \geq 1)$ winsorizes $t \in \R$ to be between $0$ and $1$. The term $\widehat{\sigma}_{\SSLS,\EIF,g}$ is the estimator of the covariance between $\widehat{\tau}_{\SSLS,g}$ and $\widehat{\tau}_{\EIF,g}$ and has the form 
\begin{align} \label{eq-consistentCov}
&
\widehat{\sigma}_{\SSLS,\EIF,g}
\\
&
=
\frac{ \frac{1}{N} \sum_{k=1}^2 \sum_{i \in \mathcal{I}_k} \widehat{\epsilon}_i^{(-k)}  \big\{ A_i - \widehat{e}^{(-k)}( \bX_i) \big\}
\big\{ \widehat{\varphi}^{(-k)} (\bO_i) - \bI(\bX_i)\T \widehat{\bT}_\EIF \big\} \ind \{ M(\bX_i) = g \}  }{ \big[ \frac{1}{N}  \sum_{i =1}^N  \ind \{ M(\bX_i) = g \} \big] \big[ \frac{1}{N} \sum_{k=1}^{2} \sum_{i \in \mathcal{I}_{k}} \big\{ A_i - \widehat{e}^{(-k)}( \bX_i) \big\}^2 \ind \big\{ M(\bX_i) = g \big\}  \big] } \ .
\nonumber
\end{align}%
By construction, the estimated variance of $\widehat{\tau}_{{\rm  W},g}$ is always less than or equal to those of $\widehat{\tau}_{\SSLS,g}$ and $\widehat{\tau}_{\EIF,g}$. Also, from simple algebra, $\widehat{\tau}_{{\rm W},g}$ collapses to $\widehat{\tau}_{\SP}$ if $\widehat{\rho}_g \geq \widehat{\sigma}_{\SP,g}/\widehat{\sigma}_{\NP,g}$ where $\widehat{\rho}_g$ is the estimator of the correlation coefficient of the two estimators, i.e., $\widehat{\rho}_g = \widehat{\sigma}_{\SP,\NP,g}/\big( \widehat{\sigma}_{\SP,g} \widehat{\sigma}_{\NP,g} )$. Similarly, $\widehat{\tau}_{{\rm W},g}$ collapses to $\widehat{\tau}_{\NP}$ if $\widehat{\rho}_g \geq \widehat{\sigma}_{\NP,g}/\widehat{\sigma}_{\SP,g}$. Finally, we remark that $\widehat{\tau}_{{\rm W},g}$
is a special case of some existing estimators that combine two estimators with respect to the squared error loss and one of the two estimators is potentially biased while the other estimator is unbiased  \citep{JS1991, JS2005, JS2005_2, Rosenman2022}.  

\section{Statistical Properties of Estimators}							\label{sec:3}

\subsection{Nonparametric and Semiparametric Models}						\label{sec:Models} 
To characterize the properties of the estimators discussed above, we first define the nonparametric model 
$\model_{\NP}$ and the semiparametric model $\model_{\SP}$:
\begin{align}
& \model_{\NP} = \big\{ P \, \big| \, 
\text{$P$ is a regular law \citep[Chapter 3]{BKRW1998}} 
\big\} \ ,
\nonumber
\\
&
\model_{\SP}
=
\big\{
P \in \model_\NP \cond 
P \text{ satisfies }  \eqref{eq:modelsp} \text{ below}
\big\} \ ,
\nonumber \\
&
\EXP \big[ \{ A_i-e^*(\bX_i) \}^2 \{ \tau^*(\bX_i) - \tau_g^* \} \cond M(\bX_i) = g \big] = 0 \ , \ g=1,\ldots,G \ .
\label{eq:modelsp} \tag{SP} 
\end{align} 
\noindent The nonparametric model $\model_{\NP}$ does not restrict the distribution of the observed data whereas the semiparametric model $\model_{\SP}$ restricts the observed data with the moment condition \eqref{eq:modelsp}. While \eqref{eq:modelsp} may seem obscure at first, in the next section, we show that \eqref{eq:modelsp} is a necessary and sufficient condition for the semiparametric estimator $\widehat{\bT}_{\SP}$ to be CAN for the $\bT^*$. Also, in Section \ref{sec:Validation}, we present a falsification test for \eqref{eq:modelsp} with the observed data. Finally, there are two familiar sufficient (but not necessary) conditions for \eqref{eq:modelsp}, which are defined as
the following two models $\model_{\PH}$ and $\model_{\SE}$:
\begin{align}
& \model_{\PH}
=
\left\{ P \in \model_\NP \, \left|
\begin{array}{ll}
1.& \hspace*{-0.2cm} \eqref{model-PLM} \text{ holds; and }
\\ 
2.& \hspace*{-0.2cm} \text{$\EXP(\epsilon_i^2 \cond A_i, \bX_i) = \sigma_g^2$ for any $(A_i, \bX_i)$}
\\
& \hspace*{-0.2cm} \text{satisfying  $M(\bX_i) =g \in \{1,\ldots,G\} $}.
\end{array}	\right.
\right\},  
\label{model-PLM-Hom}
\\
&\model_{\SE} 
=
\left\{ P \in \model_\NP \, \left|
\begin{array}{ll}
e^*(\bX_i) = e_g^* \text{ for any }  \bX_i
\\ 
\text{satisfying } M(\bX_i) = g  \in \{1,\ldots,G\} 
\end{array}	\right.
\right\}.  
\label{model-Str-Exp}
\end{align}
In words, $\model_{\PH}$ consists of all homoskedastic partially linear outcome models in \eqref{model-PLM}. In fact, regardless of whether the error is homoskedastic, all partially linear outcome models in \eqref{model-PLM} satisfy \eqref{eq:modelsp} and thus, all partially linear outcome models are nested in the semiparametric model $\model_{\SP}$. However, due to its unique theoretical properties, we focus on homoskedastic partial linear models; see Sections \ref{sec:Property-SSLS}-\ref{sec:Property-W} for details. The other model $\model_{\SE}$ consists of data from a stratified randomized experiment where study units in group $g$ are randomly assigned to treatment with probability $e_g^*$. From straightforward algebra, one can show that any distribution in $\model_{\SE}$ automatically satisfies condition \eqref{eq:modelsp} by the experimental design, implying that $\model_{\SE}$ is nested in $\model_{\SP}$. Therefore, we have the relationships $\model_{\PH} \subset \{ P \in \model_\NP \, | \, \text{$P$ satisfies \eqref{model-PLM}} \} \subset \model_{\SP} \subset \model_{\NP}$ and $\model_{\SE} \subset \model_{\SP} \subset \model_{\NP}$, and all of these inclusions are strict.

We conclude by making the following moment assumptions which we use throughout the paper.
\begin{assumption}			\label{assp:NuisFt}
There exist constants $C_{\epsilon}$ and $C_{\mu}$ so that the variance of the error $\EXP (\epsilon_i^2 \cond A_i, \bX_i)$ and the outcome regression $\mu^*(A_i, \bX_i)$ satisfy $\EXP (\epsilon_i^2 \cond A_i, \bX_i) \leq C_{\epsilon}$ and $\mu^*(A_i, \bX_i) \in [-C_{\mu},C_{\mu}]$ for all $(A_i,\bX_i)$.
\end{assumption}
\noindent Assumption \ref{assp:NuisFt} implies that the outcome has a finite variance given $(A_i, \bX_i)$, and the assumption will automatically hold if the outcome is uniformly bounded.

\subsection{Properties of the Semiparametric Estimator}						\label{sec:Property-SSLS}
We first characterize properties of the semiparametric estimator $\widehat{\bT}_{\SSLS}$. We make the following assumption to make progress:
\begin{assumption}			\label{assp:SSLS}
The nuisance functions used in Algorithm \ref{al1} satisfy the following conditions for $k=1,2$:
\begin{enumerate}[topsep=0.05cm,itemsep=0.0cm,label=(\alph*),leftmargin=0.75cm,labelsep=0.25cm]
\item  \emph{Bounded Nuisance Functions}: There exist constants $C_{\widehat{\nu}}$ and $C_{\widehat{e}}$ so that $\widehat{\nu}^{(-k)} \big( \bX_i \big)$ and $\widehat{e}^{(-k)} \big( \bX_i \big)$ satisfy $\widehat{\nu}^{(-k)} \big( \bX_i \big) \in [- C_{\widehat{\nu}}, C_{\widehat{\nu}}]$ and $\widehat{e}^{(-k)} \big( \bX_i \big) \in [C_{\widehat{e}}, 1-C_{\widehat{e}}]$ for all $\bX_i$.
\item  \emph{Consistency of $\widehat{\nu}^{(-k)}$ and $\widehat{e}^{(-k)}$}: 
$\big\| \widehat{\nu}^{(-k)}(\bX_i) - \nu^*(\bX_i) \big\|_{P,2} = O_P (r_{\nu,N} )$ and $\big\| \widehat{e}^{(-k)}(\bX_i) - e^*(\bX_i) \big\|_{P,2} = O_P (r_{e,N} )$ where $r_{\nu,N} = o(1)$, $r_{e,N} = o(N^{-1/4})$ and $r_{\nu,N} r_{e,N} = o(N^{-1/2})$ as $N \to \infty$.
\end{enumerate}
\end{assumption}
\noindent Condition (a) in Assumption \ref{assp:SSLS} is a bounded moment condition and this condition is satisfied if the outcome is uniformly bounded. Condition (b) controls how fast the estimated functions converge and this condition can be satisfied by many data-adaptive supervised learning methods such as the Nadaraya--Watson kernel regression estimator \citep{Nadaraya1964, Watson1964}, penalized generalized linear models \citep{Bickel2009}, random forests \citep{Wager2016}, and the highly-adaptive lasso \citep{Benkeser2016} under mild conditions. Also, condition (b) matches Assumption 4.1 of \citet{victor2018}, which is used to prove the asymptotic normality of partialling-out/``Robinson-style'' estimators \citep{robinson1988}. Finally,  our condition (b) is weaker than conditions in Lemma 2 of \citet{NieWager2020} since our target estimand (i.e., a $G$-dimensional parameter) is statistically simpler than \citet{NieWager2020}'s target estimand (i.e., a function).

Theorem \ref{thm:GPLM1} describes the asymptotic properties of $\widehat{\bT}_{\SSLS}$ under the semiparametric model $\model_{\SP}$.
\begin{theorem} \label{thm:GPLM1} 
Suppose that Assumptions \ref{assp:CausalAssumption}, \ref{assp:NuisFt} and \ref{assp:SSLS} hold. If the observed data $P$ belongs to $\model_{\SP}$, the following results hold:
\begin{itemize}[topsep=0.05cm,itemsep=0.0cm,label=(\alph*),leftmargin=0.0cm,labelsep=0.25cm]
\item[] (a) The estimator $\widehat{\bT}_{\SSLS}$ is CAN for the groupwise effect $\bT^*$ with a diagonal covariance $\Sigma_\SSLS = {\rm diag}(\sigma_{\SSLS,1}^2,\ldots,\sigma_{\SSLS,G}^2)$: 
\begin{align*}
\sqrt{N}  \Big( \widehat{\bT}_{\SSLS} - \bT^* \Big) \stackrel{D}{\rightarrow} N \Big(0, \Sigma_\SSLS \Big)\ , \ 
\sigma_{\SSLS,g}^2 = \frac{ \EXP \big[ \epsilon_i^2 \big\{ A_i-e^*(\bX_i) \big\}^2 \ind \{ M(\bX_i)=g \} \big] }{ \big[ \EXP \big[ \big\{ A_i-e^*( \bX_i) \big\}^2 \ind \{ M(\bX_i)=g \} \big] \big]^2 } \ .
\end{align*}
Also, $\Sigma_{\SSLS}$  can be consistently estimated by the following plug-in estimator $\widehat{\Sigma}_{\SSLS} = {\rm diag} (\widehat{\sigma}_{\SP,1}^2,\ldots,\widehat{\sigma}_{\SP,G}^2)$ where $\widehat{\sigma}_{\SSLS,g}^2$ is given in Algorithm \ref{al1}.

\item[] (b) The semiparametric estimator $\widehat{\bT}_{\SP}$ is locally efficient for $\bT^*$ in the sense that the variance $\Sigma_{\SP}$ is equal to the the semiparametric efficiency bound under model $\model_{\SP}$ at model $\model_{\PH}$.
\end{itemize}

\end{theorem}	
\noindent Part (a) of Theorem \ref{thm:GPLM1} states that the semiparametric estimator $\widehat{\bT}_{\SP}$ is CAN for $\bT^*$ under the semiparametric model $\model_{\SP}$, and the asymptotic variance can be consistently estimated. Part (b) of Theorem \ref{thm:GPLM1} states that $\widehat{\bT}_{\SP}$ is locally efficient when the outcome follows a homoskedastic partially linear outcome model in $\model_{\PH}$. In this case, $\sigma_{\SP,g}^2$ reduces to $\sigma_{\SP,g}^2 = \sigma_g^2/ \EXP \big[ \big\{ A_i-e^*( \bX_i) \big\}^2 \ind \{ M(\bX_i)=g \} \big] $, where $\sigma_g^2$ is the variance of the error associated with the $g$th subgroup; see \eqref{model-PLM-Hom}. Also, $\sigma_{\SP,g}^2$ coincides with the semiparametric efficiency bound for $\tau_g^*$ under model  $\model_{\PH}$ \citep{robinson1988, chamberlain1992}. Critically, $\widehat{\bT}_{\SP}$ is not efficient in the larger model $\model_{\SP}$, say a heteroskedastic partially linear outcome model. This phenomenon is similar to the OLS estimator being efficient under a homoskedastic linear model, but being inefficient under a larger, linear model that includes homokedastic or heteroskedastic variance; see Section \ref{sec:Property-W} for additional discussions.

Theorem \ref{coro:GPLM1} shows that if the semiparametric model $\model_{\SP}$ in Theorem \ref{thm:GPLM1} does not holds,  the semiparametric estimator $\widehat{\bT}_{\SP}$ is no longer CAN for $\bT^*$. 
\begin{theorem}	\label{coro:GPLM1}
Suppose that Assumptions \ref{assp:CausalAssumption}, \ref{assp:NuisFt}, and \ref{assp:SSLS} hold. If $P$ does not belong to $\model_{\SP}$, i.e., $P \in \model_{\NP} \cap \model_{\SP}^C$, the estimator $\widehat{\bT}_{\SSLS}$ is CAN for the overlap-weighted groupwise treatment effect $\bT_{\OV}^* =
(\tau_{\OV,1}^*, \ldots, \tau_{\OV,G}^* )\T$ \citep{Crump2006, Crump2009}:
\begin{align*}
\sqrt{N}(\widehat{\bT}_{\SSLS} - \bT_{\OV}^*) \stackrel{D}{\rightarrow} N(0,\Sigma_{\SSLS}), \
\tau_{\OV,g}^* = \frac{\EXP \big[ e^*(\bX_i) \{ 1- e^*(\bX_i) \} \tau^*(\bX_i) \cond M(\bX_i) = g \big] }{\EXP \big[ e^*(\bX_i) \{ 1- e^*(\bX_i) \} \cond M(\bX_i) = g \big] } \ .
\end{align*}	
Also, the variance $\Sigma_{\SP}$ and its consistent estimator $\widehat{\Sigma}_{\SP}$ have the same form as Theorem \ref{thm:GPLM1}.
\end{theorem}
Theorem \ref{coro:GPLM1} states that, outside of the semiparametric model $\model_{\SP}$, the semiparametric estimator $\widehat{\bT}_{\SSLS}$ is still asymptotically normal, but converges to a version of the overlap-weighted treatment effect in \citet{Crump2006, Crump2009} stratified by subgroups.  
Also, by combining Theorems \ref{thm:GPLM1} and \ref{coro:GPLM1}, we obtain the following necessary and sufficient condition for $\widehat{\bT}_{\SSLS}$ to be CAN for the original target estimand $\bT^*$.
\begin{coro}    \label{coro:CANoverSP}
Suppose that Assumptions \ref{assp:CausalAssumption}, \ref{assp:NuisFt} and \ref{assp:SSLS} hold. Then, the semiparametric estimator $\widehat{\bT}_{\SP}$ is CAN for the groupwise effect $\bT^*$ if and only if $P$ belongs to $\model_{\SP}$.    
\end{coro} 

\subsection{Properties of the Nonparametric Estimator}							\label{sec:EIF-property}
In this section, we characterize properties of the estimator $\widehat{\bT}_{\EIF}$ derived under the nonparametric approach. Consider the following set of conditions.
\begin{assumption}			\label{assp:EIF}
The nuisance functions used in Algorithm \ref{al2} satisfy the following conditions for $k=1,2$:
\begin{enumerate}[topsep=0.05cm,itemsep=0.0cm,label=(\alph*),leftmargin=0.75cm,labelsep=0.25cm]
\item  \emph{Bounded Nuisance Functions}: There exist constants $C_{\widehat{\mu}}$ and $C_{\widehat{e}}$ so that $\widehat{\mu}^{(-k)} \big(a,\bX_i \big)$ and $\widehat{e}^{(-k)} \big( \bX_i \big)$ satisfy $\widehat{\mu}^{(-k)} \big( a,\bX_i \big) \in [- C_{\widehat{\mu}}, C_{\widehat{\mu}}]$ and $\widehat{e}^{(-k)} \big( \bX_i \big) \in [C_{\widehat{e}}, 1-C_{\widehat{e}}]$ for $a=0,1$ and all $\bX_i$.
\item  \emph{Consistency of $\widehat{\mu}^{(-k)}$ and $\widehat{e}^{(-k)}$}: $\big\| \widehat{\mu}^{(-k)}(a,\bX_i) - \mu^*(a, \bX_i) \big\|_{P,2} = O_P (r_{\mu,N} )$ for $a=0,1$ and $\big\| \widehat{e}^{(-k)}(\bX_i) - e^*(\bX_i) \big\|_{P,2} = O_P (r_{e,N} )$ where $r_{\mu,N} = o(1)$, $r_{e,N} = o(1)$ and $r_{\mu,N} r_{e,N} = o(N^{-1/2})$. 
\end{enumerate}
\end{assumption} 
\noindent Conditions (a) and (b) in Assumption \ref{assp:EIF} are nearly identical to those in Assumption \ref{assp:SSLS}, except for the differences in the nuisance functions. In particular, condition (b) controls how fast the estimated functions converge and matches Assumption 5.1 of \citet{victor2018}.  

Theorem \ref{thm:EIF} shows the asymptotic properties of $\widehat{\bT}_{\EIF}$ under model $\model_{\NP}$.
\begin{theorem} \label{thm:EIF} 
Suppose that Assumptions \ref{assp:CausalAssumption}, \ref{assp:NuisFt}, and \ref{assp:EIF} hold. Then, $\widehat{\bT}_\EIF$ is CAN for $\bT^*$ with a diagonal covariance $\Sigma_\EIF = {\rm diag}(\sigma_{\EIF,1}^2,\ldots,\sigma_{\EIF,G}^2)$, i.e.,
\begin{align*}
\sqrt{N}  \Big( \widehat{\bT}_{\EIF} - \bT^* \Big) \stackrel{D}{\rightarrow} N \Big(0, \Sigma_\EIF \Big)\ , \ 
\sigma_{\EIF,g}^2 = 
\frac{ \EXP \big[ \big\{ \varphi^* (\bO_i) - \bI(\bX_i)\T \bT^* \big\}^2 \ind \{ M(\bX_i) = g \} \big] }{ \big[ \Pr \{ M(\bX_i) = g \} \big]^2}
\ .
\end{align*}
Moreover, $\Sigma_\EIF$ is the semiparametric efficiency bound of $\bT^*$ under $\model_{\NP}$ and can be consistently estimated by a plug-in estimator $\widehat{\Sigma}_\EIF = {\rm diag}( \widehat{\sigma}_{\EIF,1}^2,\ldots, \widehat{\sigma}_{\EIF,G}^2)$ where $	\widehat{\sigma}_{\EIF,g}^2 $ is given in Algorithm \ref{al2}. 
\end{theorem}
Theorem \ref{thm:EIF} states that the nonparametric estimator $\widehat{\bT}_{\NP}$ is CAN for $\bT^*$ under the nonparametric model $\model_{\NP}$. In fact, the nonparametric estimator $\widehat{\bT}_{\NP}$ is CAN for $\bT^*$ under all model spaces considered in the paper (i.e., $\model_{\NP}$,  $\model_{\SP}$, $\model_{\PH}$, $\model_{\SE}$).  Also, there is a consistent estimator of its asymptotic variance $\widehat{\Sigma}_{\EIF}$ where the asymptotic variance is equal to the semiparametric efficiency bound for $\tau^*$ under $\model_{\NP}$. 

Now, under model $\model_{\SP}$, both the semiparametric estimator $\widehat{\bT}_{\SP}$ and the nonparametric estimator $\widehat{\bT}_{\NP}$ are CAN for $\bT^*$. But, the two estimators have different asymptotic variances $\Sigma_\SP$ and $\Sigma_\NP$, respectively, and both estimators do not achieve the semiparametric efficiency bound for $\bT^*$ under model $\model_{\SP}$, even though the nonparametric estimator achieves the semiparametric efficiency bound under $\model_{\NP}$. To understand why, consider the following analogy from a toy linear model where $W$ is the outcome variable, $(V_1,V_2)$ are the two regressors, the errors are homoskedastic, and our target estimand is the regression coefficient of $V_1$, denoted as $\beta_1$. A natural estimator for $\beta_1$ is an OLS estimator where we regress $W$ on the regressors $(V_1, V_2)$. Traditional regression theory informs that the OLS estimator is CAN for $\beta_1$ and is efficient; these statistical properties are conceptually similar to the results in Theorem \ref{thm:EIF}. But, now consider a submodel of the toy model where the regression coefficient for $V_2$ is zero. Then, the OLS estimator remains CAN for $\beta_1$, but is no longer efficient. Instead, a more efficient estimator of $\beta_1$ under the submodel is another OLS estimator where we regress $W$ on $V_1$ without employing $V_2$ as a regressor. In short, even though the OLS estimator is efficient under the original toy model, the estimator is no long efficient under the submodel of the toy model where $V_2$ has no effect on $W$. Sections D.1 
and D.2
of the Supplementary Materials provides additional details, notably showing that the efficient influence function for $\bT^*$ under model $\mathcal{M}_{\SP}$ is not equal the influence functions associated with $\widehat{\bT}_{\SP}$ and $\widehat{\bT}_{\NP}$.  

\subsection{Properties of the Combined Estimator}						\label{sec:Property-W}
Theorem \ref{thm:lin} shows that the combined estimator $\widehat{\bT}_{\rm W}$ is CAN for the groupwise effects $\bT^*$ under model $\model_{\SP}$.
\begin{theorem} 							\label{thm:lin}
Suppose that Assumptions 
\ref{assp:CausalAssumption}, \ref{assp:NuisFt}, \ref{assp:SSLS}, and \ref{assp:EIF} hold, and that $P$ belongs to $\model_{\SP}$. 	Then, the combined estimator $ \widehat{\bT}_{\rm W}$ is CAN with a diagonal covariance $\Sigma_{\rm W} = {\rm diag}(\sigma_{{\rm W},1}^2,\ldots,\sigma_{{\rm W},G}^2)$, i.e.,
\begin{align*}
&
\sqrt{N}  \big( \widehat{\bT}_{\rm W} - \bT^* \big) \stackrel{D}{\rightarrow} N \big(0, \Sigma_{\rm W} \big) 
\ , \\
&
\sigma_{{\rm W},g}^2 = w_g^2 \sigma_{\SSLS,g}^2 + 2w_g (1-w_g) \sigma_{\SSLS,\EIF,g} + (1-w_g)^2 \sigma_{\EIF,g}^2
\ , \ 
\end{align*}
where $w_g= \argmin_{w \in [0,1] } \big\{  w^2 \sigma_{\SSLS,g}^2 + 2w(1-w) \sigma_{\SSLS,\EIF,g} + (1-w)^{2} \sigma_{\EIF,g}^2 \big\} $. Furthermore, $\Sigma_{{\rm W}}$ can be consistently estimated by $\widehat{\Sigma}_{{\rm W}} = {\rm diag} ( \widehat{\sigma}_{{\rm W},1}^2,\ldots,\widehat{\sigma}_{{\rm W},G}^2)$; here,  $\widehat{\sigma}_{{\rm W},g}^2 = \widehat{w}_g^2 \widehat{\sigma}_{\SSLS,g}^2 + 2\widehat{w}_g (1-\widehat{w}_g) \widehat{\sigma}_{\SSLS,\EIF,g} + (1-\widehat{w}_g)^2 \widehat{\sigma}_{\EIF,g}^2$ and $\widehat{w}_g$ is given in \eqref{eq3-linW}.
\end{theorem}
Combining the previous results, under $\model_{\SP}$, we have three estimators $\widehat{\bT}_{\NP}$, $\widehat{\bT}_{\SP}$, and $\widehat{\bT}_{\rm W}$ that are CAN for $\bT^*$. Also, while the two estimators $\widehat{\bT}_{\NP}$ and $\widehat{\bT}_{\SP}$ are not efficient for $\bT^*$ under $\model_{\SP}$, by construction of the combined estimators, the estimated variance of  $\widehat{\bT}_{\rm W}$ is no more than those from $\widehat{\bT}_{\SP}$ and $\widehat{\bT}_{\NP}$. 
Specifically, from Section \ref{sec:weighted}, if the standard errors of the $\widehat{\bT}_{\NP}$ and $\widehat{\bT}_{\SP}$ are of similar magnitude, $\widehat{\bT}_{\rm W}$ will use the information from both estimators to obtain a more efficient estimator of $\bT^*$. However, if the standard error of one of the two estimators is much larger than that of the other estimator, the combined estimator will reduce to the estimator with the smaller variance estimate. 

To illustrate the phenomena, consider the following two examples. First, consider a data generating model where the positivity assumption \hyperlink{(A3)}{(A3)} is violated with non-negligible probability (i.e., propensity score $e^*(\bX_i)$ is near 0 or 1 for many study units) while the outcome is generated from the homoskedastic partially outcome linear model. This implies that this model belongs to $\model_{\PH}$ (and, thus, $\model_{\SP}$) and that both the semiparametric estimator and the nonparametric estimator are CAN for $\bT^*$. But, the nonparametric estimator's standard error is likely to be larger than that of the semiparametric estimator due to the violation of the positivity assumption and the combined estimator is likely equal to the semiparametric estimator. Second, consider a data generating model where the treatment is randomized with probability $e_g^* \neq 0.5$ for all $g$ and $\tau^*(\bX_i)$ varies within each group, i.e., $\VAR\{ \tau^*(\bX_i) \cond M(\bX_i) = g \big\} > 0$ for all $g$. The second data generating model belongs to $\model_{\SE}$ (and, thus, $\model_{\SP}$). Like the first example, the semiparametric and nonparametric estimators obtained from the second data generating model are CAN for $\bT^*$. Nonetheless, as discussed in Section 3 of \citet{Hahn1998}, the variance of the semiparametric estimator is strictly larger than that of the nonparametric estimator, suggesting that the combined estimator is likely equal to the nonparametric estimator. Of note, if $e_g^*$ is exactly equal to $0.5$ or $\VAR\{ \tau^*(\bX_i) \cond M(\bX_i) = g \big\} = 0$, both estimators have the same variance. In other words, regardless of whether the conditional variance of $\tau^*(\bX_i)$ is zero or non-zero, if the within-group propensity score is constant (i.e., under model $\model_{\SE}$), $\widehat{\bT}_{\NP}$ is at least as efficient as $\widehat{\bT}_{\SP}$.

Figure \ref{Figure-Models} provides a graphical summary of all the results in Section 3 along with a recommendation on which of the three estimators to use. First, if model $\model_{\SP}$ does not hold (i.e., Case 1 in Figure   \ref{Figure-Models}), we recommend $\widehat{\bT}_{\NP}$ as it is the only CAN estimator for $\bT^*$ among the three estimators discussed in the paper and, in the absence of any additional assumptions, is a semiparametric efficient estimator of $\bT^*$. Second, if model $\model_{\SP}$ holds (i.e., Cases 2-5 in Figure \ref{Figure-Models}), we recommend $\widehat{\bT}_{{\rm W}}$ as it is not only CAN for $\bT^*$, but also more efficient than the other two estimators $\widehat{\bT}_{\NP}$ and $\widehat{\bT}_{\SP}$; as mentioned before, none of the three estimators achieve the semiparametric efficiency bound under $\model_{\SP}$. Third, if model $\model_{\SE}$ holds (i.e., Case 3 in Figure \ref{Figure-Models}), we again recommend $\widehat{\bT}_{\rm W}$ for the same reason as before; we remark that $\widehat{\bT}_{\NP}$ is at least as efficient as $\widehat{\bT}_{\SP}$ as discussed in the second example of the previous paragraph. Fourth, if model $\model_{\PH}$ holds, (i.e., Case 4 in Figure \ref{Figure-Models}), we recommend $\widehat{\bT}_{\rm W}$ and $\widehat{\bT}_{\SP}$; they are asymptotically equivalent, CAN for $\bT^*$, and efficient in the sense that they attain the semiparametric efficiency bound for $\bT^*$ derived under model $\model_{\SP}$ (i.e., part (b) of Theorem \ref{thm:GPLM1}). Finally, if model $\model_{\PH}$ and $\model_{\SE}$ jointly hold, the three estimators are asymptotically identical and efficient, i.e., again, they attain the semiparametric efficiency bound for $\bT^*$ derived under model $\model_{\SP}$. Therefore, we recommend any one of them. For additional explanations and derivations, see Section C.7 
of the Supplementary Material.

\begin{figure}[!htp]
\centering
\scalebox{0.7}{\begin{tikzpicture}
\begin{scope}[fill opacity = 1,text opacity=1]
\draw[draw = black, even odd rule, line width=0.25mm] (-3, -3) rectangle (5, 3);
\node[fill = white] at (4.6, 3) (M0) {$\mathcal{M}_{\NP}$};   
\draw[draw = black,even odd rule, line width=0.25mm] (1,0) ellipse (3.5cm and 2.6cm);
\node[fill = white] at (3.5, 2) (M1) {$\mathcal{M}_{\SP}$};   
\draw[draw = black,even odd rule, line width=0.25mm] (-0,0) ellipse (2.0cm and 1.5cm);
\node[fill = white] at (-0.5, 1.5) (M2) {$\mathcal{M}_{\SE}$};   
\draw[draw = black,even odd rule, line width=0.25mm] ( 2,0) ellipse (2.0cm and 1.5cm);
\node[fill = white] at ( 2.5, 1.5) (M3) {$\mathcal{M}_{\PH}$};   
\node[fill = none] at (-2.25,2.5) (C1) {Case 1};
\node[fill = none] at (1,2.1) (C2) {Case 2};
\node[fill = none] at (-1,0.25) (C3) {Case 3};
\node[fill = none] at (3,0.25) (C4) {Case 4};
\node[fill = none] at (1,-0.25) (C5) {Case 5};

\draw[draw = none, even odd rule] (8, 1.5+0.5) rectangle (5.1, 3+0.5)
node[below right=1ex] { 
$ \displaystyle{ \begin{array}{l}
\bullet  \ 
\text{Consistent and asymptotically normal (CAN) for $\bT^*$}
\\[0.25cm]
\makebox[1.7cm][l]{\text{Case 1: }}
\makebox[4.5cm][l]{\text{Only $\widehat{\bT}_{\NP}$ is CAN for $\bT^*$}}
\\
\makebox[1.7cm][l]{\text{Cases 2-5: }}
\makebox[4.5cm][l]{\text{$\widehat{\bT}_{\NP},\widehat{\bT}_{\SP},\widehat{\bT}_{{\rm W}}$ are CAN for $\bT^*$}}
\end{array} }$
};    

\draw[draw = none, even odd rule] (8, -1.25+0.25) rectangle (5.1, 1.25+0.25)
node[below right=1ex] { 
$ \displaystyle{ \begin{array}{l}
\bullet  \ 
\text{Asymptotic variances under cases 2-5}
\\[0.25cm]
\makebox[1.7cm][l]{\text{Case 2: }}
\makebox[4.5cm][l]{\text{$\AVAR(\widehat{\tau}_{{\rm W},g}) \leq \AVAR(\widehat{\tau}_{\SP,g}), \ \AVAR(\widehat{\tau}_{\NP,g})$}}
\\
\makebox[1.7cm][l]{\text{Case 3: }}
\makebox[4.5cm][l]{\text{$\AVAR(\widehat{\tau}_{{\rm W},g}) \leq \AVAR(\widehat{\tau}_{\NP,g}) \leq \AVAR(\widehat{\tau}_{\SP,g})$}}
\\
\makebox[1.7cm][l]{\text{Case 4: }}
\makebox[4.5cm][l]{\text{$\AVAR(\widehat{\tau}_{{\rm W},g}) = \AVAR(\widehat{\tau}_{\SP,g}) \leq \AVAR(\widehat{\tau}_{\NP,g})$}}
\\
\makebox[1.7cm][l]{\text{Case 5: }}
\makebox[4.5cm][l]{\text{$\AVAR(\widehat{\tau}_{{\rm W},g}) = \AVAR(\widehat{\tau}_{\SP,g}) = \AVAR(\widehat{\tau}_{\NP,g})$}}
\\
\makebox[6cm][l]{\text{Here, $\AVAR(\widehat{\tau}_g)$ is the asymptotic variance of $\widehat{\tau}_g$.}}
\end{array} }$
};
\draw[draw = none, even odd rule] (8, -3-0.4) rectangle (5.1, -1.65)
node[below right=1ex] { 
$ \displaystyle{ \begin{array}{l}
\bullet  \ 
\text{Recommended estimator} 
\\[0.25cm]
\makebox[1.7cm][l]{\text{Case 1: }}
\makebox[2cm][l]{\text{$\widehat{\bT}_{\NP} $}}
\makebox[1.7cm][l]{\text{Cases 2,\,3: }}
\makebox[2cm][l]{\text{$\widehat{\bT}_{\rm W} $}}
\\
\makebox[1.7cm][l]{\text{Case 4: }}
\makebox[2cm][l]{\text{$\widehat{\bT}_{{\rm W}}, \ \widehat{\bT}_{\SP}$ }}
\makebox[1.7cm][l]{\text{Case 5: }}
\makebox[2cm][l]{\text{$\widehat{\bT}_{{\rm W}}, \ \widehat{\bT}_{\SP}, \ \widehat{\bT}_{\NP}$}}
\end{array} } $
};   
\end{scope}
\end{tikzpicture}}
\caption{\small Graphical Visualization of Models and Recommended Estimators. The models $\model_{\PH}$ and $\model_{\SE}$ are defined in \eqref{model-PLM-Hom} and \eqref{model-Str-Exp}, respectively. Cases 1-5 refer to submodels of $\model_{\NP}$ as follows: Case 1$=\model_{\NP} \cap \model_{\SP}^C$; 
Case 2$= \model_{\SP} \cap \{ \model_{\SE} \cup \model_{\PH}\}^C$; \\
Case 3$= \model_{\SE} \cap \model_{\PH}^C$; 
Case 4$= \model_{\SE}^C \cap \model_{\PH}$;  
Case 5$= \model_{\SE} \cap \model_{\PH}$.	}
\label{Figure-Models}
\end{figure}

\section{Some Useful Tools When Using the Estimators}

\subsection{A Falsification Test for the Semiparametric Model}						\label{sec:Validation}
 
We propose a falsification test for condition \eqref{eq:modelsp} in model $\model_{\SP}$. To recap, Corollary \ref{coro:CANoverSP} showed that condition \eqref{eq:modelsp} is a sufficient and necessary condition for the semiparametric estimator $\widehat{\bT}_{\SP}$ to be CAN for the groupwise effect $\bT^*$. Consequently, knowing whether condition \eqref{eq:modelsp} holds or not would help researchers decide which estimator to use. 

The proposed test is, in spirit, related to the Durbin-Wu-Hausman (DWH) test in econometrics \citep{Durbin1954,Wu1973,Hausman1978} where we compare two different estimators to see if they yield similar estimates. Specifically, consider the following null hypothesis $H_{0g}: \tau_{\OV,g}^* = \tau_g^*$ for each $g$; note that by Corollary \ref{coro:CANoverSP}, the null hypothesis $H_{0g}$ would be true if and only if condition \eqref{eq:modelsp} holds for the $g$th subgroup. Under the null hypothesis $H_{0g}$, the difference between the semiparametric and nonparametric estimators is CAN and centered around zero, i.e., $
\sqrt{N}
\big( \widehat{\tau}_{\SP,g} - \widehat{\tau}_{\NP,g} \big)
\stackrel{D}{\rightarrow}
N \big( 0, \sigma_{\SP}^2 + \sigma_{\NP}^2 - 2 \sigma_{\SP,\NP} \big)$. Also, from equations \eqref{eq-consistentVarSP}, \eqref{eq-consistentVarNP}, and \eqref{eq-consistentCov}, we can consistently estimate the variance of this difference via the estimator $\widehat{\sigma}_{\SP}^2 + \widehat{\sigma}_{\NP}^2 - 2 \widehat{\sigma}_{\SP,\NP}$. Then, we can use a Wald test statistic of the form $Z_g^2$  which asymptotically follows a chi-square distribution with one degree of freedom under $H_{0g}$:
\begin{align*}
Z_g^2 =  
\frac{ 
N (
\widehat{\tau}_{\SP,g} - \widehat{\tau}_{\NP,g}
)^2
}{  \widehat{\sigma}_{\SP}^2 + \widehat{\sigma}_{\NP}^2 - 2 \widehat{\sigma}_{\SP,\NP}  }  \stackrel{D}{\rightarrow}
\chi_1^2 \ .
\end{align*}
If $Z_g^2$ is larger than $\chi_{1,1-\alpha}^2$ where $\chi_{1,1-\alpha}^2$ is the $1-\alpha$ percentile of a chi-square distribution with one degree of freedom, we would reject $H_{0g}$ and the condition \eqref{eq:modelsp} is rejected for that subgroup $g$; in other words, model $\model_{\SP}$ is not plausible for the observed data and we would use the nonparametric estimator for inferring the groupwise effect. However, if $Z_g^2$ is smaller than $\chi_{1,1-\alpha}^2$, we would retain the null and use the semiparametric or weighted estimator for inferring the groupwise effect. Similar to existing falsification tests for model specification, retaining the null does not mean that model $\model_{\SP}$ is true. But, it may suggest that using the semiparametric estimator or, more importantly, the combined estimator could be appropriate since the combined estimator's standard error is no worse than that of the nonparametric estimator (i.e., Cases 2-5 in Figure \ref{Figure-Models}).

\subsection{Cluster-Robust Variances}						\label{sec:CRV}
In some observational studies, study units may be grouped into non-overlapping clusters and exhibit correlation among units in the same clusters. These clusters could be the same as the subgroups for the groupwise effects or the subgroups could partially overlap with the clusters. When the observations are correlated within clusters,  the variance estimators we discussed above will often be smaller than the true variance and may lead to misleadingly narrow confidence intervals. To address this concern, we present cluster-robust variance estimators in \citet{Cameron2015} adapted to our setting.

Formally, let there be $C$ non-overlapping clusters with $N_c$ study units in each cluster $c \in \{1,\ldots, C\}$. For each cluster $c$, let $\mathcal{C}_c \subset \{1,\ldots,N\}$ be the subset of study units in cluster $c$. For both the semiparametric and the nonparametric estimators, we modify the cross-fitting procedures in Algorithms \ref{al1} and \ref{al2} so that study units in the same cluster belong to the same split sample, i.e., $\mathcal{C}_c \subset \mathcal{I}_1$ or $\mathcal{C}_c \subset \mathcal{I}_2$ for each cluster $c \in \{1,\ldots,C\}$. Then, using the generalized estimating equation theory \citep{GEE}, we can arrive at the following variance estimators for $\widehat{\bT}_{\SSLS}$ and $\widehat{\bT}_{\EIF}$ when there is potential concern for clustering:
\begin{align*}
& \widetilde{\rm Var}(\widehat{\bT}_{\SSLS}, \widetilde{\bT}_{\EIF}) 
= \widetilde{V}_{1}^{-1} \widetilde{V}_{2} \widetilde{V}_{1}^{-1}= \frac{1}{N} 
\bigg(
\begin{array}{ll}
\widetilde{\Sigma}_{\SSLS,\CR} & \widetilde{\Sigma}_{\SSLS,\EIF,\CR}
\\
\widetilde{\Sigma}_{\SSLS,\EIF,\CR}\T & \widetilde{\Sigma}_{\EIF,\CR}
\end{array}
\bigg)
\in \R^{2G \times 2G}, \\
& \widetilde{V}_{1} = 
\frac{1}{N}
\sum_{k=1}^{2} \sum_{i \in \mathcal{I}_{k}}
\begin{bmatrix}
\big\{ A_i - \widehat{e}^{(-k)}( \bX_i) \big\} \bI(\bX_i) 
\\
\bI(\bX_i)
\end{bmatrix}^{\otimes 2}
\in \R^{2G \times 2G},
\\
& \widetilde{V}_2 = 
\frac{1}{N}
\sum_{k=1}^2 \sum_{c: \mathcal{C}_c \subset \mathcal{I}_k }  
\begin{bmatrix}
\sum_{i \in \mathcal{C}_c}
\big[
\widehat{\epsilon}_{i}^{(-k)} \big\{ A_i - \widehat{e}^{(-k)}( \bX_i) \big\}	\bI(\bX_i)
\big]
\\
\sum_{i \in \mathcal{C}_c}
\big[
\big\{ \widehat{\varphi}^{(-k)} (\bO_i) - \bI(\bX_i)\T \widehat{\bT}_{\EIF} \big\}	\bI(\bX_i)
\big]
\end{bmatrix}^{\otimes 2}
\in \R^{2G \times 2G} \ .
\end{align*}
Unlike the original variance estimators without clustering, the cluster-robust variance estimators $ \widetilde{\Sigma}_{\SSLS}$, $\widetilde{\Sigma}_{\EIF}$, and $\widetilde{\Sigma}_{\SSLS,\EIF} $ are not diagonal matrices unless study units in the same cluster belongs to the same subgroup, i.e., $M(\bX_i) = M(\bX_j)$ for any $i,j \in \mathcal{C}_c$. This means that the estimators across different subgroup effects are no longer asymptotically independent from each other when study units are clustered. Also, similar to the combined estimator without clustering, we can use the cluster-robust variance estimators above to obtain the cluster-robust combined estimator of $\widehat{\bT}_\SSLS$ and $\widehat{\bT}_\EIF$.

\subsection{Simultaneous Inference} \label{sec:simult}
When studying groupwise effects, investigators often conduct multiple hypothesis tests across subgroups. Thankfully, once we have an asymptotically normal estimator of $\bT^*$, simultaneous testing is straightforward and we briefly illustrate this with the nonparametric estimator; the procedure for the semiparametric and the combined estimators are similar so long as the necessary and sufficient condition \eqref{eq:modelsp} holds.

Formally, for $\alpha \in (0,1)$ and subgroup $g$, consider the hypothesis $H_{0g} : \tau_{g}^* = \tau_{0g}$ versus $H_{1g} : \tau_{g}^* \neq \tau_{0g}$. Based on the asymptotic normality of $\widehat{\bT}_{\EIF}$, each $H_{0g}$ can be tested by the usual t-test with the test statistic $T_g = \sqrt{N} \big( \widehat{\tau}_{\EIF,g} - \tau_{0g} \big) /\widehat{\sigma}_{\EIF,g} $ where we reject $H_{0g}$ if $|T_g| > z_{1-\alpha/2}$. Also, we can obtain the usual $1-\alpha$ confidence interval of $\tau_g^*$ via  $\widehat{\tau}_{\EIF,g} \pm z_{1-\alpha/2} \widehat{\sigma}_{\EIF,g}/\sqrt{N}$.

Now, to test $G$ hypotheses $H_{01},\ldots,H_{0G}$ simultaneously, we can use a wide array of multiple testing procedures in the literature and we review one such procedure here. 
First, if there is no clustering and all study units are independent, $\widehat{\tau}_{\EIF,g}$ are asymptotically independent and follow a standard normal distribution across subgroups. In this case, we can use Sidak's correction \citep{Sidak1967} or the ``maxT'' method where a new, stringent critical value $q_{1-\alpha/2}$ based on the maximum of $G$ independent standard normals (i.e., $q_{1-\alpha/2} = z_{1- \{ 1- (1-\alpha)^{1/G} \}/2}$) is used. 
By rejecting $H_{0g}$ if $|T_g|$ exceeds $q_{1-\alpha/2}$, the familywise error rate (FWER) is less than or equal to $\alpha$. Relatedly, we can construct a simultaneous two-sided $1-\alpha$ confidence interval of $\tau_g^*$ by replacing the critical value $z_{1-\alpha/2}$ with $q_{1-\alpha/2}$, say $\widehat{\tau}_{\EIF,g} \pm q_{1-\alpha/2} \widehat{\sigma}_{\EIF,g}/\sqrt{N}$. Also, if $\widehat{\bT}_\EIF$ is normally distributed in finite samples and has known variance $\Sigma_\EIF$, Sidak's correction is optimal in the sense that it controls the FWER exactly at level $\alpha$ and is the least conservative simultaneous, bounded two-sided $1-\alpha$ confidence interval; see Section 7.1 of \citet{dunn1958}. 

Second, if there is clustering between study units where study units' data are correlated within clusters, we can use the same maxT statistic as before, but compute a different critical value $q_{1-\alpha/2}$ base on the maximum of $G$ correlated standard normal distributions; see Section 2.3 of \citet{Westfall1993}
for details. 
Note that this procedure may not have the same optimality guarantees as the non-clustered setup in the previous paragraph.

\section{Simulation}						\label{sec:simulation}	
\subsection{Setup}						\label{sec:simDGP}
We conduct simulation studies to study the finite sample performance of the three estimators $\widehat{\bT}_{\SSLS}$, $\widehat{\bT}_{\EIF}$, and $\widehat{\bT}_{\rm W}$. Our simulation scenario consists of $N = 2,000$ study units of $(Y_{i}, A_{i}, \bX_{i}) \in \reals \otimes \{0,1\} \otimes \reals^4$. We choose the cluster size $N_c$ and the number of clusters $C$ as either (i) $(N_c,C)=(1,2000)$ where there is no clustering or (ii) $(N_c,C)=(10,200)$ where study units are correlated within clusters. The covariate vector $\bX_{i}=(X_{i1}, X_{i2}, X_{i3}, X_{i4})$ consists of three continuous random variables, $X_{i1}$, $X_{i2}$, and $X_{i3}$, each from a standard normal distribution, and one binary variable, $X_{i4}$, from a Bernouilli distribution with $p = 0.5$. All four variables are mutually independent of each other and $X_{i3}$ are the same for all study units in the same cluster. From the covariates $\bX_{i}$, we define $G = 4$ mutually exclusive subgroups via  $M(\bX_{i1}) = \ind (X_{i1} \leq -1) + 2\ind (-1 < X_{i1} \leq 0) + 3\ind (0 < X_{i1} \leq 1 ) + 4 \ind (1 \leq X_{i1})$.

We consider the following propensity score models 
\begin{align}
& \text{(Constant, CPS)} 
&&
A_{i} \sim {\rm Ber} \big\{  \text{expit}(0.5 +  V_c) \big\}
\label{eq-sim-PS}
\\
& \text{(Varying, VPS)}
&&
A_{i} \sim {\rm Ber} \big\{ \text{expit} \big( 0.25 X_{i1} + 0.3 X_{i2}^2 - 0.3 | X_{i3} | X_{i4} + V_c \big) \big\}
\nonumber
\end{align}
where $\text{expit}(x) = \exp(x)/\{1+ \exp(x) \}$ and $V_c$ follows $N(0,\sigma_A^2)$ independently from the other variables. When $N_c = 1$, we set $\sigma_A=0$. On the other hand, when $N_c = 10$, we set $\sigma_A=5$ for the constant propensity score (CPS) model and $\sigma_A=2.5$ for the varying propensity score model (VPS). 
The constant propensity score model belongs to model $\model_{\SE}$ so that $\widehat{\bT}_{\SSLS}$ and the $\widehat{\bT}_{\rm W}$ are consistent for the groupwise effects, irrespective of the outcome model. In contrast, the varying propensity score model does not belong to $\model_{\SE}$ and $\widehat{\bT}_{\SSLS}$ and $\widehat{\bT}_{\rm W}$ can be consistent if they satisfy \eqref{eq:modelsp}.

We consider the outcome model $Y_i^{(a)} = \mu^*(0,\bX_i) + a \tau^*(\bX_i) + U_c + \epsilon_i$ where
\begin{align*}
&
\tau^* (\bX_{i})
=
1 + 
0.125 \beta \big[ \{ - 2X_{i1} - 1 \} \ind\{ X_{i 1} \leq -1 \} +
X_{i1}^2 \ind \{ -1 < X_{i1}  \} \big]
\\
&
\mu^*  (0,\bX_{i})
=
X_{i1} - 0.5 X_{i2}^2 + X_{i3} X_{i4}  \ .
\end{align*}
Here, $\beta \in \{0,1,2,3\}$ controls the effect heterogeneity within a subgroup where a larger $\beta$ leads to more variation in the treatment effect within each subgroup $g$. In other words, when $\beta = 0$, \eqref{model-PLM} holds, and when $\beta \neq 0$, \eqref{model-PLM} fails to hold. Also, $\epsilon_{i} \sim N(0,1)$ and $U_c \sim N(0,\sigma_Y^2)$ are the study unit and cluster-level random effects, respectively. When $N_c = 1$, we set $\sigma_Y=0$. When $N_c = 10$, we set $\sigma_Y=0.75$. In total, we consider 16 scenarios based on different combinations of $N_c$, propensity score models, and the effect heterogeneity parameter $\beta$. For each simulation scenario, we repeat the simulation $500$ times. 


To estimate the nuisance functions $\nu^*$, $\mu^*$, and $e^*$ inside the three estimators,  we use ensembles of machine learning methods via the superlearner algorithm \citep{SL2007, Polley2010}. Also, to alleviate the impact of a particular random split in cross-fitting, we use the median-adjustment of \citet{victor2018} with five cross-fitting repetitions; see Section A.2 
of the Supplementary Materials for details.  

Finally, for comparison, we use causal forests \citep{WA2018,GRF} implemented in the \texttt{grf} R-package \citep{grfpackage} to estimate the groupwise effects. Specifically, we use the \texttt{subset} and \texttt{clusters} options in the \texttt{average\_treatment\_effect} function to accommodate clustering; see \citet{grfvig} for additional details. We denote this estimator as GRF in the the results below.

\subsection{Results}

We only report the results associated with the first subgroup effect $\tau_1^*$, but the results associated with other subgroup effects are similar; see Section A.3 
of the Supplementary Material. We measure (a) bias (in $z$-score units), (b) ratio of standard errors where the denominator of this ratio is the standard error of the combined estimator $\widehat{\bT}_{\rm W}$, and (c) coverage of 95\% confidence intervals (CIs).  

\begin{figure}[!htb]
\centering
\includegraphics[width=\textwidth]{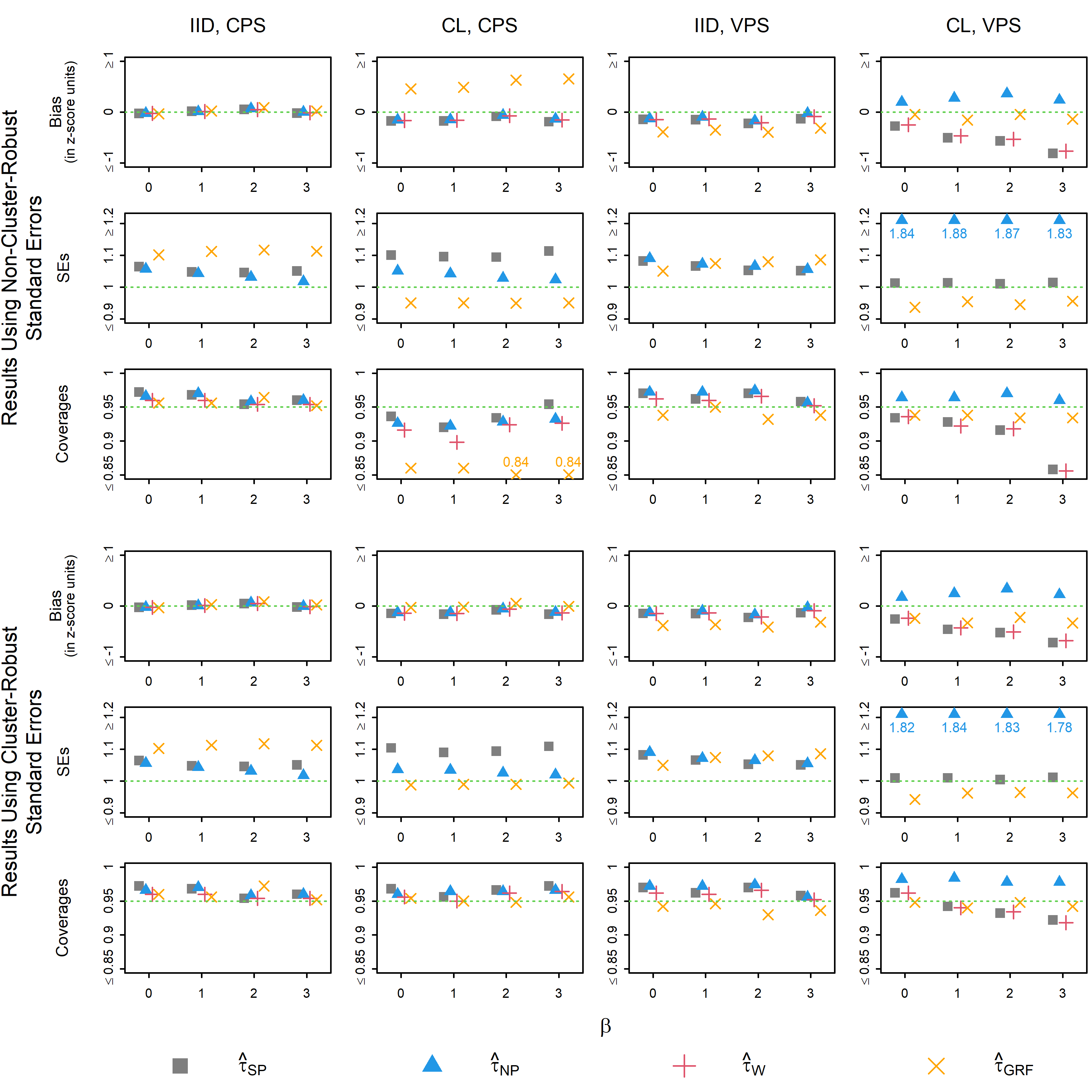}
\caption{\small 
A Graphical Summary of the Simulation Results  Associated with the First Groupwise Effect $\tau_1^*$. Each column shows different combination of propensity score models (CPS, VPS) in equation \eqref{eq-sim-PS} and whether the data is i.i.d. (IID) or clustered (CL). The top three rows use non-cluster-robust standard errors and the bottom three rows use cluster-robust standard errors.
The $x$-axes indicate the effect heterogeneity parameter $\beta$ with $\beta = 0$ satisfying model $\model_{\PH}$. The first and fourth rows show bias in z-score units. The second and fifth rows show ratio of standard errors where the denominator of the ratio is the standard error of the combined estimator $\widehat{\bT}_{\rm W}$. The third and sixth rows show coverage of 95\% CIs. Green dashed lines are drawn for reference.} 	
\label{fig-Simulation}

\end{figure}

Figure \ref{fig-Simulation} summarizes the result. In terms of bias, when $\beta=0$ so that $\tau^*(\bX_i)$ is constant within each subgroup and 
model $\model_{\PH}$ holds, all estimators presented in Section \ref{sec:2} have negligible bias, as expected from our theoretical results. Similarly, when the propensity score is constant (i.e., $\model_{\SE}$), all estimators have little to no bias, even if $\beta \neq 0$. However, when neither the model $\model_{\PH}$ nor $\model_{\SE}$ hold, the semiparametric estimator and the combined estimator show some bias. On the other hand, the nonparametric estimator  shows little to no bias in this setting. 

In terms of standard errors, as expected, the standard errors of the semiparametric estimator and the nonparametric estimator are always larger than the standard errors of the combined estimator across all scenarios, with the ratios of standard errors always exceeding $1$. This also implies that the corresponding confidence intervals from the semiparametric estimator and the nonparametric estimator are larger than that from the combined estimator. For example, when $\beta = 0$ and the true propensity score model is the VPS model, the average length of confidence interval from the nonparametric estimator is roughly 1.84 times longer than that from the combined estimator. In general, when condition \eqref{eq:modelsp} holds, 
inference based on the combined estimator is much more efficient than that based on the nonparametric or the semiparametric estimators.  

For coverage, so long as the three estimators are consistent, using non-cluster-robust standard errors produce nominal coverage when the study units are not clustered. On the other hand, when the units are clustered, the three estimators using non-clustered standard errors may fail to achieve nominal coverage. This is most noticeable in the constant propensity score setting with clustered data where using cluster-robust standard errors help all three estimators achieve closer to nominal coverage compared to not using cluster-robust standard errors. Also, we see that using clustered standard errors when the underlying data is actually independent does not hurt coverage.

\section{Applications: Early Childhood Longitudinal Study} \label{sec:data2}
\subsection{Background, Defining Subgroups, and Clustering}						\label{sec:databackground2}

We apply the methods discussed above to infer groupwise treatment effects of center-based pre-school programs on children's academic achievement from the Early Childhood Longitudinal Study's Kindergarten (ECLSK) Class of 1998-1999 dataset \citep{ECLSK2009}. 
Briefly, the ECLSK dataset consists of children's longitudinal histories from kindergarten through eighth grade in the United States. We consider that a child is treated $(A_i=1)$ if he/she received center-based care before entering kindergarten; otherwise, he/she is considered to be untreated ($A_i=0$), which is usually parental care. We consider the outcome $Y_i$ as the standardized reading score of each child, which was measured during the 1998 Fall semester and after treatment assignment. The outcome is continuously distributed over an interval of $ [22.62,93.95]$. As pre-treatment covariates, we include the following 8 variables in the analysis: census region (northeast/midwest/south/west), living location (central city/urban/rural), child's gender (male/female), age, race/ethnicity (white/black/Hispanic/Asian/other), family type (intact/non-intact), parental education (has college graduate or not), and economic status. We restrict our analysis to 15,980 children with complete data on the outcome, treatment, and pre-treatment covariates. Finally, we remark that our analysis differs from that in \citet{Lee2021} where they focused on students' math scores as the outcome and did not conduct a comparison between the three estimators presented in the paper.

Prior works \citep{ECLSK1, ECLSK2} conjectured that the effect of early child development interventions on children's academic and social skills are heterogeneous across region and location. Motivated from these works, we consider three subgroups defined by location to define groupwise effects. Also, since children's performances in the same kindergarten are likely correlated, we account for this correlation by using the cluster-robust variances introduced in Section \ref{sec:CRV} where kindergartens are chosen as clusters. Also, to estimate the three nuisance functions $\nu^*$, $\mu^*$, and $e^*$, we use the same procedure in Section \ref{sec:simDGP} except we repeat cross-fitting 100 times. 

\subsection{Results}
We first assessed covariate balance and overlap. For covariate balance, we compare the means of covariates in the treated and control groups before and after propensity score adjustment. We observe that covariate balance is dramatically improved after adjusting with the propensity score. We also visually found sufficient overlap between the treated and control groups; see Section A.4 
of the Supplementary Material for details. 

After checking covariate balance and overlap, we run the falsification test for the condition \eqref{eq:modelsp}. The falsification test statistics under $H_{0g}: \tau_{\OV,g}^*=\tau_g^*$ for $g=1,2,3$ are $(Z_1^2,Z_2^2,Z_3^2) = ( 0.592, 0.021, 0.522)$, which is significantly smaller than 3.84, the 95\% percentile of a chi-square distribution with one degree of freedom. Therefore, the result suggests that the model $\model_{\SP}$ is reasonable for all three subgroups.  

Next, we estimated the groupwise effects and Table \ref{tab:ECLSK} summarizes the results. Overall, the three proposed estimators show very similar effect estimates. In particular, estimates obtained from the semiparametric estimator and the combined estimator are the same in Central City and Rural subgroups with weights $\widehat{w}_{\text{Central City}}=\widehat{w}_{\text{Rural}}=1$. In the Urban subgroup, the estimate obtained from the combined estimator lies between the estimates obtained from the semiparametric estimator and the nonparametric estimator with a weight $\widehat{w}_{\text{Urban}}=0.359$. Also, as expected, the standard error of the combined estimator was no larger than the standard errors from the semiparametric estimator or the nonparametric estimator. Notably, in the Urban subgroup, the standard error of the combined estimator is strictly smaller than those of the semiparametric and nonparametric estimators. The GRF-based estimators also yield similar effect sizes across the three subgroups, but are associated with larger standard errors compared to the combined estimator. 

In terms of statistical significance at level $\alpha = 0.05$, the effect estimates from all the estimators are significant in the Urban and Central City subgroups whereas the effect estimates from all four estimators are insignificant in the Rural subgroup. The estimates remain statistically significant even after accounting for multiple testing across the three subgroups. Consequently, we conclude that the effect of center-based care before kindergarten on 1st year reading scores differs across living locations.

\begin{table}[!htp]
\renewcommand{\arraystretch}{1.2} \centering
\small
\setlength{\tabcolsep}{3pt}
\begin{tabular}{|c|c|c|c|c|c|}
\hline
Subgroup                           & Statistic      & $\widehat{\bT}_\SSLS$ & $\widehat{\bT}_\EIF$ & $\widehat{\bT}_{\rm W}$ & $\widehat{\bT}_{\rm GRF}$ \\ \hline
\multirow{4}{*}{\begin{tabular}[c]{@{}c@{}}Central\\ City\end{tabular}}  &   Estimate  &           2.119  &           2.236  &           2.119  &           2.047 \\ \cline{2-6}
&         SE  &           0.292  &           0.344  &           0.292  &           0.333 \\ \cline{2-6}
&   95\% CI  &   (1.547,2.691)  &   (1.561,2.912)  &   (1.547,2.691)  &   (1.395,2.700) \\ \cline{2-6}
&  95\% SCI  &   (1.422,2.816)  &   (1.414,3.059)  &   (1.422,2.816)  &   (1.253,2.842)      \\ \hline
\multirow{4}{*}{Urban}  &   Estimate  &           2.033  &           2.104  &           2.079  &           2.164 \\ \cline{2-6}
&         SE  &           0.325  &           0.321  &           0.319  &           0.370 \\ \cline{2-6}
&   95\% CI  &   (1.395,2.670)  &   (1.476,2.733)  &   (1.454,2.703)  &   (1.438,2.890) \\ \cline{2-6}
&  95\% SCI  &   (1.256,2.809)  &   (1.338,2.870)  &   (1.317,2.840)  &   (1.280,3.049)      \\ \hline
\multirow{4}{*}{Rural}  &   Estimate  &           0.591  &           0.606  &           0.591  &           0.585 \\ \cline{2-6}
&         SE  &           0.361  &           0.379  &           0.361  &           0.366 \\ \cline{2-6}
&   95\% CI  &  (-0.115,1.298)  &  (-0.137,1.349)  &  (-0.115,1.298)  &  (-0.132,1.303) \\ \cline{2-6}
&  95\% SCI  &  (-0.270,1.452)  &  (-0.300,1.511)  &  (-0.270,1.452)  &  (-0.289,1.459)      \\ \hline
\end{tabular}
\caption{\small Summary of the Data Analysis. Each row shows the statistics of interest for each subgroup. Each column shows the different estimators. SE, CI, and SCI stand for standard error, confidence interval, and simultaneous confidence interval, respectively. }
\label{tab:ECLSK}
\end{table}

\section{Conclusion} \label{sec:conclusion}
This paper compares two different approaches of estimating groupwise treatment effects $\tau_g^*$, the nonparametric approach and the semiparametric approach. We state the assumptions underlying each approach, compare their statistical properties, and present a combined approach that has favorable efficiency properties in some settings. We also present some useful tools while using the estimators discussed in the paper, notably a falsification test for the semiparametric model and cluster-robust variance estimators when the study units' data exhibit clustering. We demonstrate each approach through simulation and empirical studies. 

For practice, our work suggests using the combined estimator if the model $\model_{\SP}$ is satisfied by a study design (e.g., stratified experiment in $\model_{\SE}$) or is not severely violated based on the falsification test in Section \ref{sec:Validation}. In this case, the combined estimator is the most efficient among the three estimators considered here. But, if the semiparametric model $\model_{\SP}$ is not plausible, we recommend the nonparametric approach to estimate groupwise treatment effects. 

Lastly, we end the paper by clarifying the relationship between the partially linear outcome model \eqref{model-PLM} and the proposed semiparametric model $\model_{\SP}$. Efficient estimation of $\bT^*$ under the partially linear outcome model has been well-established for both homoskedastic and heteroskedastic error cases; see \citet{robinson1988}, \citet{chamberlain1992}, \citet{RMN1992}, \citet{Newey1994}, \citet{BZ1997}, \citet{Hardle2000}, \citet{Qi2000}, \citet{RR2001}, and \citet{MCW2006} for related discussions. However, we again highlight that the semiparametric model $\model_{\SP}$ is a strictly larger model than the partially linear model. Therefore, it is plausible that the efficient influence function for $\bT^*$ under the partially linear outcome model (allowing for heteroskedasticity) and that under the semiparametric model $\model_{\SP}$ can be different. For future research, it would be useful to derive the semiparametric efficiency bound for $\bT^*$ under $\model_{\SP}$ and construct an estimator of $\bT^*$ that attains this bound rather than rely on the combined estimator to obtain a relatively efficient estimator.

\newpage

\appendix

\section*{Supplementary Material}

	This document contains supplementary materials for ``A Groupwise Approach for Inferring Heterogeneous Treatment Effects in Causal Inference.'' 
	Section \ref{sup-detail:main} discusses the details about the results of the main paper. 
	Section \ref{sup-section:1} presents useful lemmas used in the proofs. 
	Section \ref{sup-section:3} contains the proofs of the lemmas in Section \ref{sup-section:1}.
	Lastly, Section \ref{sup-sec:main-thms} contains the proof of the theorems in the main paper.

\section{Details of the Main Paper} \label{sup-detail:main}

\subsection{A Visual Illustration of the Three Estimators}      \label{sec:supp:threeIFvisual}

We provide some rationales for why the combined estimator can perform better than the other two. 
The influence functions of the estimators $\widehat{\bT}_{\SP}$ and $\widehat{\bT}_{\NP}$ (denoted by $\bIF_{\SP}$ and $\bIF_{\NP}$, respectively) are valid in that the expectations of the products between these influence functions and the score function of the law are equal to the pathwise derivative of the groupwise effect, i.e.,
\begin{align}       \label{eq-diffpara}
    \frac{ \partial \bT(\eta) }{\partial \eta} \bigg|_{\eta=\eta^*}
    =
    \EXP \big\{ \bIF_{\SP} (\bO_i) \cdot s^*(\bO_i) \big\}
    =
    \EXP \big\{ \bIF_{\NP} (\bO_i) \cdot s^*(\bO_i) \big\} \ ,
\end{align}
where $\bT(\eta)$ is the groupwise effect under the parametric submodel of $\model_{\SP}$ parametrized by 1-dimensional parameter $\eta$ which recovers the true law at $\eta^*$ and $s^*(\bO_i)$ is the corresponding score function of the observed data $\bO_i=(Y_i,A_i,\bX_i)$ under model $\model_{\SP}$; see \ref{proof:SSLS} and \ref{proof:EIF} for details. Therefore, both $\widehat{\bT}_{\SP}$ and $\widehat{\bT}_{\NP}$ are CAN estimators for the groupwise effect under model $\model_{\SP}$. 

To become an efficient estimator under model $\model_{\SP}$, an influence function must belong to the tangent space of model $\model_{\SP}$ (denoted by $\mathcal{T}_{\SP}$; see \eqref{eq-tangentspace} for the exact form). Since the laws in the semiparametric model $\model_{\SP}$ must satisfy condition \eqref{eq:modelsp}, this imposes a restriction on the tangent space $\mathcal{T}_{\SP}$, indicating that $\mathcal{T}_{\SP}$ is not equal to the entire Hilbert space of $\bO$. From some algebra, we can show that the two influence functions $\bIF_{\SP}$ and $\bIF_{\NP}$ do not belong to $\mathcal{T}_{\SP}$ in general; again, see \ref{proof:SSLS} and \ref{proof:EIF} for details. This implies that there exist functions $\bm{r}_{\SP}$ and $\bm{r}_{\NP}$ that belong to the orthocomplement space of $\mathcal{T}_{\SP}$ (denoted by $\mathcal{T}_{\SP}^{\perp}$) so that
\begin{align*}
    &
    \bIF_{\SP}(\bO_i) = \bIF_{\SP}^{\EFF} (\bO_i) + \bm{r}_{\SP} (\bO_i)
    \quad , \quad
    \bIF_{\NP}(\bO_i) = \bIF_{\SP}^{\EFF} (\bO_i) + \bm{r}_{\NP} (\bO_i)
    \quad , \quad \bm{r}_{\SP} , \bm{r}_{\NP} \in \mathcal{T}_{\SP}^{\perp}
\end{align*}
where $\bIF_{\SP}^{\EFF}$ is the efficient influence function for $\bT^*$ under model $\model_{\SP}$. As a result, any linear combinations of the two influence functions $\bIF_{\SP}(\bO_i)$ and $\bIF_{\NP}(\bO_i)$ must have a form of
\begin{align*}
    \bIF_{{\rm W}} (\bO_i)
    : 
    & 
    \hspace*{-0.1cm}
    =
    W \cdot  \bIF_{\SP}(\bO_i)
    +
    (I - W)  \cdot \bIF_{\NP}(\bO_i)
    \\
    &
    =
    \bIF_{\SP}^{\EFF}(\bO_i)
    +
    \underbrace{ 
    W \cdot  \bm{r}_{\SP} (\bO_i)
    + (I - W) \cdot \bm{r}_{\NP} (\bO_i) 
    }_{=: \bm{r}_{{\rm W}}(\bO_i)}
\end{align*}
where $W = \text{diag}(w_1,\ldots,w_G)$ is a diagonal weight matrix. Here, the weighted residual function $r_{{\rm W}}(\bO_i)$ also belongs to the tangent space $\mathcal{T}_{\SP}$, and the weighted influence function $\bIF_{{\rm W}}$ also satisfies the differentiable parameter condition \eqref{eq-diffpara}, indicating that the estimator based on the weighted influence function (which is in fact the combined estimator $\widehat{\bT}_{{\rm W}}$) is also CAN for the groupwise effect $\bT^*$. 

\begin{figure}[!htb]
	\centering
	\includegraphics[width=0.4\textwidth]{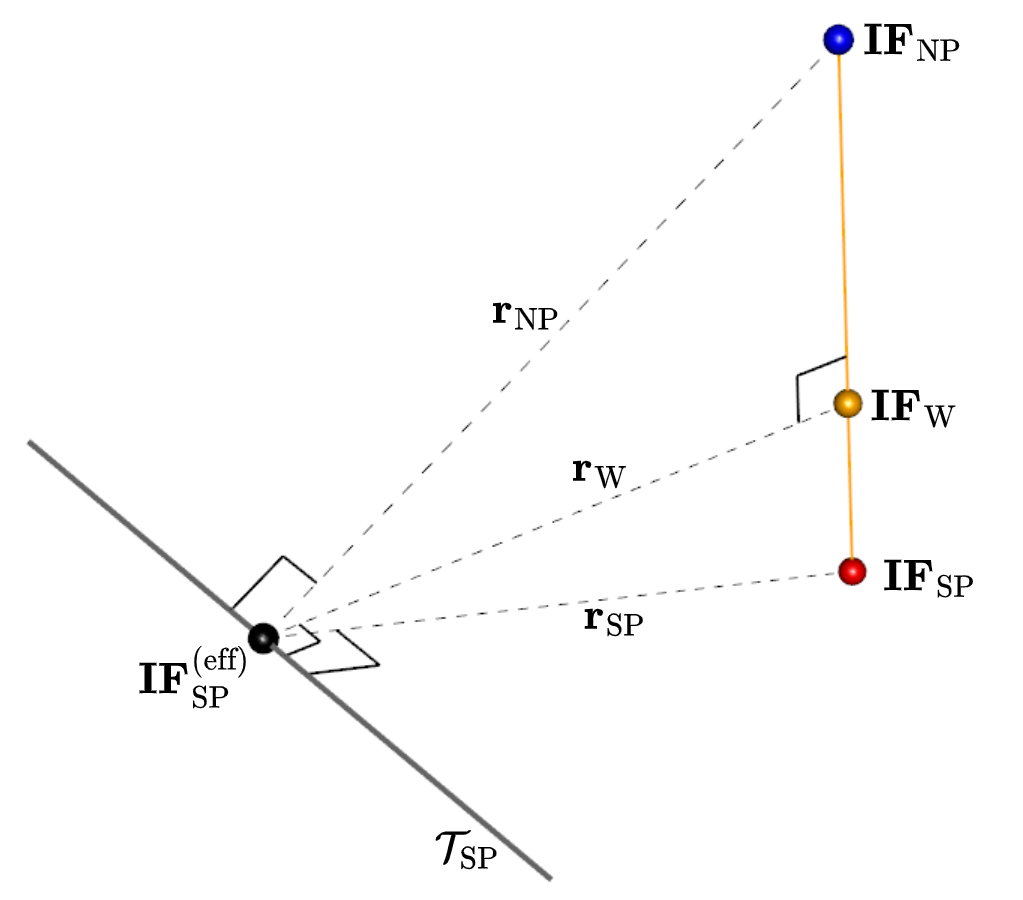}
	\caption{\small An Illustration of the Influence Functions. 
        Black, blue, red, and orange dots depict $\bIF_{\SP}^{\EFF}$, $\bIF_{\NP}$, $\bIF_{\SP}$, and $\bIF_{{\rm W}}$, respectively.
        The gray line depicts the tangent space $\mathcal{T}_{\SP}$.
        Three dashed lines depict the residual functions $\bm{r}_{\NP}$, $\bm{r}_{\SP}$, and $\bm{r}_{{\rm W}}$. The orange line is the collection of the weighted influence functions of the form $W \cdot  \bIF_{\SP} 
    + (I - W) \cdot \bIF_{\NP}$. Black corners (\scalebox{0.8}{$\overline{ {\color{white}{.A}} } \hspace*{-0.05cm} \big|$}) depicts that two segments are orthogonal.}
	\label{fig-VisualIll}
\end{figure}

Based on the results above, we find that three regular, asymptotic linear estimators $\widehat{\bT}_{\SP}$, 
 $\widehat{\bT}_{\NP}$, and $\widehat{\bT}_{{\rm W}}$) are CAN for $\bT^*$ with the corresponding influence functions $\bIF_{\SP}$, $\bIF_{\NP}$, and $\bIF_{{\rm W}}$, respectively. If the first two influence functions are associated with non-zero residual functions $\bm{r}_{\SP}$ and $\bm{r}_{\NP}$, none of the estimators is efficient. Consequently, if the weight matrix $W$ is chosen appropriately, the weighted residual $\bm{r}_{{\rm W}}$ may have a smaller variance than $\bm{r}_{\SP}$ and $\bm{r}_{\NP}$. In fact, the combined estimator $\widehat{\bT}_{{\rm W}}$ is constructed to make the variance of $\bm{r}_{{\rm W}}$ as small as possible. Figure \ref{fig-VisualIll} depicts how the weighted influence function (and the corresponding estimator) can be more efficient than the other two influence functions (and the corresponding estimators). 

\subsection{Details of the Superlearner Library and Cross-fitting Procedure} \label{sup-detail:CF}

We include the following methods and the corresponding R packages in our super learner library: linear regression via \texttt{glm}, lasso/elastic net via \texttt{glmnet} \citep{glmnet}, spline via \texttt{earth} \citep{earth} and \texttt{polspline} \citep{polspline}, generalized additive model via \texttt{gam} \citep{gam}, boosting  via \texttt{xgboost} \citep{xgboost} and \texttt{gbm} \citep{gbm}, random forest via \texttt{ranger} \citep{ranger}, and neural net via \texttt{RSNNS} \citep{RSNNS}.

Let $\widehat{\bT}_\SSLS^{(s)}$ and  $\widehat{\bT}_\EIF^{(s)}$ be the estimators from $s$th cross-fitting procedure and let $\widehat{\Sigma}_\SSLS^{(s)}$, $\widehat{\Sigma}_\EIF^{(s)}$, and $\widehat{\Sigma}_{\SSLS,\EIF}^{(s)}$ be the associated variance estimates. Afterwards, we compute the component-wise medians of the estimators, i.e.
\begin{align*}
	\begin{pmatrix}
	\widehat{\bT}_\SSLS^{\text{med}} \\
	\widehat{\bT}_\EIF^{\text{med}} 
	\end{pmatrix}
	=
	\median_{s=1,\ldots,S}
	\begin{pmatrix}
	\widehat{\bT}_\SSLS^{(s)} \\
	\widehat{\bT}_\EIF^{(s)}
	\end{pmatrix}
	\ .
\end{align*}
Also, the following variance estimators are used:
\begin{align*}
	\begin{bmatrix}
		\widehat{\Sigma}_\SSLS^{\text{med}} & \widehat{\Sigma}_{\SSLS,\EIF}^{\text{med}}
		\\
		\widehat{\Sigma}_{\SSLS,\EIF}^{\text{med}, \intercal} & \widehat{\Sigma}_\EIF^{\text{med}}
	\end{bmatrix}
	=
	\median_{s=1,\ldots,S}
	\left[
		\begin{bmatrix}
		\widehat{\Sigma}_\SSLS^{(s)} & \widehat{\Sigma}_{\SSLS,\EIF}^{(s)}
		\\
		\widehat{\Sigma}_{\SSLS,\EIF}^{(s), \intercal} & \widehat{\Sigma}_\EIF^{(s)}
	\end{bmatrix}
	+
	\begin{pmatrix}
	\widehat{\bT}_\SSLS^{(s)} - \widehat{\bT}_\SSLS^{\text{med}} \\
	\widehat{\bT}_\EIF^{(s)} - \widehat{\bT}_\EIF^{\text{med}} 
	\end{pmatrix}^{\otimes 2}
	\right]
\end{align*}
where the median is evaluated based on the matrix 2-norm. As shown in Theorem 3.3 of \citet{victor2018}, the established results extend to the median-adjusted estimators.

\newpage

\subsection{Additional Results of the Simulation} \label{sup-detail:Simulation}

\begin{figure}[!htb]
	\centering
	\includegraphics[width=\textwidth]{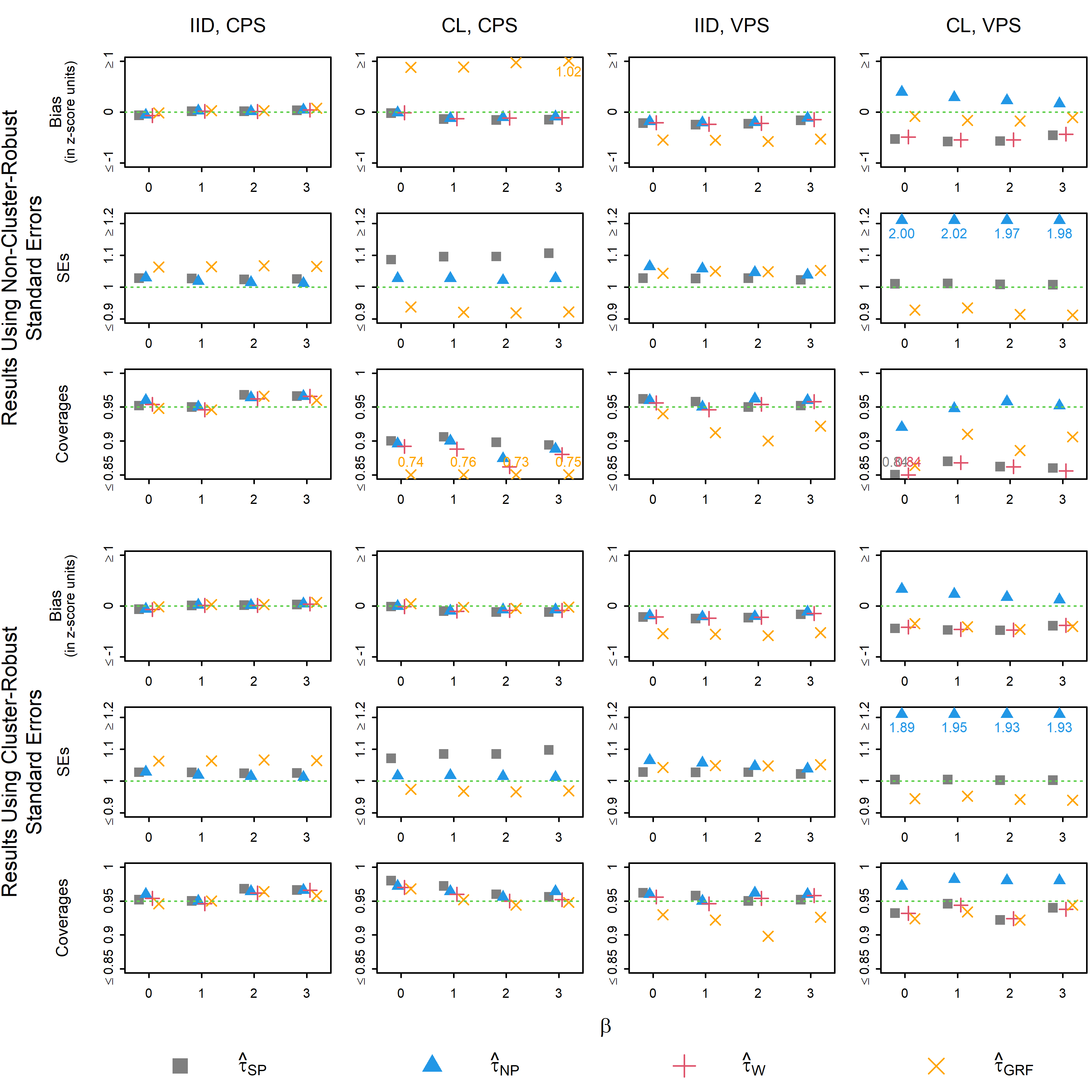}
	\caption{\small Graphical Summary of the Simulation Results Associated with the Second Groupwise Effect $\tau_2^*$. Each column shows a different combination of propensity score (PS) models in \eqref{eq-sim-PS} in the main paper and clustering. The top three rows use non-cluster-robust standard errors and the bottom three rows use cluster-robust standard errors.
The $x$-axes of the plots indicate the effect heterogeneity parameter $\beta$. The first and fourth rows show bias in z-score units. The second and fifth rows show the ratio of standard errors where the denominator is the combined estimator's standard error. The third and sixth rows show coverage of 95\% CIs. Green dashed lines are drawn for reference.} 	
\end{figure}		

\newpage

\begin{figure}[!htb]
	\centering
	\includegraphics[width=\textwidth]{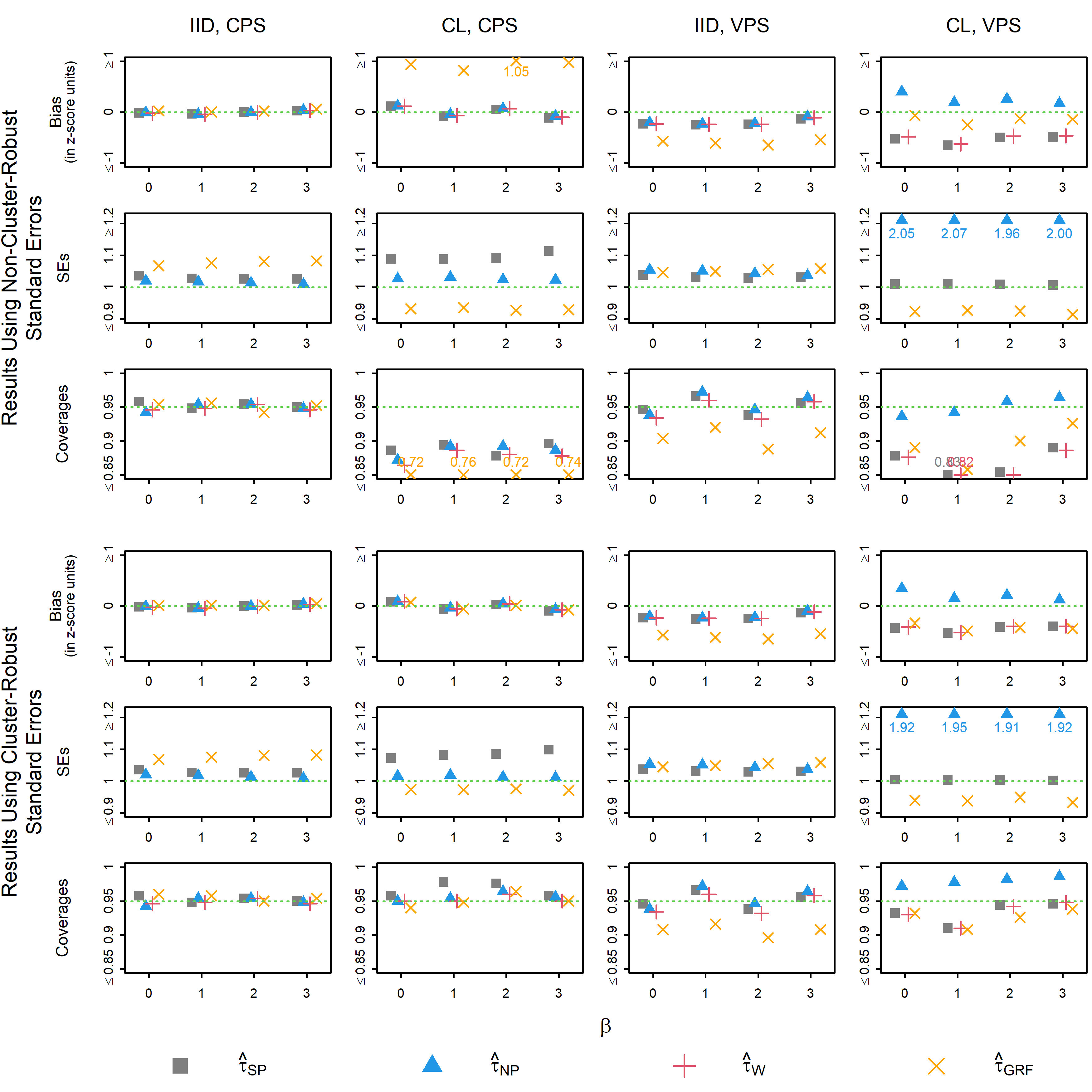}
	\caption{\small Graphical Summary of the Simulation Results Associated with the Third Groupwise Effect $\tau_3^*$. Each column shows a different combination of propensity score (PS) models in \eqref{eq-sim-PS} in the main paper and clustering. The top three rows use non-cluster-robust standard errors and the bottom three rows use cluster-robust standard errors.
The $x$-axes of the plots indicate the effect heterogeneity parameter $\beta$. The first and fourth rows show bias in z-score units. The second and fifth rows show the ratio of standard errors where the denominator is the combined estimator's standard error. The third and sixth rows show coverage of 95\% CIs. Green dashed lines are drawn for reference.} 	
\end{figure}	

\newpage	

\begin{figure}[!htb]
	\centering
	\includegraphics[width=\textwidth]{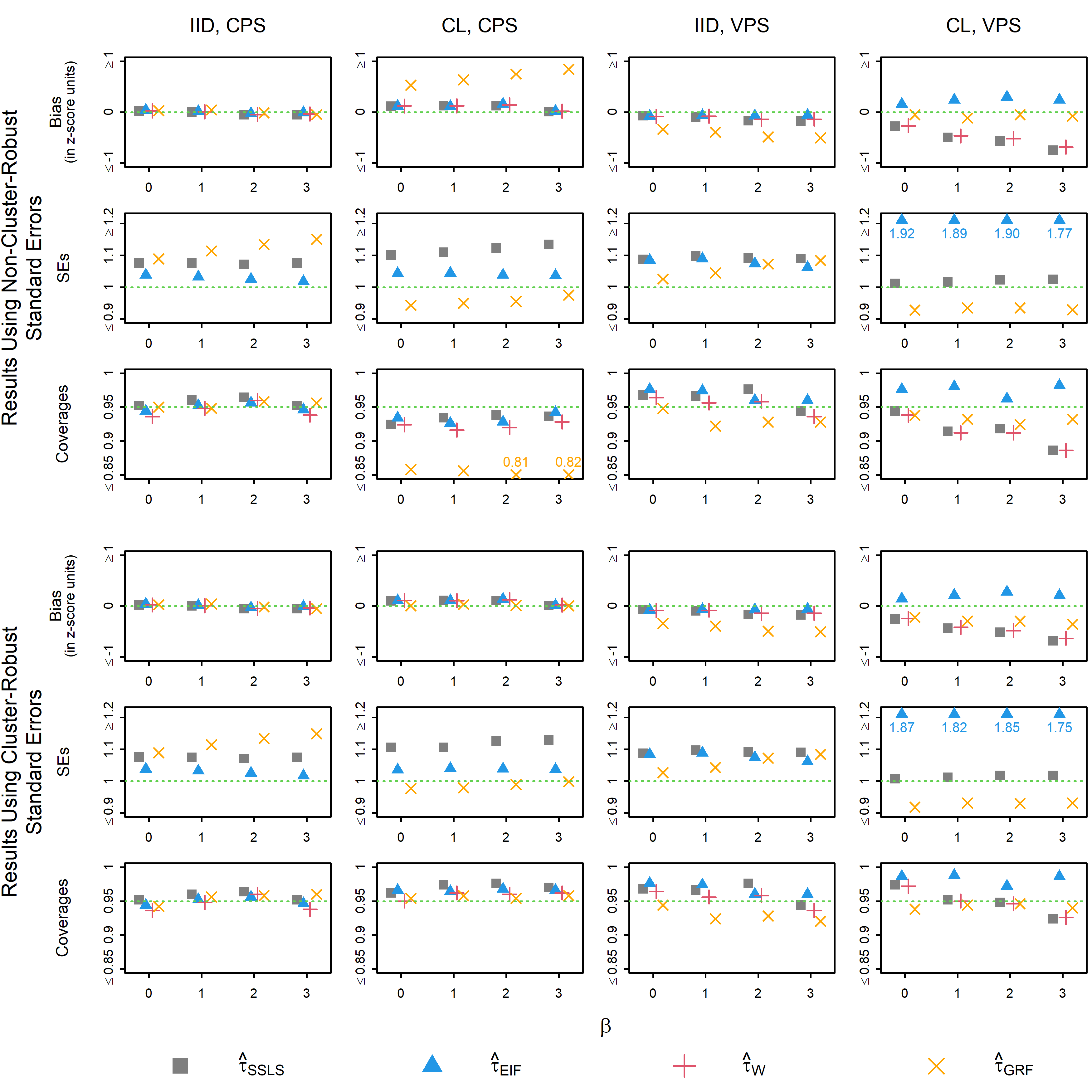}
	\caption{\small Graphical Summary of the Simulation Results Associated with the Fourth Groupwise Effect $\tau_4^*$. Each column shows a different combination of propensity score (PS) models in \eqref{eq-sim-PS} in the main paper and clustering. The top three rows use non-cluster-robust standard errors and the bottom three rows use cluster-robust standard errors.
The $x$-axes of the plots indicate the effect heterogeneity parameter $\beta$. The first and fourth rows show bias in z-score units. The second and fifth rows show the ratio of standard errors where the denominator is the combined estimator's standard error. The third and sixth rows show coverage of 95\% CIs. Green dashed lines are drawn for reference.} 	
\end{figure}		

\newpage

\subsection{Assessment of Assumptions (A2) and (A3) in the Main Paper}		\label{sup-detail:assumption}

To assess assumption (A2), we assessed covariate balance as follows. Let $\widetilde{X}_i^{{\rm PS.nonAdj}}$ and $\widetilde{X}_i^{{\rm PS.Adj}}$ be
\begin{align*}
	\widetilde{X}_{ip}^{{\rm PS.nonAdj}}
	=
	\frac{A_i X_{ip}}{\frac{1}{N} \sum_{i=1}^N A_i}
	-
	\frac{(1-A_i) X_{ip}}{\frac{1}{N} \sum_{i=1}^N (1-A_i)}
	\ , \
	\widetilde{X}_{ip}^{{\rm PS.Adj}}
	=
	\frac{A_i X_{ip}}{\widehat{e}^{\rm Med}(\bX_i)}
	-
	\frac{(1-A_i) X_{ip}}{1-\widehat{e}^{\rm Med}(\bX_i)} \ ,
\end{align*}
where $X_{ip}$ is the $p$th covariate of unit $i$ and $\widehat{e}^{\rm Med}(\bX_i)$ is the median value of the propensity score estimate obtained from 100 sample split. To address the correlation within each cluster, we consider the following generalized linear mixed effect models (GLMMs):
\begin{align*}
	\widetilde{X}_{c(j), p}^{{\rm PS.nonAdj}} \sim \mu_p^{{\rm PS.nonAdj}} + u_{cp}^{{\rm PS.nonAdj}} + \epsilon_{c(j), p}^{{\rm PS.nonAdj}} \ , \
	\widetilde{X}_{c(j), p}^{{\rm PS.Adj}} \sim \mu_p^{{\rm PS.Adj}} + u_{cp}^{{\rm PS.Adj}} + \epsilon_{c(j), p}^{{\rm PS.Adj}}
\end{align*}
where $u_{cp}$ is the cluster-level random effect for cluster $c$ and $\epsilon_{c(j), p}$ is the unit-level error for the $j$th unit in cluster $c$. Using the GLMMs, we test $H_{0p}^{{\rm PS.nonAdj}}: \mu_p^{{\rm PS.nonAdj}}  = 0$ and $H_{0p}^{{\rm PS.Adj}}: \mu_p^{{\rm PS.nonAdj}} = 0$. For $H_{0p}^{{\rm PS.nonAdj}}$, a larger (smaller) test statistic suggests that covariate balance is achieved (violated) without adjusting the propensity score. Similarly, for $H_{0p}^{{\rm PS.Adj}}$, a larger (smaller) test statistic suggests that covariate balance is achieved (violated) with adjusting the propensity score. 

Table \ref{tab:AssessA1} summarizes covariate balance assessment. We find that covariate balance is dramatically improved after adjusting the propensity score and there is no significant Wald statistics that rejects $H_{0p}^{{\rm PS.Adj}}$ other than socioeconomic status. This concludes assumption (A2) is not severely violated.

\begin{table}[!htp]
		\renewcommand{\arraystretch}{1.25} \centering
		\footnotesize
		\setlength{\tabcolsep}{3pt}
\begin{tabular}{|c|c|c|c|c|c|}
\hline
Variable                    & $H_{0p}^{{\rm PS.nonAdj}}$ & $H_{0p}^{{\rm PS.Adj}}$ & Variable                                                                                             & $H_{0p}^{{\rm PS.nonAdj}}$ & $H_{0p}^{{\rm PS.Adj}}$ \\ \hline
 Census Region = Northeast  &  -0.71  &  -1.07  &                           Race = Asian  &  -2.81  &  -1.89  \\ \hline 
     Census Region = South  &   1.88  &   0.46  &                           Race = Black  &   6.47  &   0.98  \\ \hline 
      Census Region = West  &  -4.87  &  -1.53  &                           Race = White  &   2.11  &   0.75  \\ \hline 
           Location = City  &   -0.8  &  -0.52  &                        Race = Hispanic  &  -7.42  &  -1.13  \\ \hline 
          Location = Rural  &  -3.62  &  -1.46  &                        Family = Intact  &  -1.76  &  -0.46  \\ \hline 
             Gender = Male  &  -0.59  &  -0.29  &  Parental Education = College Graduate  &   7.26  &   0.74  \\ \hline 
                       Age  &   2.25  &   0.98  &                   Socioeconomic Status  &   16.5  &   2.59  \\ \hline 
\end{tabular}
\caption{Result of Covariate Balance Assessment. The numbers show the Wald statistics obtained from testing $H_{0p}^{{\rm PS.nonAdj}}$ and $H_{0p}^{{\rm PS.Adj}}$.}
\label{tab:AssessA1}
\end{table}

\begin{figure}[!htb]
	\centering
	\includegraphics[width=1\textwidth]{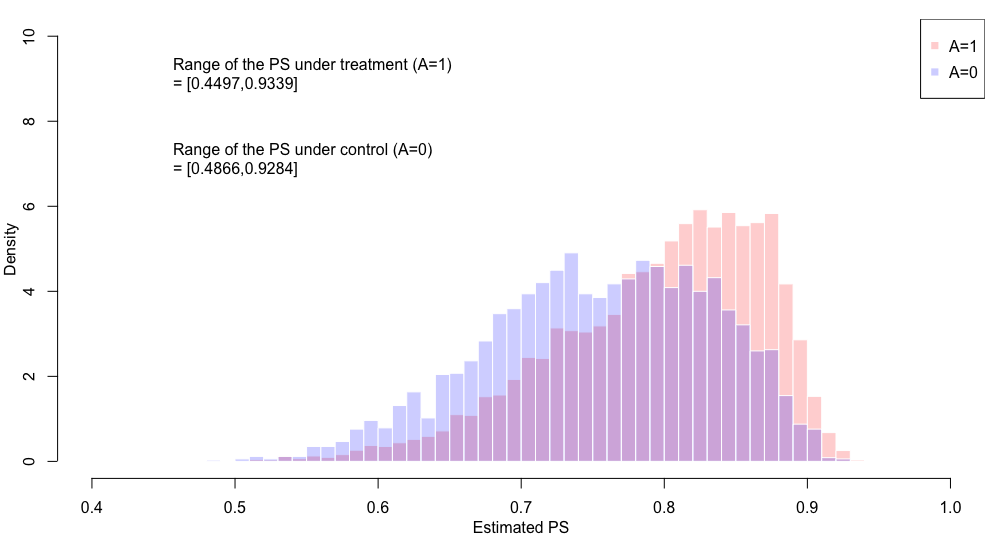}
	\caption{Histograms of the Propensity Score Estimates.}
	\label{fig:Overlap}
\end{figure}

Next, to assess assumption (A3), we plot $\widehat{e}^{\rm Med}(\bX_i)$ according to the treatment status. Figure \ref{fig:Overlap} shows the result and the overlap assumption is not severely violated.

\newpage

\section{Lemma} \label{sup-section:1}

We introduce useful lemmas for the proof of theorems in the main paper.

	\begin{lemma}			\label{lem1-001}
		Let $\mathbf{U}$ and $\mathbf{V}$ be the random vectors. Then, 
		\begin{align*}
			\big\|E (\mathbf{U}\mathbf{V}^\intercal ) \big\|_2 \leq \EXP \big(\| \mathbf{U}\mathbf{V}^\intercal \|_2 \big) \leq \sqrt{ \EXP \big(\| \mathbf{U} \|_2^{2} \big) \EXP \big(\| \mathbf{V} \|_2^{2} \big) }  \ .
		\end{align*}
				Furthermore, suppose $\mathbf{U}$ is a bounded random vector, i.e., $\norm{\mathbf{U}}_2\leq M$ for some $M$. Then, 
		\begin{align*}
			\EXP \big(\norm{\mathbf{U}\mathbf{V}^\intercal}_2 \big) \leq M \cdot \EXP \big(\norm{\mathbf{V}}_2 \big) \ .
		\end{align*}
		Also, for $r\geq 1$, we have
		\begin{align*}
			\EXP \big(\norm{\mathbf{U}\mathbf{V}^\intercal}_2^r \big) \leq  \sqrt{ \EXP \big(\| \mathbf{U} \|_2^{2r} \big) \EXP \big(\| \mathbf{V} \|_2^{2r} \big) } \ .
		\end{align*}

	\end{lemma}
	
	\begin{proof}
		The proof is in Section \ref{proof-Lem-1}.
	\end{proof}

	\begin{lemma}		\citep{victor2018} 	\label{lem1-002}
			Let $a_N$  be a sequence of positive numbers for $N = 1,2,\ldots$. If $\norm{\mathbf{X}_N} = O_P(a_N)$ conditional on $\mathbf{Y}_N$, then $\norm{\mathbf{X}_N} = O_P(a_N)$ unconditionally.
	\end{lemma}
	
		\begin{proof}
		The proof is in Section \ref{proof-Lem-2}.
	\end{proof}

%

	\begin{lemma}  	\label{lem-assumption-SSLS}
	Let $\bV_i^* = \{ A_i - e^*(\bX_i) \} \bI(\bX_i) $, $Z_i^* = Y_i - \nu^*(\bX_i) $, $\widehat{\bV}_i\kk = \{ A_i - e\kk (\bX_i) \} \bI(\bX_i) $, $\widehat{Z}_i\kk =  Y_i - \nu\kk(\bX_i) $, $\xi_i = Z_i^* - \bV_i\sT \bT_\OV^* $, and $\xi_i\kk = \widehat{Z}_i\kk - \widehat{\bV}_i\kT \bT_\OV^* $. Under Assumptions (A1)-(A4) and Assumptions \ref{assp:NuisFt} and \ref{assp:SSLS} in the main paper, the following conditions hold.	
		\begin{enumerate}[label=(\alph*),leftmargin=1cm]
			\item $\EXP ( \bV_i^{\sT} \xi_i )=0$.		
			\item $ \big\| \bV_i^* \big\|_{P,4}$, $\big\| \widehat{\bV}_i\kk \big\|_{P,4}$, $\big\| \bV_i\sT \xi_i \big\|_{P,2}$, and $\EXP \big( \xi_i \xi_i\T  \cond \bV_i^* \big)$ are bounded.
			\item $\EXP \big( \bV_i^* \bV_i\sT \big)$ and $\EXP \big\{ \bV_i\kk \bV_i\kT \cond \mathcal{I}_k^C \big\}$ are full rank.
			\item $\| \widehat{Z}_i\kk - Z_i^* \|_{P,2} = o_P(1) $ and $ \| \widehat{\bV}_i\kk - \bV_i^*  \|_{P,2}  = o_P(1)$ as $N \rightarrow \infty$. 
			\item  $\sqrt{N} \, \big\| \EXP \big\{ 
				\widehat{\bV}_i\kT \xi_i\kk \, \big| \, \mathcal{I}_k^C
			\big\} \big\|_{2}  = o_P(1)$  and 
			 $ \big\| \widehat{\bV}_i\kT \xi_i\kk - \bV_i\sT \xi_i \big\|_{P,2}  = o_P(1) $ as $N \to \infty$.  
		\end{enumerate}
		\end{lemma}

		\begin{proof}
		The proof is in Section \ref{proof-Lem-4}.
	\end{proof}

	\begin{lemma}		\label{lem-EIF}
		Let $\mathcal{M} = \big\{ P \cond P(\bO_i) \text{ satisfies Assumptions (A1)-(A4) in the main maper}  \}$. The efficient influence function of $\bT^*$ in $\mathcal{M}$ is $\bm{\phi}(\bO_i) = \big( \phi_1(\bO_i), \ldots, \phi_G(\bO_i) \big)\T$ where
		\begin{align*}
			&
			\phi_g (\bO_i)
			=
			\frac{ \ind \big\{ M(\bX_i) = g \big\} \big\{
				\varphi(\bO_i) - \tau_g^*
			\big\} }{p_g^*}
			\ , \ 
			p_g^* = P \big\{ M(\bX_i) = g \big\} \ , \\
			&
			\varphi(\bO_i)
			=
			\frac{A_i \big\{ Y_i - \mu^*(1,\bX_i) \big\} }{e^* (\bX_i)}
				-
				\frac{(1-A_i) \big\{ Y_i - \mu^*(0,\bX_i) \big\} }{1-e^* (\bX_i)}
				+
				\mu^*(1,\bX_i) - \mu^*(0,\bX_i) \ .
		\end{align*}
		Therefore, the semiparametric efficiency bound of $\bT^*$ is $\Sigma_\EIF = {\rm diag}(\sigma_{\EIF,1}^2,\ldots,\sigma_{\EIF,G}^2)$ where 
		\begin{align*}
			\sigma_{\EIF,g}^2 = 
		\frac{ \EXP \big[ \big\{ \varphi(\bO_i) - \tau_g^* \big\}^2 \ind \{ M(\bX_i) = g \} \big] }{ ( p_g^* )^2}
		\end{align*}
	Moreover, the estimated slope coefficients from regressing $\varphi(\bO_i)$ on $\bI(\bX_i)$ without the intercept term achieve the bound.

	\end{lemma}

		\begin{proof}
		The proof is in Section \ref{sec:proof:lem-EIF}.
	\end{proof}

	\begin{lemma}  	\label{lem-assumption-EIF}
	Let $\bV_i^* = \widehat{\bV}_i\kk = \bI(\bX_i) $, $Z_i^* = \varphi^* (\bO_i) $, $\widehat{Z}_i\kk =  \widehat{\varphi}\kk (\bO_i) $, $\xi_i=\epsilon_i = Z_i^* - \bV_i\sT \bT^* $, and $\xi_i\kk = \epsilon_i\kk = \widehat{Z}_i\kk - \widehat{\bV}_i\kT \bT^* $. Under Assumptions (A1)-(A4) and Assumptions \ref{assp:NuisFt} and \ref{assp:EIF} in the main paper, the conditions in Lemma \ref{lem-assumption-SSLS} hold.

\end{lemma}

		\begin{proof}
		The proof is in Section \ref{proof-Lem-6}.
	\end{proof}

	\begin{lemma}  	\label{lem-assumption-Joint}
	Let us denote
	\begin{align*}
	&
	\bV_i^* 
	=
	 \begin{bmatrix}
	 \bV_{\SP,i}^* \\
	 \bV_{\NP,i}^* 
	 \end{bmatrix}
	=
	\begin{bmatrix} \{ A_i -e^*(\bX_i)  \bI(\bX_i) \\ \bI(\bX_i) 	\end{bmatrix}	 
	&&
	\widehat{\bV}_i\kk 
	=
	\begin{bmatrix}
	\widehat{\bV}_{\SP,i}\kk 
	\\
	\widehat{\bV}_{\NP,i}\kk 
	\end{bmatrix}
	= \begin{bmatrix} \{ A_i - \widehat{e}\kk(\bX_i)  \bI(\bX_i) \\ \bI(\bX_i) 	\end{bmatrix}	 
	\\
	&
	\bZ_i^* 
	=
	\begin{bmatrix}
	Z_{\SP,i}^*
	\\
	Z_{\NP,i}^*
	\end{bmatrix}
	= \begin{bmatrix} Y_i - \nu^*(\bX_i) \\ \varphi^* (\bO_i)  \end{bmatrix} 
	&& \widehat{\bZ}_i\kk 
	=
	\begin{bmatrix}
	\widehat{Z}_{\SP,i}\kk 
	\\
	\widehat{Z}_{\NP,i}\kk 
	\end{bmatrix}
	= \begin{bmatrix} Y_i - \widehat{\nu}\kk(\bX_i) \\ \widehat{\varphi}\kk (\bO_i)  \end{bmatrix} 
	\\
	&
	\bxi_i = 
	\begin{bmatrix}
		\xi_{\SP,i} 
		\\
		\xi_{\NP,i} 
	\end{bmatrix}
	=
	\begin{bmatrix}
		 {Z}_{\SP,i}^* - {\bV}_{\SP,i}\sT \bT_{\OV}^*
		 \\
		 {Z}_{\SP,i}^* - {\bV}_{\SP,i}\sT \bT^*
	\end{bmatrix}
	&&
	\widehat{\bxi}_i\kk
	=
	\begin{bmatrix}
		\widehat{\xi}_{\SP,i} \kk
		\\
		\widehat{\xi}_{\NP,i} \kk 
	\end{bmatrix}
	=
	\begin{bmatrix}
		 \widehat{Z}_{\SP,i}\kk - \widehat{\bV}_{\SP,i}\kT \bT_{\OV}^*
		 \\
		 \widehat{Z}_{\SP,i}\kk - \widehat{\bV}_{\SP,i}\kT \bT^*
	\end{bmatrix}
\end{align*}	
	Under Assumptions (A1)-(A4) and Assumptions \ref{assp:NuisFt}, \ref{assp:SSLS}, and \ref{assp:EIF} in the main paper, the conditions in Lemma \ref{lem-assumption-SSLS} hold.

\end{lemma}

		\begin{proof}
		The proof is in Section \ref{proof-Lem-assumption-Joint}.
	\end{proof}

	\begin{lemma} 						\label{lem-comparisionSSLSEIF}
		Let $\Sigma_\SSLS$, $\Sigma_\EIF$, and $\Sigma_{\SSLS,\EIF}$ be the asymptotic variances of $\widehat{\bT}_\SSLS$, $\widehat{\bT}_\EIF$, and the asymptotic covariance of $\widehat{\bT}_\SSLS$ and $\widehat{\bT}_\EIF$, respectively. Suppose model \eqref{model-PLM} in the main paper is true and the error $\epsilon_i$ is homoscedastic within each subgroup (i.e., $\VAR(\epsilon_i \cond A_i, \bX_i) = \sigma_{\epsilon,g}^2$ for all $(A_i,\bX_i)$ satisfying $M(\bX_i) = g$).  Then, $\Sigma_\EIF - \Sigma_\SSLS$ is positive semi-definite and $\Sigma_{\SSLS,\EIF} = \Sigma_\SSLS$. Moreover, if the propensity score $e^*(\bX_i)$ is constant within each subgroup (i.e., condition (b) of Theorem \ref{coro:GPLM1} in the main paper), we have $\Sigma_\EIF = \Sigma_\SSLS$.
	
	\end{lemma}

		\begin{proof}
		The proof is in Section \ref{sec:proof:lem-comparisionSSLSEIF}.
	\end{proof}

\newpage

\section{Proof of Lemmas in Section \ref{sup-section:1}} \label{sup-sec:supp-lemmas} 		\label{sup-section:3}

\subsection{Proof of Lemma \ref{lem1-001}}					\label{proof-Lem-1}

		For the first result, we observe that the matrix spectral norm is convex and induced by the vector $2$-norm. As a result, the Jensen's inequality, the submultiplicavity of the spectral norm, and H\"older's inequality gives the result
		\begin{align}								\label{pf:lemma}
			 \bigg\| \int \mathbf{U}\mathbf{V}^\intercal dP \bigg\|_2 & \leq  \int \norm{ \mathbf{U}\mathbf{V}^\intercal}_2 \, dP   
			 \leq \int \norm{\mathbf{U}}_2 \norm{\mathbf{V}}_2 \, dP  
			 \leq \sqrt{\int \norm{\mathbf{U}}_2^2 \, dP \int \norm{\mathbf{V}}_2^2 \, dP }  \ .
		\end{align}
		Here, $P$ is the law of $(\mathbf{U},\mathbf{V})$. The second result  is trivial by replacing $\| \mathbf{U} \|_2$ with $M$ in the third integral of \eqref{pf:lemma}. The last result holds via analogous steps:
		\begin{align*}
			\int \norm{ \mathbf{U} \mathbf{V}^\intercal}_2^r \, dP \leq \int \norm{\mathbf{U}}_2^r \norm{\mathbf{V}}_2^r \, dP \leq \sqrt{ \int \norm{\mathbf{U}}_2^{2r} dP \int \norm{\mathbf{V}}_2^{2r} \, dP }  \ .
		\end{align*}

\subsection{Proof of Lemma \ref{lem1-002}}				\label{proof-Lem-2}

See Lemma 6.1 of \citet{victor2018}.

\subsection{Proof of Lemma \ref{lem-assumption-SSLS}}						\label{proof-Lem-4}
	
We proof the claim of the lemma in following \textbf{Step 1} -- \textbf{Step 6}.   \\

\noindent \textbf{Step 1}: We find
\begin{align*}
	\EXP \big[ \epsilon_i \{ A_i - e^*(\bX_i) \} \bI(\bX_i) \big]
	=
	\EXP \big[ \bV_i^* \big\{ Z_i^* -  \bV_i\sT \bT_\OV^* \big\} \big]
	=
	\EXP ( \bV_i^*  Z_i^* ) - \EXP  ( \bV_i^* \bV_i\sT ) \bT_\OV^* \ .
\end{align*}
The $g$th component of $\EXP ( \bV_i^*  Z_i^* )$ is,
\begin{align*}
	&
	\EXP \big[ \{Y_i - \nu^*(\bX_i) \} \{ A_i - e^*(\bX_i) \} \ind \{ M(\bX_i)  = g \} \big]
	\\
	&
	=
	\EXP \big[  \{ A_i - e^*(\bX_i) \}^2 \ind \{ M(\bX_i)  = g \} \tau^*(\bX_i) 
	+
	\epsilon_i  \{ A_i - e^*(\bX_i) \} \ind \{ M(\bX_i)  = g \} \big]
	\\
	&
	=
	\EXP \big[ e^*(\bX_i) \{ 1 - e^*(\bX_i) \} \tau^*(\bX_i) \cond M(\bX_i)  = g  \big] \Pr \{ M(\bX_i) = g \}
	\ .
\end{align*}
Also, $ \EXP  ( \bV_i^* \bV_i\sT ) $ is a diagonal matrix of which $g$th component is $\EXP \big[  \{ A_i - e^*(\bX_i) \}^2 \ind \{ M(\bX_i)  = g \} \big] = \EXP \big[ e^*(\bX_i)\{ 1 - e^*(\bX_i) \} \cond  M(\bX_i)  = g \big] \Pr\{ M(\bX_i) = g \}$. Therefore, each component of $\EXP \big[ \xi_i \{ A_i - e^*(\bX_i) \} \bI(\bX_i) \big]$ is zero:
\begin{align*}
	&
		\Big[ \EXP \big[ e^*(\bX_i) \{ 1 - e^*(\bX_i) \} \tau^*(\bX_i) \cond M(\bX_i)  = g  \big] - \EXP \big[  e^*(\bX_i) \{ 1 - e^*(\bX_i) \} \cond  M(\bX_i)  = g \big]  \tau_{\OV,g}^* \Big] \Pr\{ M(\bX_i) = g \}
		=0 \ .
\end{align*}
The equality holds from the definition of $\bT_{\OV,g}^* = \EXP \big[ e^*(\bX_i) \{ 1- e^*(\bX_i) \} \tau^*(\bX_i) \cond M(\bX_i) = g \big] /\EXP \big[ e^*(\bX_i) \{ 1- e^*(\bX_i) \} \cond M(\bX_i) = g \big] $. 	\\
	
\noindent \textbf{Step 2} :  We show condition (b) holds. Since $A_i$ and $e^*(\bX_i)$ are trivially bounded,  we obtain
	\begin{align*}
		& \norm{\bV_i}_{P,4}^* \leq \norm{A_i}_{P,4} + \norm{ e^*(\bX_i) }_{P,4} \leq 2  \quad , \quad
		 \norm{\widehat{\bV}_i\kk }_{P,4} \leq \norm{A_i}_{P,4} + \norm{ \widehat{e}\kk (\bX_i) }_{P,4} \leq 2 \ .
	\end{align*}
	For $\norm{ \bV_i^* \xi_i }_{P,2}$, we observe:
	\begin{align*}
			&  \| \bV_i^* \xi_i \|_{P,2} \leq \| A_i \xi_i \|_{P,2}  + \| e^*(\bX_i) \xi_i \|_{P,2} \leq 2 \| \xi_i \|_{P,2} < \infty \ , 
		\end{align*}
		Therefore, it suffices to show that $ \| \xi_i \|_{P,2}$ is finite. We find
		\begin{align*}
			\xi_i
			=
			Z_i^* - \bV_i\sT \bT_\OV^*
			& =
			Y_i - \nu^*(\bX_i) - \big\{ A_i - e^*(\bX_i) \big\} \bI\T(\bX_i) \bT_{\OV}^*
			\\
			&
			=
			\underbrace{
			\mu^*(A_i,\bX_i) - \nu^*(\bX_i) - \big\{ A_i - e^*(\bX_i) \big\} \bI\T(\bX_i) \bT_{\OV}^*}_{\text{bounded}}
			+ \epsilon_i
		\end{align*}
		In Step 6, we establish that $\bT_\OV^*$ is bounded. Therefore, since the underbraced term and $\EXP ( \xi^2 \cond A_i, \bX_i)$ is bounded, we find $\EXP \big( \xi^2 \cond A_i,\bX_i) $ is bounded as well. Therefore, 
		\begin{align*}
			\| \xi_i \|_{P,2}^2
			=
			\EXP \big( \xi_i^2 \big)
			=
			\EXP \big\{ \EXP (\xi_i^2 \cond A_i, \bX_i) \big\}
			< \infty \ .
		\end{align*} 
	Similarly, we find $\EXP \big( \xi_i^2 \cond \bV_i^* \big)$ is bounded as follows:
	\begin{align*}
		\EXP \big( \xi_i^2 \cond \bV_i^* \big)
		=
		\EXP \big\{ \EXP( \xi_i^2 \cond \bV_i^*, A_i, \bX_i) \cond \bV_i^* \big\}
		=
		\EXP \big\{ \EXP( \xi_i^2 \cond A_i, \bX_i) \cond \bV_i^* \big\}
		< \infty \ .
	\end{align*} 
	
\noindent \textbf{Step 3} :  We establish condition (c). First, to show the full rank condition of $\EXP \big( \bV_i^* \bV_i\sT \big)$, we observe that $\EXP \big( \bV_i^* \bV_i\sT \big) $ is a diagonal matrix with the $g$th diagonal entry $\EXP \big[ \big\{A_i-e^*(\bX_i)\big\}^2 \ind \big\{ M(\bX_i)=g \big\} \big]$. As a result, it suffices to show that every diagonal entry is bounded between two positive constants. Note that $g$th diagonal entry is
	\begin{align*}
		\int \big\{a- e^*(\bx) \big\}^2 I\big\{ M(\bx) = g \big\} \, dP(a,\bx) = \int_{\{0,1\} \times \mathcal{X}_g} \hspace*{-0.2cm} \big\{ a - e^*(\bx) \big\}^2  \, dP(a,\bx)  = \int_{\mathcal{X}_g} e^*(\bx) \big\{ 1-e^*(\bx) \big\} \, dP(\bx) \ ,
	\end{align*}
	where $\mathcal{X}_g = \{ \bx \cond M(\bx) = g \}$ and $P(\cdot)$ is the distribution of the corresponding random variable(s). The second equality is straightforward from $e^*(\bx) = P\big( A_i=1 \cond \bX_i = \bx \big)$. Note that the integral is strictly positive from Assumption  (A2) and further makes $\EXP \big( \bV_i^* \bV_i\sT \big)$ full rank.  The full rank condition of $\EXP \big\{ \bV_i\kk \bV_i\kT \cond \mathcal{I}_k^C \big\}$ is similarly established using $\widehat{e}\kk(\bX_i) \in [e_c, 1-e_c]$. \\

\noindent \textbf{Step 4} :  We establish condition (d). 
	\begin{align*}
		& \| \widehat{Z}_i\kk - Z_i^* \|_{P,2}
		= \norm{\nu^*( \bX )  - \widehat{\nu}\kk ( \bX ) }_{P,2}  \ , \\
		& \norm{  \widehat{\bV}_i\kk - \bV_i^* }_{P,2} = \big\| \big\{ e^*(\bX_i) - \widehat{e}\kk (\bX_i)\big\} \bI(\bX) \big\|_{P,2} \leq \norm{ e(\bX_i) - \widehat{e}\kk (\bX_i) }_{P,2} \ .
	\end{align*}
	As a result, $\| \widehat{Z}_i\kk - Z_i^* \|_{P,2}  = o_P(1)$ and $ \| \widehat{\bV}_i\kk - \bV_i^*  \|_{P,2}   = o_P(1)$ converge to 0 from Assumption \ref{assp:SSLS} in the main paper. \\

\noindent \textbf{Step 5} : We show that condition (e) holds. First, $\widehat{\bV}_i\kk \xi_i\kk$ is represented as
	\begin{align*}
		&
		\widehat{\bV}_i\kk \xi_i\kk
		\\
		& = \big\{
			A_i - \widehat{e}\kk(\bX_i)
		\big\} \bI(\bX_i) \Big[
			\big\{
			Y_i - \widehat{\nu}\kk (\bX_i)
		\big\} - \big\{
			A_i - \widehat{e}\kk(\bX_i)
		\big\} \bI(\bX_i) \T \bT_\OV^*
		\Big] \\
		& = \big\{
			A_i - \widehat{e}\kk(\bX_i)
		\big\} \bI(\bX_i)  \Big[
			\big\{
			\nu^*(\bX_i) - \widehat{\nu}\kk (\bX_i)
		\big\} - \big\{
			e^* (\bX_i)- \widehat{e}\kk (\bX_i)
		\big\} \bI(\bX_i) \T  \bT_\OV^* + \xi_i
		\Big] \ .		
	\end{align*}
	From the moment condition of $\EXP \big[ \{ A_i - e^*(\bX_i) \} \bI(\bX_i) \xi_i \big]=0$, we find
	\begin{align*}
		\Big\|
		 \EXP \big[ 
			\big\{
			A_i - \widehat{e}\kk(\bX_i)
		\big\} \bI(\bX_i)  \xi_i \cond \mathcal{I}_k^C
		\big]  
		\Big\|_2
		& =
		\Big\|
		\EXP \big[ 
			\big\{
			e^*(\bX_i) - \widehat{e}\kk(\bX_i)
		\big\} \bI(\bX_i)  \xi_i \cond \mathcal{I}_k^C
		\big]
		\Big\|_2
		\\
		& 
		\leq 
		\big\| \big\{ e^*(\bX_i) - \widehat{e}\kk(\bX_i)
		\big\} \bI(\bX_i)  \big\|_{P,2}
		\big\| \xi_i
		\big\|_{P,2}
		\\
		&
		\precsim
		\big\| e^*(\bX_i) - \widehat{e}\kk(\bX_i) \big\|_{P,2}
		=
		O_P (r_{e,N} ) \ .
	\end{align*}
	
	Therefore, we obtain
	\begin{align*}
		\sqrt{N}
		\Big\|
		\EXP \big[ \widehat{\bV}_i\kk \xi_i\kk  \cond \mathcal{I}_k^C
		\big]  
		\Big\|_2
		& \leq 
		\sqrt{N}
		\Big\|
		 \EXP \big[ 
			\big\{
			e^*(\bX_i)- \widehat{e}\kk(\bX_i)
		\big\} \big\{
			\nu^*(\bX_i) - \widehat{\nu}\kk(\bX_i)
		\big\} \bI(\bX_i)  \cond \mathcal{I}_k^C 
		\big] \Big\|_2
		\\
		&
		\hspace*{1cm}
		+
		\sqrt{N} \Big\|
		 \EXP \big[
		\big\{ e^*(\bX_i)- \widehat{e}\kk (\bX_i)
		\big\}^2 \bI(\bX_i) \bI(\bX_i)^\intercal \bT_\OV^* \cond \mathcal{I}_k^C
		\big] \Big\|_2
		+
		\sqrt{N} O_P(r_{e,N})
		 \\
		 &
		 \leq
		 \sqrt{N} \, \big\| \widehat{e}\kk (\bX_i)- e^*(\bX_i) \big\|_{P,2} \big\| \nu^*(\bX_i) - \widehat{\nu}\kk(\bX_i) \big\|_{P,2} \\
		 & \hspace*{1cm} 
		 + \sqrt{N} \, \big\| e^*(\bX_i)- \widehat{e}\kk (\bX_i)\big\|_{P,2}^2 \norm{\bT_\OV^*}_2 +\sqrt{N} 	O_P(r_{e,N}) \ ,
	\end{align*}
	which is $o_P(1)$ as $N \rightarrow \infty$ because of Assumption \ref{assp:SSLS} and the finite $ \norm{\bT_\OV^*}_2$ shown in \textbf{Step 6}. This concludes the first part of condition (e). For the second part, we observe that 
	\begin{align*}
		&
		 \widehat{\bV}_i\kk \xi_i\kk - \bV_i^*\xi_i
		 \\
		 & = \frac{1}{2}\Big[
	 	( \widehat{\bV}_i\kk -\bV_i^*)\big\{ ( \widehat{Z}_i\kk +Z_i^* ) - ( \widehat{\bV}_i\kk  + \bV_i^*) \T  \bT_\OV^* \big\}
	 	+ ( \widehat{\bV}_i\kk + \bV_i^*)\big\{ ( \widehat{Z}_i\kk -Z_i^* ) - ( \widehat{\bV}_i\kk  - \bV_i^*) \T  \bT_\OV^* \big\}
	 \Big]
	 \\
	 &
	 =\frac{1}{2}\Big[
	  \big\{
			e^* (\bX_i)- \widehat{e}\kk (\bX_i)
		\big\} \bI(\bX) \big[
			 \big\{
			\nu^*(\bX_i) - \widehat{\nu}\kk (\bX_i)
		\big\}  -  \big\{
			e^* (\bX_i)- \widehat{e}\kk (\bX_i)
		\big\} \bI(\bX)\T \bT_\OV^* + 2 \xi_i
		\big]
		\\
		&
		\hspace*{2cm}
		+		
		 \big\{
			2A_i - e^*(\bX_i)- \widehat{e}\kk(\bX_i)
		\big\} \bI(\bX) \big[
			 \big\{
			\nu^*(\bX_i) - \widehat{\nu}\kk(\bX_i)
		\big\}  - \big\{
			e^*(\bX_i)- \widehat{e}\kk(\bX_i)
		\big\} \bI(\bX)\T \bT_\OV^*
		\big]
		\Big] \ .
	\end{align*}
	From Assumption \ref{assp:SSLS}, we have the following result for any square integrable function $\zeta$ and for some positive constant $C$.
	\begin{align}				\label{proof2-01-01} 
		\norm{\zeta(\bX_i) \xi_i}_{P,2}^2 = E\big\{ \|\zeta(\bX_i)\|_{2}^2 \EXP \big(\xi_i^2 \cond A_i , \bX_i \big) \big\} \leq C \norm{\zeta(\bX_i)}_{P,2}^2 \ .
	\end{align}
	Also, since $A_i$ is binary,  $\big\| \big\{ e^*(\bX_i)- \widehat{e}\kk(\bX_i)
		\big\} \bI(\bX_i) \big\|_2$ and $\big\| \big\{ 2A_i-e^*(\bX_i)- \widehat{e}\kk(\bX_i)
		\big\} \bI(\bX_i) \big\|_2$ are bounded by a constant. As a result, by Lemma \ref{lem1-001} and \eqref{proof2-01-01}, we have
		\begin{align*}
			\big\|  \widehat{\bV}_i\kT \xi_i\kk - \bV_i^*\xi_i \big\|_{P,2} 
			\leq K_1 \big\|
				\nu^*(\bX_i) - \widehat{\nu}\kk (\bX_i)
			\big\|_{P,2} + \big( K_2 \norm{\bT_\OV^*}_2 +  K_3 \big) \big\|
				e^*(\bX_i)- \widehat{e}\kk(\bX_i)
			\big\|_{P,2} 
		\end{align*}
		for some constants $K_1$, $K_2$, and $K_3$, and the quantity is $o_P(1)$  as $N\rightarrow \infty$ from the same reasons above.  This shows that the second part of condition (e). \\

\noindent \textbf{Step 6} :  We show that $\norm{\bT_\OV^*}_2$ is bounded above. From \textbf{Step 2}, we obtain that $\EXP \big( \bV_i^* \bV_i\sT \big) $ is invertible matrix so its singular values are positive. As a result, we find the finite upper bound of $\|\bT_\OV^*\|_2$ with Lemma \ref{lem1-001} in the third inequality:
	\begin{align*}
		\norm{\bT_\OV^*}_2 \leq \big\|  \EXP \big( \bV_i^* \bV_i\sT \big) ^{-1} \big\|_2 \norm{ \EXP \big( \bV_i^* Z_i^* \big) }_2  \leq \frac{1}{\sigma_{\rm min}}   \norm{ \EXP \big( \bV_i^* Z_i^* \big) }_2	 \leq \frac{1}{\sigma_{\rm min}}  
		\|  \bV_i^* \|_{P,2} \| Z_i^* \|_{P,2}
		 \ ,
	\end{align*}
	where $\sigma_{\rm min}$ is the smallest singular value of $\EXP \big( \bV_i^* \bV_i\sT \big)$. Note that $\|  \bV_i^* \|_{P,2}$ is bounded as established above. Additionally, $\| Z_i^* \|_{P,2}$ is bounded as follows:
	\begin{align*}
		\| Z_i^* \|_{P,2}
		=
		\big\| Y_i - \nu^*(\bX_i) \big\|_{P,2}
		& 
		\leq 
		\big\| Y_i \big\|_{P,2} + \big\| \nu^*(\bX_i) \big\|_{P,2}
		\\
		&
		\leq
		\EXP \big\{
		\EXP \big( Y_i^2 \cond A_i, \bX_i \big) \big\}^{1/2}
		+ C
		\\
		&
		\leq
		\EXP \big\{
		\EXP \big( \xi_i^2 \cond A_i, \bX_i \big) + \mu(A_i,\bX_i)^2 \big\}^{1/2}
		+ C
		\\
		&
		< \infty
	\end{align*}
	In the second inequality, we use $| \nu^*(\bX_i) | = |e^*(\bX_i) \mu^*(1,bX_i) + \{ 1-e^*(\bX_i) \} \mu^*(0,\bX_i) | \leq C$.

\subsection{Proof of Lemma \ref{lem-EIF}}					\label{sec:proof:lem-EIF}

		We follow the proof technique laid out in \citet{Newey1990} and \citet{Hahn1998}. First, the density of $\bO_i$ with respect to some $\sigma$-finite measure is
		\begin{align*}
			f_\bO^* (y,a,\bx)
			=
			f_Y^* (y \cond a, \bx)
			f_A^* (a \cond \bx)
			f_\bX^* (\bx)
		\end{align*}
		where the smoothness and regularity conditions are given in Definition A.1 of the appendix in \citet{Newey1990}. We assume that the density of parametric submodel $f_\bO(y,a,\bx \con \eta)$ equals to the true density $f_\bO^*(y,a,\bx)$ at $\eta=\eta^*$. The corresponding score function is
		\begin{align*}
			s_\bO ( y, a, \bx \con \eta )
			=
			s_Y (y , a , \bx \con \eta )
			+
			s_A (a , \bx \con \eta )
			+
			s_\bX (\bx \con \eta )
		\end{align*}
		where
		\begin{align*}
			&
			s_Y ( y , a , \bx \con \eta )
			=
			\frac{ \partial \log f_Y ( y , a , \bx \con \eta) }{\partial \eta} 
			\ , 
			&&
			s_A ( a , \bx \con \eta )
			=
			\frac{ \partial \log f_A ( a , \bx \con \eta) }{\partial \eta} 
			\ , 
			&&
			s_\bX ( \bx \con \eta )
			=
			\frac{ \partial \log f_\bX ( \bx \con \eta) }{\partial \eta} \ .
		\end{align*}
		From the parametric submodel, we obtain the $G$-dimensional tangent space which is the mean closure of all $G$-dimensional linear combinations of scores, i.e.
		\begin{align}						\label{proof-lemma-EIF-1}
			\mathcal{T}
			=
			\Big\{
				\bm{S} ( y , a , \bx )
				\, \Big| \, 
				& 
				\bm{S} ( y , a , \bx )
				=
				\big( S_1 (y , a , \bx) , \ldots, S_G (y , a, \bx ) \big) \T \in \R^G \ ,
				\\
				& 
				S_g ( y, a, \bx)
				= 
				S_{gY} (y , a , \bx)
				+
				S_{gA}(a , \bx)
				+
				S_{g\bX} (\bx) \ , 
				\nonumber \\
				&
				\EXP \big\{ S_{gY} (Y , a , \bx) \cond A_i = a , \bX_i = \bx \big\} = 0 \ , 
				\nonumber \\
				&
				\EXP \big\{ S_{gA} (A , \bx) \cond \bX_i = \bx \big\} = 0 \ , \
				\EXP \big\{ S_{g\bX} (\bX) \big\} = 0 
				\Big\} \ . \nonumber 
		\end{align}
		The estimand is represented as $\bT(\eta) = ( \tau_1 (\eta), \ldots, \tau_G (\eta) )\T$ where
		\begin{align*}
			\tau_g (\eta)
			=
			\frac{\int \ind \{ M(\bx) = g \big\} \big\{ \mu (1, \bx \con \eta) - \mu (0, \bx \con \eta) \big\} f_\bX (\bx \con \eta) \, d\bx }{\int \ind \{ M(\bx) = g \big\}  f_\bX (\bx \con \eta) \, d\bx  }
			\ , \
			\mu(a , \bx \con \eta)
			=
			\int y f_Y ( y \cond a , \bx \con \eta) \, dy \ .
		\end{align*}
		Note that $\tau_g^* = \tau_g(\eta^*)$ and $\mu ^* (a, \bx) = \mu(a, \bx \con \eta^*)$. The derivative of $\bT (\eta)$ evaluated at true $\eta^*$ has each component as
		\begin{align}							\label{proof-lemma-EIF-2}
			\frac{ \partial \tau_g (\eta^*) }{\partial \eta}
			& = 
			\frac{ \iint  y \ind \{ M(\bx)=g \big\} \big\{ s_Y^*(y,1,\bx) f_Y^*(y \cond 1,\bx) - s_Y^*(y,0,\bx) f_Y^*(y \cond 0,\bx) \big\} f_\bX^*(\bx) \, dy dx }{p_g^*}
			\\
			& \hspace*{0.5cm} + 
			\frac{ \int  \ind \{ M(\bx)=g \big\} \big\{ \mu^*(1,\bx) - \mu^*(0,\bx) \big\} s_\bX^*(\bx) f_\bX^*(\bx) \, dx }{p_g^*}
			-
			\frac{\tau_g^* \int  \ind \{ M(\bx)=g \big\} s_\bX^*(\bx) f_\bX^*(\bx) \, dx }{p_g^*} \ .
			\nonumber
		\end{align}
		Next, we show that $\bm{\phi}(\bO_i)$ is the EIF of $\bT^*$. We first show that $\bT(\eta)$ is a differentiable parameter, i.e.
		\begin{align*}
			\frac{\partial \bT(\eta^*)}{\partial \eta}
			=
			\EXP 
			\big\{
				\bm{\phi}(\bO_i) \cdot s_\bO (Y_i, A_i, \bX_i \con \eta^*)
			\big\} \ , 
		\end{align*}
		which is sufficient to show 
		\begin{align*}
			\frac{\partial \tau_g(\eta^*)}{\partial \eta}
			=
			\EXP 
			\big\{
				\phi_g( \bO_i) \cdot s_\bO (Y_i, A_i, \bX_i \con \eta^*)
			\big\} \ ,
		\end{align*}
		where $\partial \tau_g (\eta^*) / \partial \eta$ has the form in \eqref{proof-lemma-EIF-2}. We expand $\phi_g( \bO_i) \cdot s_\bO (Y_i, A_i, \bX_i \con \eta^*)$ as follows.
		\begin{align}							\label{proof-lemma-EIF-3}
			&
			\phi_g( \bO_i) \cdot s_\bO (Y_i, A_i, \bX_i \con \eta^*)
			\\
			&
			=
			\underbrace{
			\frac{ \ind \big\{ M(\bX_i) = g \big\} }{p_g^*}
			\bigg[
				\frac{A_i \big\{ Y_i - \mu^*(1,\bX_i) \big\} }{e^* (\bX_i)}
				-
				\frac{(1-A_i) \big\{ Y_i - \mu^*(0,\bX_i) \big\} }{1-e^* (\bX_i)}
			\bigg] }_{F_1 (\bO_i) }  \cdot s_\bO (Y_i, A_i, \bX_i \con \eta^*) 
			\nonumber
			\\
			& 
			\hspace*{1cm} + 
			\underbrace{
			\frac{ \ind \big\{ M(\bX_i) = g \big\} }{p_g^*}
			\big\{
				\mu^*(1,\bX_i) - \mu^*(0,\bX_i) - \tau_g^*
			\big\} }_{F_2 (\bX_i) }  \cdot s_\bO (Y_i, A_i, \bX_i \con \eta^*) \ .
			\nonumber
		\end{align}
		Later, we show that $\EXP \{ F_1(\bO_i) \cond A_i, \bX_i \big\} = 0$ and $\EXP \{ F_2(\bX_i) \big\} = 0$. The expectation of $F_1 (\bO_i) \cdot s_\bO (Y_i, A_i, \bX_i \con \eta^*)$ is 
		\begin{align*}
			&
			\EXP \big\{ F_1 (\bO_i) \cdot s_\bO (Y_i, A_i, \bX_i \con \eta^*) \big\}
			\\
			&
			=
			\EXP \big[ \EXP \big\{ F_1 (\bO_i) \cdot s_Y(Y_i, A_i, \bX_i \con \eta^*) \cond A_i, \bX_i \big\} \big]
			+
			\EXP \big[ \underbrace{ \EXP \big\{ F_1 (\bO_i) \cond A_i, \bX_i \big\} }_{=0} \cdot \big\{ s_A ( A_i, \bX_i \con \eta^*) + s_\bX (\bX_i) \big\} \big]
			\\
			&
			=
			\EXP \bigg[
				\frac{ \ind \big\{ M(\bX_i) = g \big\} }{p_g^*}
				\EXP \bigg[
					\bigg\{ \frac{A_i}{e^*(\bX_i)	} - \frac{1-A_i}{1-e^*(\bX_i)	} \bigg\} Y_i s_Y (Y_i, A_i, \bX_i \con \eta^*) \, \bigg| \, A_i, \bX_i
				\bigg]
				\\
				& \hspace*{2cm}
				- \frac{ \ind \big\{ M(\bX_i) = g \big\} }{p_g^*} \bigg\{ \frac{A_i \mu^*(1,\bX_i)}{e^*(\bX_i)} -  \frac{(1-A_i) \mu^*(0,\bX_i)}{1-e^*(\bX_i)} \bigg\} \underbrace{ \EXP \big\{ s_Y (Y_i, A_i, \bX_i \con \eta^*) \cond A_i, \bX_i \big\} }_{=0}
			\bigg]
			\\
			&
			=
			\EXP \bigg[
			\frac{ \ind \big\{ M(\bX_i) = g \big\} }{p_g^*}
				\Big[
				\EXP \big\{
				Y_i s_Y (Y_i, 1, \bX_i \con \eta^*) \cond A_i = 1 , \bX_i \big\}
				-
				\EXP \big\{
				Y_i s_Y (Y_i, 0, \bX_i \con \eta^*) \cond A_i = 0 , \bX_i \big\}
				\Big]
				\bigg]
				\\
				& 
				=
				\frac{ \iint  y \ind \{ M(\bx)=g \big\} \big\{ s_Y^*(y,1,\bx) f_Y^*(y \cond 1,\bx) - s_Y^*(y,0,\bx) f_Y^*(y \cond 0,\bx) \big\} f_\bX^*(\bx) \, dy dx }{p_g^*} \ .
		\end{align*}
		The first equality is from the law of total expectation. The second equality is from the form of $F_1(\bO_i)$. The third equality is from $E(A_i \cond \bX_i) = e^*(\bX_i)$. The last equality is trivial.
		
		The expectation of $F_2(\bX_i) \cdot s_\bO(Y_i, A_i, \bX_i \con \eta^*)$ is
		\begin{align*}
			&
			\EXP \big\{ F_2 (\bX_i) \cdot s_\bO (Y_i, A_i, \bX_i \con \eta^*) \big\}
			\\
			&
			=
			\EXP \big[ F_2 (\bX_i) \big[ \underbrace{ \EXP \big\{ s_Y(Y_i, A_i, \bX_i \con \eta^*) \cond A_i, \bX_i \big\} + \EXP \big\{ s_A(A_i, \bX_i \con \eta^*) \cond \bX_i \big\} \big] }_{=0} \big]
			+
			\EXP \big\{ F_2(\bX_i) s_\bX(\bX_i \con \eta^*) \big\}
			\\
			&
			=
			\EXP \bigg[
				\frac{ \ind \big\{ M(\bX_i) = g \big\} }{p_g^*}
			\big\{
				\mu^*(1,\bX_i) - \mu^*(0,\bX_i) - \tau_g^*
			\big\} s_\bX (\bX_i \con \eta^*)
			\bigg]
			\\
			&
			=
			\frac{ \int  \ind \{ M(\bx)=g \big\} \big\{ \mu^*(1,\bx) - \mu^*(0,\bx) - \tau_g^* \big\} s_\bX^*(\bx) f_\bX^*(\bx) \, dx }{p_g^*} \ .
		\end{align*}
		The first equality is from the law of total expectation. The second equality is from the form of $F_2(\bX_i)$. The third equality is trivial. Combining the above results and \eqref{proof-lemma-EIF-2}, we find
		\begin{align*}
			&
			\EXP \big\{ \phi_g( \bO_i) \cdot s_\bO (Y_i, A_i, \bX_i \con \eta^*) \big\}
			\\
			&
			=
			\EXP \big\{ F_1(\bO_i) \cdot s_\bO (Y_i, A_i, \bX_i \con \eta^*) \big\}
			+
			\EXP \big\{ F_2(\bX_i) \cdot s_\bO (Y_i, A_i, \bX_i \con \eta^*) \big\}
			\\
			&
			=
			\frac{ \partial \tau_g (\eta^*) }{\partial \eta} \ .
		\end{align*}
		This concludes that $\bT(\eta)$ is a differentiable parameter.
		
		Second, we show that $\bm{\phi}(\bO_i)$ belongs to the tangent space $\mathcal{T}$ in \eqref{proof-lemma-EIF-1}, which suffices to show $\phi_g(\bO_i)$ satisfies the elementry-wise conditions on $\bm{S}(y,a,\bx)$. For $F_1(\bO_i)$ and $F_2(\bX_i)$ in \eqref{proof-lemma-EIF-3}, we find
		\begin{align*}
			 \EXP \big\{ F_1(\bO_i) \cond A_i, \bX_i  \big\}
			 &
			 =
			 \frac{ \ind \big\{ M(\bX_i) = g \big\} }{p_g^*}
			 \EXP
			\bigg[
				\frac{A_i \big\{ Y_i - \mu^*(1,\bX_i) \big\} }{e^* (\bX_i)}
				-
				\frac{(1-A_i) \big\{ Y_i - \mu^*(0,\bX_i) \big\} }{1-e^* (\bX_i)}
				\, \bigg| \, A_i, \bX_i
			\bigg]
			\\
			& = 
			\frac{ \ind \big\{ M(\bX_i) = g \big\} }{p_g^*}
			\big[
			\big\{
				\EXP (Y_i \cond A_i=1,\bX_i) - \mu^*(1,\bX_i)
				\big\}
				-
				\big\{
				\EXP (Y_i \cond A_i=0,\bX_i) - \mu^*(0,\bX_i)
				\big\} 
			\big]
			\\
			& = 0 \ , \\
			 \EXP \big\{ F_2(\bX_i) \big\}
			 & = 
			 \EXP \bigg[
			 \frac{ \ind \big\{ M(\bX_i) = g \big\} }{p_g^*}
			\big\{
				\mu^*(1,\bX_i) - \mu^*(0,\bX_i) - \tau_g^*
			\big\} 
			\bigg]
			\\
			&
			=
			\frac{\int  \ind \big\{ M(\bx) = g \big\} \big\{ \mu^*(1,\bx) - \mu^*(0,\bx) \big\} f_\bX^*(\bx) \, d\bx  }{p_g^*} - \frac{\tau_g^* \int \ind \big\{ M(\bx) = g \big\} f_\bX^*(\bx) \, d\bx  }{p_g^*}
			\\
			& = 0 \ .
		\end{align*}
		That is, $\phi_g(\bO_i) = F_1(\bO_i) + F_2(\bX_i) \in \text{span} \big\{ S_{gY} (Y_i, A_i, \bX_i) , S_{g\bX}(\bX_i) \big\} \subset \mathcal{T} $. This concludes that $\bm{\phi}(\bO_i)$ is the EIF of $\bT^*$.
		
		The semiparametric efficiency bound of $\bT^*$ is the variance of $\bm{\phi}(\bO_i)$. Observing the form of $\bm{\phi}(\bO_i)$, it is trivial that $\VAR \{ \bm{\phi}(\bO_i) \}$ is a diagonal matrix of which $g$th diagonal is
		\begin{align*}
			\VAR \big\{ \phi_g(\bO_i) \big\}
			& =
			\EXP \big[  \big\{ \phi_g (\bO_i)  \big\}^2 \big]
			=
			\frac{ \EXP \big[ \ind \big\{ M(\bX_i) = g \big\} \big\{
				\varphi(\bO_i) - \tau_g^*
			\big\} ^2 \big] }{(p_g^*)^2} \ .
		\end{align*}
		
	Lastly, the estimated slope coefficients from regressing $\varphi(\bO_i)$ on $\bI(\bX_i)$ without the intercept term are
		\begin{align*}
			\widetilde{\bT}
			=
			(\widetilde{\tau}_1,\ldots,\widetilde{\tau}_G)\T
			\ , \
			\widetilde{\tau}_g
			=
			\frac{ 
			\frac{1}{N} \sum_{i=1}^N \varphi(\bO_i) \ind \{ M(\bX_i) = g \}
			}{
			\frac{1}{N} \sum_{i=1}^N \ind \{ M(\bX_i) = g \}
			} \ .
		\end{align*}		
		The denominator converges to $p_g^*$ in probability, i.e.
		\begin{align*}
			\frac{1}{N} \sum_{i=1}^N \ind \{ M(\bX_i) = g \} \stackrel{P}{\rightarrow} p_g^* \ .
		\end{align*}
		From the central limit theorem, the numerator is asymptotically Normal, i.e.
		\begin{align*}
			&
			\sqrt{N}
			\bigg[
			\frac{1}{N} \sum_{i=1}^N \varphi(\bO_i) \ind \{ M(\bX_i) = g \}
			-
			p_g^*
			\tau_g^*
			\bigg]
			\\
			&
			\stackrel{D}{\rightarrow} 
			N \Big( 0 , \VAR \big[ \varphi(\bO_i) \ind \{ M(\bX_i) = g \} \big] 
			\Big) 
			\stackrel{D}{=}
			N \Big( 0 , \EXP \big[ \ind \big\{ M(\bX_i) = g \big\} \big\{
				\varphi(\bO_i) - \tau_g^*
			\big\} ^2 \big] \Big)
		\end{align*}
		From the Slutsky's theorem, we have
		\begin{align*}
			\sqrt{N}
			\big( \widetilde{\tau}_g - \tau_g^* \big)
			\stackrel{D}
			\rightarrow
			N \bigg( 0 ,
			\frac{ \EXP \big[ \ind \big\{ M(\bX_i) = g \big\} \big\{
				\varphi(\bO_i) - \tau_g^*
			\big\} ^2 \big] }{(p_g^*)^2} 
			\bigg) 
			\stackrel{D}{=}
			N \Big( 0, \EXP \big[  \big\{ \phi_g (\bO_i)  \big\}^2 \big] \Big) \ .
		\end{align*}
		Since $\widetilde{\tau}_g$ and $\widetilde{\tau}_{g'}$ are independent, this implies
		\begin{align*}
			\sqrt{N} \big( \widetilde{\bT} - \bT^* \big)
			\stackrel{D}{\rightarrow}
			N \Big( 0, \EXP \big\{ \bm{\phi}(\bO_i)   \bm{\phi}(\bO_i) \T \big\} \Big) \ .
		\end{align*}

		\subsection{Proof of Lemma \ref{lem-assumption-EIF}}						\label{proof-Lem-6}
	
We proof the claim of the lemma in following \textbf{Step 1} -- \textbf{Step 6}.   \\

\noindent \textbf{Step 1}: We find
\begin{align*}
	\EXP \big\{ \epsilon_i \bI(\bX_i) \big\}
	=
	\EXP \big[ \bV_i^* \big\{ Z_i^* -  \bV_i\sT \bT^* \big\} \big]
	=
	\EXP ( \bV_i^*  Z_i^* ) - \EXP  ( \bV_i^* \bV_i\sT ) \bT^* \ .
\end{align*}
The $g$th component of $\EXP ( \bV_i^*  Z_i^* )$ is
\begin{align*}
	&
	\EXP \big[  \ind \{ M(\bX_i)  = g \}  \varphi^* (\bO_i) \big]
	=
	\EXP \big\{  \varphi^* (\bO_i) \cond M(\bX_i)  = g \big\} \Pr \{ M(\bX_i) = g \}
	=
	\tau_g \cdot \Pr \{ M(\bX_i) = g \}
	\ .
\end{align*}
Also, $ \EXP  ( \bV_i^* \bV_i\sT ) $ is a diagonal matrix of which $g$th component is $\EXP \big[ \ind \{ M(\bX_i)  = g \} \big] = \Pr \{ M(\bX_i) = g \}$. Therefore, each component of $\EXP \big\{ \epsilon_i \bI(\bX_i) \big\}$ is zero. \\
	
\noindent \textbf{Step 2} :  We show condition (b) holds. Since $\bI(\bX_i)$ is trivially bounded, we obtain
	\begin{align*}
		& \norm{\bV_i}_{P,4}^* =  \norm{\widehat{\bV}_i\kk }_{P,4} \leq 1  \ .
	\end{align*}
	For $\norm{ \bV_i^* \epsilon_i }_{P,2}$, we observe:
	\begin{align*}
			&  \| \bV_i^* \epsilon_i \|_{P,2} \leq \| A_i \epsilon_i \|_{P,2}  + \| e^*(\bX_i) \epsilon_i \|_{P,2} \leq 2 \|\epsilon_i \|_{P,2} < \infty \ , 
		\end{align*}
		Therefore, it suffices to show that $ \|\epsilon_i \|_{P,2}$ is finite. Since $\EXP (\epsilon_i^2 \cond A_i, \bX_i)$ is bounded, we find
		\begin{align*}
			\|\epsilon_i \|_{P,2}^2
			=
			\EXP \big( \epsilon_i^2 \big)
			=
			\EXP \big\{ \EXP (\epsilon_i^2 \cond A_i, \bX_i) \big\}
			< \infty \ .
		\end{align*}
	Similarly, we find $\EXP \big( \epsilon_i^2 \cond \bV_i^* \big)$ is bounded as follows:
	\begin{align*}
		\EXP \big( \epsilon_i^2 \cond \bV_i^* \big)
		=
		\EXP \big\{ \EXP( \epsilon_i^2 \cond \bV_i^*, A_i, \bX_i) \cond \bV_i^* \big\}
		=
		\EXP \big\{ \EXP( \epsilon_i^2 \cond A_i, \bX_i) \cond \bV_i^* \big\}
		< \infty \ .
	\end{align*}
	
	\noindent \textbf{Step 3} :  We establish condition (c). Note that $ \EXP  ( \bV_i^* \bV_i\sT ) = \EXP \big\{ \bV_i\kk \bV_i\kT \cond \mathcal{I}_k^C \big\} $ is a diagonal matrix of which $g$th component is $\EXP \big[ \ind \{ M(\bX_i)  = g \} \big] = \Pr\{ M(\bX_i) = g \}$. Therefore, $ \EXP  ( \bV_i^* \bV_i\sT ) = \EXP \big\{ \bV_i\kk \bV_i\kT \cond \mathcal{I}_k^C \big\} $ are full rank.\\

\noindent \textbf{Step 4} :  We establish condition (d). 
	\begin{align*}
		& \| \widehat{Z}_i\kk - Z_i^* \|_{P,2}
		= \norm{\varphi^*( \bX )  - \widehat{\varphi}\kk ( \bX ) }_{P,2}  \ , \
		\norm{  \widehat{\bV}_i\kk - \bV_i^* }_{P,2} = 0 \ .
	\end{align*}
	Therefore, it suffices to show $ \norm{ \varphi^*( \bX )  - \widehat{\varphi}\kk ( \bX ) }_{P,2} =o_P(1)$. After some algebra, we find
	\begin{align*}
		&
		\big| 
		\varphi^*( \bX )  - \widehat{\varphi}\kk ( \bX ) 
		\big|
		\\
		&
		=
		\bigg|
		\frac{A_i \{ Y_i - \mu^*(1, \bX_i) \} }{e^*(\bX_i)}
		-
		\frac{(1-A_i) \{ Y_i - \mu^*(0, \bX_i) \} }{1-e^*(\bX_i)}
		+
		\mu^*(1,A_i) - \mu^*(0, A_i)
		\\
		& \hspace*{1cm}
		-
		\frac{A_i \{ Y_i - \widehat{\mu}\kk (1, \bX_i) \} }{ \widehat{e}\kk (\bX_i)}
		+
		\frac{(1-A_i) \{ Y_i - \widehat{\mu}\kk (0, \bX_i) \} }{1 - \widehat{e}\kk (\bX_i)}
		-
		\widehat{\mu}\kk (1,A_i) + \widehat{\mu}\kk (0, A_i)
		\bigg|
		\\
		&
		\leq
		\frac{A_i }{e^*(\bX_i) \widehat{e}\kk(\bX_i)}
		\bigg[
			 | Y_i | | \widehat{e}\kk(\bX_i) - e^*(\bX_i) |
			 +
			 0.5  | \widehat{e}\kk(\bX_i) - e^*(\bX_i) |  | \widehat{\mu}\kk(1,\bX_i) + \mu^*(1,\bX_i) |
			 \\
			 & \hspace*{4cm}
			 +
			 0.5  | \widehat{e}\kk(\bX_i) + e^*(\bX_i) |  | \widehat{\mu}\kk(1,\bX_i) - \mu^*(1,\bX_i) |
		\bigg]
		\\
		&
		\hspace*{0.5cm}
		+
		\frac{1-A_i }{\{ 1-e^*(\bX_i) \} \{1- \widehat{e}\kk(\bX_i) \}}
		\bigg[
			 | Y_i | | \widehat{e}\kk(\bX_i) - e^*(\bX_i) |
			 +
			 0.5  | \widehat{e}\kk(\bX_i) - e^*(\bX_i) |  | \widehat{\mu}\kk(0,\bX_i) + \mu^*(0,\bX_i) |
			 \\
			 & \hspace*{4cm}
			 +
			 0.5  | \widehat{e}\kk(\bX_i) + e^*(\bX_i) |  | \widehat{\mu}\kk(0,\bX_i) - \mu^*(0,\bX_i) |
		\bigg]
		\\
		& \hspace*{0.5cm}
		+
		| \widehat{\mu}\kk(1,\bX_i) - \mu^*(1,\bX_i) |
		+
		| \widehat{\mu}\kk(0,\bX_i) - \mu^*(0,\bX_i) |
		\\
		&
		\leq
		K_1 |Y_i|  | \widehat{e}\kk(\bX_i) - e^*(\bX_i) |
		+
		K_2  | \widehat{e}\kk(\bX_i) - e^*(\bX_i) |
		\\
		&\hspace*{1cm}
		+
		K_3  | \widehat{\mu}\kk(1,\bX_i) - \mu^*(1,\bX_i) |
		+
		K_4  | \widehat{\mu}\kk(0,\bX_i) - \mu^*(0,\bX_i) |
		\ .
	\end{align*}
	In the first inequality, we use $aa'-bb' = 0.5(a-a')(b+b') + 0.5(a+a')(b-b')$. In the second inequality, we use that $e^*(\bX_i)$ and $\widehat{e}\kk (\bX_i)$ are between 0 and 1. Therefore, from the H\"older's inequality, we find
	\begin{align}						\label{eq-proof-2002}
		&
		\big\|
		\varphi^*( \bX )  - \widehat{\varphi}\kk ( \bX ) 
		\big\|_{P,2}
		\nonumber
		\\
		&
		\leq 
		\big\{
		K_1
		\| Y_i \|_{P,2}
		+ K_2 \big\}
		\| \widehat{e}\kk(\bX_i) - e^*(\bX_i) \|_{P,2}
		\nonumber
		\\
		& \hspace*{1cm}
		+
		K_3 \| \widehat{\mu}\kk(1,\bX_i) - \mu^*(1,\bX_i) \|_{P,2}
		+
		K_4 \| \widehat{\mu}\kk(0,\bX_i) - \mu^*(0,\bX_i) \|_{P,2}
		\nonumber
		\\
		&
		\leq 
		K_1'
		\| \widehat{e}\kk(\bX_i) - e^*(\bX_i) \|_{P,2}
		+
		K_3 \| \widehat{\mu}\kk(1,\bX_i) - \mu^*(1,\bX_i) \|_{P,2}
		+
		K_4 \| \widehat{\mu}\kk(0,\bX_i) - \mu^*(0,\bX_i) \|_{P,2}
		\nonumber
		\\
		&
		=
		o_P(1) 
	\end{align}
	where $\big\| Y_i \big\|_{P,2} \leq \EXP \big\{
		\EXP \big( Y_i^2 \cond A_i, \bX_i \big) \big\}^{1/2} \leq \EXP \big\{
		\EXP \big( \epsilon_i^2 \cond A_i, \bX_i \big) + \mu(A_i,\bX_i)^2 \big\}^{1/2} < \infty$. \\
	
	\noindent \textbf{Step 5} : We show that condition (e) holds. First, $\widehat{\bV}_i\kk \epsilon_i\kk$ is represented as
	\begin{align*}
		&
		\widehat{\bV}_i\kk \epsilon_i\kk
		= 
		\bI(\bX_i) \Big\{
			\widehat{\varphi}\kk (\bO_i) - \bI(\bX_i) \T \bT^*
		\Big\}
		=
		\bI(\bX_i) \Big\{
			\widehat{\varphi}\kk (\bO_i) + \epsilon_i - \varphi^*(\bO_i)
		\Big\} \ .
	\end{align*}
	From the moment condition of $\EXP \big[ \bI(\bX_i) \epsilon_i \big]=0$, we find
	\begin{align*}
		\Big\|
		 \EXP \big[ 
			\big\{
			A_i - \widehat{e}\kk(\bX_i)
		\big\} \bI(\bX_i)  \epsilon_i \cond \mathcal{I}_k^C
		\big]  
		\Big\|_2
		& =
		\Big\|
		\EXP \big[ 
			\bI(\bX_i) \big\{
			\widehat{\varphi}\kk (\bO_i) - \varphi^*(\bO_i)
		\big\}
		\cond \mathcal{I}_k^C
		\big]
		\Big\|_2
		\\
		& 
		\leq 
		\Big|
		\EXP \big\{
			\widehat{\varphi}\kk (\bO_i) - \varphi^*(\bO_i)
		\cond \mathcal{I}_k^C
		\big\} \Big|_2 \ .
	\end{align*}
	Here $\big| \EXP \big\{
			\widehat{\varphi}\kk (\bO_i) - \varphi^*(\bO_i)
		\cond \mathcal{I}_k^C
		\big\} \big| $ is bounded above as follows.
		\begin{align*}
			&
			\Big|
			\EXP \big\{
			\widehat{\varphi}\kk (\bO_i) - \varphi^*(\bO_i)
		\cond \mathcal{I}_k^C
		\big\}  \Big|
		\\
		&
		=
		\bigg|
		\EXP \bigg[
			\frac{A_i \{ Y_i - \widehat{\mu}\kk(1,\bX_i) \}}{\widehat{e}\kk(\bX_i)}
			-
			\frac{(1-A_i) \{ Y_i - \widehat{\mu}\kk(0,\bX_i) \}}{1-\widehat{e}\kk(\bX_i)}
			-
			\frac{A_i \{ Y_i - \mu^*(1,\bX_i) \}}{e^*(\bX_i)}
			+
			\frac{(1-A_i) \{ Y_i - \mu^*(0,\bX_i) \}}{1-e^*(\bX_i)}
			\\
			& \hspace*{1cm}
			+ 
			\widehat{\mu}\kk(1,\bX_i) - \widehat{\mu}\kk(0,\bX_i) 
			-
			\mu^*(1,\bX_i) + \mu^*(0,\bX_i)
			\, \bigg| \, \mathcal{I}_k^C
		\bigg] \bigg|
		\\
		& 
		\leq
		\EXP \bigg[
		\frac{ | \widehat{e}\kk(\bX_i) - e^*(\bX_i) |  | \widehat{\mu}\kk(1,\bX_i) - \mu^*(1,\bX_i) | }{\widehat{e}\kk(\bX_i) }
		+
		\frac{ | \widehat{e}\kk(\bX_i) - e^*(\bX_i) |  | \widehat{\mu}\kk(0,\bX_i) - \mu^*(0,\bX_i) | }{1-\widehat{e}\kk(\bX_i) }
		\, \bigg| \, \mathcal{I}_k^C
		\bigg]
		\\
		&
		\leq 
		c_e^{-1}
		\| \widehat{e}\kk(\bX_i) - e^*(\bX_i) \|_{P,2}
		\Big\{
		\| \widehat{\mu}\kk(1,\bX_i) - \mu^*(1,\bX_i) \|_{P,2}
		+
		\| \widehat{\mu}\kk(0,\bX_i) - \mu^*(0,\bX_i) \|_{P,2}
		\Big\} \ .
		\end{align*}
		The first inequality is from $\EXP( Y_i \cond A_i, \bX_i) = \mu^*(A_i, \bX_i)$ and $\EXP(A_i \cond \bX_i) = e^*(\bX_i)$. The second inequality is from $\widehat{e}\kk(\bX_i) \in [c_e, 1-c_e]$ and the H\"older's inequality. Therefore, we obtain
	\begin{align*}
		&
		\sqrt{N}
		\Big\|
		\EXP \big[ \widehat{\bV}_i\kk \epsilon_i\kk  \cond \mathcal{I}_k^C
		\big]  
		\Big\|_2
		\\
		& \leq 
		\sqrt{N}
		\Big|
		\EXP \big\{
			\widehat{\varphi}\kk (\bO_i) - \varphi^*(\bO_i)
		\cond \mathcal{I}_k^C
		\big\} \Big|_2
		\\
		&
		\leq
		\sqrt{N}
		c_e^{-1}
		\| \widehat{e}\kk(\bX_i) - e^*(\bX_i) \|_{P,2}
		\Big\{
		\| \widehat{\mu}\kk(1,\bX_i) - \mu^*(1,\bX_i) \|_{P,2}
		+
		\| \widehat{\mu}\kk(0,\bX_i) - \mu^*(0,\bX_i) \|_{P,2}
		\Big\}
	\end{align*}
	which is $o_P(1)$ as $N \rightarrow \infty$ because of Assumption \ref{assp:EIF}. This concludes the first part of condition (e). For the second part, we observe that 
	\begin{align*}
		&
		 \widehat{\bV}_i\kk \epsilon_i\kk - \bV_i^*\epsilon_i
		 =
		 \bI(\bX_i) ( \epsilon_i\kk - \epsilon_i )
		 =
		 \bI(\bX_i) \Big\{
			\widehat{\varphi}\kk (\bO_i) - \varphi^*(\bO_i)
		\Big\} \ .
	\end{align*}
	Therefore, $\| \widehat{\bV}_i\kk \epsilon_i\kk - \bV_i^*\epsilon_i \|_{P,2} \leq \| \widehat{\varphi}\kk (\bO_i) - \varphi^*(\bO_i) \|_{P,2} = o_P(1)$ where the last result is from \eqref{eq-proof-2002}. This shows that the second part of condition (e). \\

		\noindent \textbf{Step 6} :  We show that $\norm{\bT^*}_2$ is bounded above. From \textbf{Step 2}, we obtain that $\EXP \big( \bV_i^* \bV_i\sT \big) $ is invertible matrix so its singular values are positive. As a result, we find the finite upper bound of $\|\bT^*\|_2$ with Lemma \ref{lem1-001} in the third inequality:
	\begin{align*}
		\norm{\bT^*}_2 \leq \big\|  \EXP \big( \bV_i^* \bV_i\sT \big) ^{-1} \big\|_2 \norm{ \EXP \big( \bV_i^* Z_i^* \big) }_2  \leq \frac{1}{\sigma_{\rm min}}   \norm{ \EXP \big( \bV_i^* Z_i^* \big) }_2	 \leq \frac{1}{\sigma_{\rm min}}  
		\|  \bV_i^* \|_{P,2} \| Z_i^* \|_{P,2}
		 \ ,
	\end{align*}
	where $\sigma_{\rm min}$ is the smallest singular value of $\EXP \big( \bV_i^* \bV_i\sT \big)$. From the assumption, we have  $\|  \bV_i^* \|_{P,2} < \infty$ and 
	\begin{align*}
		& \big\| Z_i^* \big\|_{P,2}^2
		\leq C \cdot
		\EXP
		\Bigg[
			\frac{\EXP(\epsilon_i^2 \cond A_i=1,\bX_i)}{e^*(\bX_i)}
			+
			\frac{\EXP(\epsilon_i^2 \cond A_i=0,\bX_i)}{1-e^*(\bX_i)}
			+
			\big\{ \mu^*(1,\bX_i) - \mu^*(0,\bX_i) \big\}^2
		\Bigg]
		< \infty
	\end{align*}
	Therefore, we fine $\big\| \bT^* \big\|_2 < \infty$.

\subsection{Proof of Lemma \ref{lem-assumption-Joint}}					\label{proof-Lem-assumption-Joint}

We denote 
\begin{align*}
	&
	\bV_i^* 
	=
	 \begin{bmatrix}
	 \bV_{\SP,i}^* \\
	 \bV_{\NP,i}^* 
	 \end{bmatrix}
	=
	\begin{bmatrix} \{ A_i -e^*(\bX_i)  \bI(\bX_i) \\ \bI(\bX_i) 	\end{bmatrix}	 
	&&
	\widehat{\bV}_i\kk 
	=
	\begin{bmatrix}
	\widehat{\bV}_{\SP,i}\kk 
	\\
	\widehat{\bV}_{\NP,i}\kk 
	\end{bmatrix}
	= \begin{bmatrix} \{ A_i - \widehat{e}\kk(\bX_i)  \bI(\bX_i) \\ \bI(\bX_i) 	\end{bmatrix}	 
	\\
	&
	\bZ_i^* 
	=
	\begin{bmatrix}
	Z_{\SP,i}^*
	\\
	Z_{\NP,i}^*
	\end{bmatrix}
	= \begin{bmatrix} Y_i - \nu^*(\bX_i) \\ \varphi^* (\bO_i)  \end{bmatrix} 
	&& \widehat{\bZ}_i\kk 
	=
	\begin{bmatrix}
	\widehat{Z}_{\SP,i}\kk 
	\\
	\widehat{Z}_{\NP,i}\kk 
	\end{bmatrix}
	= \begin{bmatrix} Y_i - \widehat{\nu}\kk(\bX_i) \\ \widehat{\varphi}\kk (\bO_i)  \end{bmatrix} 
	\\
	&
	\bxi_i = 
	\begin{bmatrix}
		\xi_{\SP,i} 
		\\
		\xi_{\NP,i} 
	\end{bmatrix}
	=
	\begin{bmatrix}
		 {Z}_{\SP,i}^* - {\bV}_{\SP,i}\sT \bT_{\OV}^*
		 \\
		 {Z}_{\SP,i}^* - {\bV}_{\SP,i}\sT \bT^*
	\end{bmatrix}
	&&
	\widehat{\bxi}_i\kk
	=
	\begin{bmatrix}
		\widehat{\xi}_{\SP,i} \kk
		\\
		\widehat{\xi}_{\NP,i} \kk 
	\end{bmatrix}
	=
	\begin{bmatrix}
		 \widehat{Z}_{\SP,i}\kk - \widehat{\bV}_{\SP,i}\kT \bT_{\OV}^*
		 \\
		 \widehat{Z}_{\SP,i}\kk - \widehat{\bV}_{\SP,i}\kT \bT^*
	\end{bmatrix}
\end{align*}
We remark that the terms with subscript $\SP$ and  $\NP$ are identical to terms used in Lemmas \ref{lem-assumption-SSLS} and \ref{lem-assumption-EIF}, respectively.

We proof the claim of the lemma in following \textbf{Step 1} -- \textbf{Step 6}.   \\

\noindent \textbf{Step 1}: We find $\EXP \big\{  \bV_i\sT \bxi_i \big\}
	= \bm{0}$ from Lemmas \ref{lem-assumption-SSLS} and \ref{lem-assumption-EIF}.
	
\noindent \textbf{Step 2} :  We show condition (b) holds. Since $\bV_i^*$ and $\widehat{\bV}_i\kk$ are trivially bounded, we obtain $\norm{\bV_i}_{P,4}^* < \infty$ and $\norm{\widehat{\bV}_i\kk }_{P,4} < \infty$. 	For $\norm{ \bV_i\sT \bxi_i }_{P,2}$, we observe $\| \bV_i\sT \bxi_i \|_{P,2} \leq 
			\| \bV_{\SP,i}\sT \xi_{\SP,i} \|_{P,2}
			+
			\| \bV_{\NP,i}\sT \xi_{\NP,i} \|_{P,2} 
			< \infty$  from Lemmas \ref{lem-assumption-SSLS} and \ref{lem-assumption-EIF}. 		Similarly, we find $\EXP \big( \bxi_i \bxi_i\T \cond \bV_i^* \big)$ is bounded as follows:
	\begin{align*}
		\EXP \big( \bxi_i \bxi_i\T \cond \bV_i^* \big)
		=
		\EXP \big\{ \EXP( \bxi_i \bxi_i\T \cond \bV_i^*, A_i, \bX_i) \cond \bV_i^* \big\}
		=
		\EXP \big\{ \EXP( \bxi_i \bxi_i\T \cond A_i, \bX_i) \cond \bV_i^* \big\}
		< \infty \ .
	\end{align*}
	The latter term follows from $\EXP \big( \xi_{\SP,i}^2 \cond A_i, \bX_i \big) < \infty$, $\EXP \big( \xi_{\NP,i}^2 \cond A_i, \bX_i \big) < \infty$, and $\EXP \big( \xi_{\SP,i}\xi_{\NP,i} \cond A_i, \bX_i \big)$ where the last results is obtained from the Cauchy-Schwarz inequality.
	
	\noindent \textbf{Step 3} :  We establish condition (c). Note that $ \EXP  ( \bV_i^* \bV_i\sT )$ is a block matrix with
	\begin{align*}
		 \EXP  ( \bV_i^* \bV_i\sT )
		 =
		 \begin{bmatrix}
		 	 \EXP  ( \bV_{\SP,i}^* \bV_{\SP,i}\sT ) &  \EXP  ( \bV_{\SP,i}^* \bV_{\NP,i}\sT )
		 	 \\
		 	  \EXP  ( \bV_{\NP,i}^* \bV_{\SP,i}\sT ) &  \EXP  ( \bV_{\NP,i}^* \bV_{\NP,i}\sT )
		 \end{bmatrix}
	\end{align*}
	Note that $\det\{ \EXP  ( \bV_{\SP,i}^* \bV_{\SP,i}\sT )\}>0$ and $\det\{  \EXP  ( \bV_{\NP,i}^* \bV_{\NP,i}\sT ) \} >0$, and 
	\begin{align*}
	& \det\{ \EXP  ( \bV_i^* \bV_i\sT ) \}
	\\
	& = \det\{ \EXP  ( \bV_{\SP,i}^* \bV_{\SP,i}\sT )\} \det\{\EXP  ( \bV_{\NP,i}^* \bV_{\NP,i}\sT ) -   \EXP  ( \bV_{\NP,i}^* \bV_{\SP,i}\sT )  \{ \EXP  ( \bV_{\SP,i}^* \bV_{\SP,i}\sT ) \}^{-1}
	  \EXP  ( \bV_{\SP,i}^* \bV_{\NP,i}\sT ) \}
	\end{align*}
	From straightforward algebra, we find the $g$th diagonal of	$\bV_{\NP,i}^* \bV_{\SP,i}\sT$ is $\{A_i -e^*(\bX_i) \} \ind\{ M(\bX_i) =g \}$. Therefore, we find
	\begin{align*}
	& \det\{ \EXP  ( \bV_i^* \bV_i\sT ) \}
	\\
	& = \det\{ \EXP  ( \bV_{\SP,i}^* \bV_{\SP,i}\sT )\} \det\{ \EXP  ( \bV_{\NP,i}^* \bV_{\NP,i}\sT ) -   \EXP  ( \bV_{\NP,i}^* \bV_{\SP,i}\sT )  \{ \EXP  ( \bV_{\SP,i}^* \bV_{\SP,i}\sT ) \}^{-1}
	  \EXP  ( \bV_{\SP,i}^* \bV_{\NP,i}\sT ) \}
	  \\
	  &
	  =
	  \det\{ \EXP  ( \bV_{\SP,i}^* \bV_{\SP,i}\sT )\} \det\{\EXP  ( \bV_{\NP,i}^* \bV_{\NP,i}\sT ) \} > 0 \ .
	\end{align*}
	Similarly, we find
	\begin{align*}
		&
		\EXP  ( \bV_{\NP,i}\kk \bV_{\NP,i}\kT ) -  \EXP  ( \bV_{\NP,i}\kk \bV_{\SP,i}\kT )  \{ \EXP  ( \bV_{\SP,i}\kk \bV_{\SP,i}\kT ) \}^{-1}
	  \EXP  ( \bV_{\SP,i}\kk \bV_{\NP,i}\kT )
	  \\
	  &
	  =
	  \text{diag} \Bigg[
		\EXP \Big[ \underbrace{ 1- \big[ e^*(\bX_i)	\big\{ 1 - \widehat{e}\kk(\bX_i) \big\} + \big\{ 1-e^*(\bX_i) \big\} \widehat{e}\kk(\bX_i) \big]^2 }_{\in (0,1)} \Cond M(\bX_i) =g \Big]
	  \Bigg]_{g=1,\ldots,G} \ .
	\end{align*}
	The underbraced terms are between 0 and 1. Therefore, $\det\{ \EXP  ( \bV_i\kk \bV_i\kT ) \}>0$.

\noindent \textbf{Step 4-5} :  This is trivial from Lemmas \ref{lem-assumption-SSLS} and \ref{lem-assumption-EIF}; for instance, 
\begin{align*}
	\big\| \widehat{\bZ}_i\kk - \bZ_i^* \big\|_{P,2}
	\leq
	\big\| \widehat{Z}_{\SP,i}\kk - Z_{\SP,i}^* \big\|_{P,2}
	+
	\big\| \widehat{Z}_{\NP,i}\kk - Z_{\NP,i}^* \big\|_{P,2}
	=o_P(1) \ ,
\end{align*}
and the other results can be established from similar manners.

	\subsection{Proof of Lemma \ref{lem-comparisionSSLSEIF}}													\label{sec:proof:lem-comparisionSSLSEIF}

From Theorem \ref{thm:GPLM1} and \ref{thm:EIF}, we find
\begin{align*}
	&
	\sqrt{N}
	\begin{pmatrix}
		\widehat{\bT}_\SSLS - \bT^* 
		\\
		\widehat{\bT}_\EIF - \bT^* 
	\end{pmatrix}
	=
	\frac{1}{\sqrt{N}}
	\sum_{i=1}^N
	\begin{pmatrix}
		\Big[ \EXP \big[ \big\{ A_i - e^*(\bX_i)  \big\}^2 \bI(\bX_i) \bI(\bX_i)\T \big] \Big] ^{-1} \big\{ Y_i - \nu^*(\bX_i) \big\} \big\{ A_i - e^*(\bX_i) \big\} \bI(\bX_i)
		\\
		\varphi^* (\bO_i) \bI(\bX_i)
	\end{pmatrix}		
	+ o_P(1)
 \ .
\end{align*}
Using the Lindberg-Feller central limit theorem and the Cramer-Wold theorem, we obtain the asymptotic normality as follows:
\begin{align*}
	\sqrt{N}
	\begin{pmatrix}
		\widehat{\bT}_\SSLS - \bT^* 
		\\
		\widehat{\bT}_\EIF - \bT^* 
	\end{pmatrix}
	\stackrel{D}{\rightarrow}
	N 
	\Bigg(
		\bigg(
			\begin{matrix}
			0 \\ 0
			\end{matrix}
		\bigg)
		 \ , \
		 \bigg(
	\begin{matrix}
			\Sigma_\SSLS & \Sigma_{\SSLS,\EIF}
			\\
			\Sigma_{\SSLS,\EIF}\T & \Sigma_\EIF
		\end{matrix} \bigg)
	\Bigg)
\end{align*}
where
\begin{align*}
	\Sigma_\SSLS
	&
	=
	\Big[
	\EXP \big[ \big\{ A_i - e^* (\bX_i) \big\}^2 \bI(\bX_i)^{\otimes 2} \big]
	\Big]^{-1}
	\Big[
	\EXP \big[
	\epsilon_i^2
	\big\{ A_i - e^* (\bX_i) \big\}^2 \bI(\bX_i)^{\otimes 2} \big]
	\Big]
	\Big[
	\EXP \big[ \big\{ A_i - e^* (\bX_i) \big\}^2 \bI(\bX_i)^{\otimes 2} \big]
	\Big]^{-1}
	\\
	\Sigma_\EIF
	&
	=	
	\Big[
	\EXP \big\{ \bI(\bX_i)^{\otimes 2} \big\}
	\Big]^{-1}
	\Big[
	\EXP \big[
	\big\{ \varphi^*(\bO_i) - \bI(\bX_i)\T\bT^* \big\}^2
	\bI(\bX_i)^{\otimes 2} \big]
	\Big]
	\Big[
	\EXP \big\{ \bI(\bX_i)^{\otimes 2} \big\}
	\Big]^{-1}
	\\
	\Sigma_{\SSLS, \EIF}
	& =
	\Big[
	\EXP \big[ \big\{ A_i - e^* (\bX_i) \big\}^2 \bI(\bX_i)^{\otimes 2} \big]
	\Big]^{-1}
	\Big[
	\EXP \big[
	\epsilon_i
	\big\{ A_i - e^* (\bX_i) \big\} \big\{ \varphi^*(\bO_i) - \bI(\bX_i)\T\bT^* \big\} \bI(\bX_i)^{\otimes 2} \big]
	\Big]
	\Big[
	\EXP \big\{ \bI(\bX_i)^{\otimes 2} \big\}
	\Big]^{-1}
\end{align*}
If model \eqref{model-PLM} in the main paper is true, we find
\begin{align*}
\varphi^*(\bO_i) - \bI(\bX_i)\T \bT^* 
=
\epsilon_i \bigg\{ \frac{A_i}{e^*(\bX_i)} - \frac{1-A_i}{1-e^*(\bX_i)} \bigg\} \ , 
\end{align*}
and $\EXP(\epsilon_i^2 \cond A_i, \bX_i) = \sum_{g=1}^G \sigma_{\epsilon,g}^2 \ind \{ M(\bX_i) = g \}$. Combining all, we find
\begin{align*}
	&
	\EXP \big[ \big\{ A_i - e^* (\bX_i) \big\}^2 \bI(\bX_i)^{\otimes 2} \big]
	=
	{\rm diag} \Big(
		p_g^* \EXP \big[ e^*(\bX_i) \{ 1-e^*(\bX_i) \} \cond M(\bX_i)= g \big]
	\Big)_{g=1,\ldots,G}
	\\
	&
	\EXP \big[ \epsilon_i^2 \big\{ A_i - e^* (\bX_i) \big\}^2 \bI(\bX_i)^{\otimes 2} \big]
	=
	{\rm diag} \Big(
		p_g^* \sigma_{\epsilon,g}^2 \EXP \big[ e^*(\bX_i) \{ 1-e^*(\bX_i) \} \cond M(\bX_i)= g \big]
	\Big)_{g=1,\ldots,G}
	\\
	&
	\EXP \big\{ \bI(\bX_i)^{\otimes 2} \big\}
	=
	{\rm diag} \Big(
		p_g^* 
	\Big)_{g=1,\ldots,G}
	\\
	&
	\EXP \big[
	\big\{ \varphi^*(\bO_i) - \bI(\bX_i)\T\bT^* \big\}^2
	\bI(\bX_i)^{\otimes 2} \big]
	=
	{\rm diag} \Bigg(
		 p_g^* \EXP\bigg[
		\frac{1}{e^*(\bX_i) \{ 1- e^*(\bX_i) \}}
		\, \bigg| \, M(\bX_i) = g
	\bigg]
	\Bigg)_{g=1,\ldots,G}
	\\
	&
	\EXP \big[
	\epsilon_i
	\big\{ A_i - e^* (\bX_i) \big\} \big\{ \varphi^*(\bO_i) - \bI(\bX_i)\T\bT^* \big\} \bI(\bX_i)^{\otimes 2} \big]
	\Big]
	=
	{\rm diag} \Big(
		p_g^*  \sigma_{\epsilon,g}^2
	\Big)_{g=1,\ldots,G} \ .
\end{align*}
Therefore, the variances $\Sigma_\SSLS$, $\Sigma_\EIF$, and $\Sigma_{\SSLS,\EIF}$ are
\begin{align*}
	&
	\Sigma_\SSLS
	=
	{\rm diag} (\sigma_{\SSLS,1}^2, \ldots, \sigma_{\SSLS,G}^2 ) 
	\ , \
	\sigma_{\SSLS,g}^2 
	=
	\frac{\sigma_{\epsilon,g}^2}{p_g^* } \frac{1}{\EXP \big[ e^*(\bX_i) \{ 1- e^*(\bX_i) \} \cond M(\bX_i) = g \big]}
	\\
	&
	\Sigma_\EIF
	=
	{\rm diag} (\sigma_{\EIF,1}^2, \ldots, \sigma_{\EIF,G}^2 ) 
	\ , \
	\sigma_{\EIF,g}^2
	=
	\frac{\sigma_{\epsilon,g}^2}{p_g^*} \EXP\bigg[
		\frac{1}{e^*(\bX_i) \{ 1- e^*(\bX_i) \}}
		\, \bigg| \, M(\bX_i) = g
	\bigg]
	\\
	&
	\Sigma_{\SSLS,\EIF}
	=
	{\rm diag} (\sigma_{\SSLS,\EIF,1}, \ldots, \sigma_{\SSLS,\EIF,G} ) 
	\ , \
	\sigma_{\SSLS,\EIF,g}^
	=
	\frac{\sigma_{\epsilon,g}^2}{p_g^* } \frac{1}{\EXP \big[ e^*(\bX_i) \{ 1- e^*(\bX_i) \} \cond M(\bX_i) = g \big]} \ .
\end{align*}
Therefore, $\Sigma_\SSLS = \Sigma_{\SSLS,\EIF}$. Moreover, from the Jensen's inequality, we find
\begin{align*}
	\frac{1}{\EXP \big[ e^*(\bX_i) \{ 1- e^*(\bX_i) \} \cond M(\bX_i) = g \big]} 
	\leq
	\EXP\bigg[
		\frac{1}{e^*(\bX_i) \{ 1- e^*(\bX_i) \}}
		\, \bigg| \, M(\bX_i) = g
	\bigg] \ ,
\end{align*}
where the equality holds if and only if $e^*(\bX_i)$ is constant for $\bX_i$ in the $g$th subgroup. That is, $\Sigma_\EIF - \Sigma_\SSLS$ is positive semi-definite and $\Sigma_\EIF = \Sigma_\SSLS$ if $e^*(\bX_i)$ is constant  within each subgroup. This concludes the proof.

\newpage

\section{Proof of Theorems in the Main Paper} \label{sup-sec:main-thms}

\subsection{Proof of Theorem \ref{thm:GPLM1} and Theorem \ref{coro:GPLM1} in the Main Paper}			\label{proof:SSLS}

	The proof follows from Theorem 3.1 and 3.2 of \citet{victor2018}. For completeness, we provide a full exposition tailored to our context below. For notational brevity, we define $\bV_i^* = \{ A_i - e^*(\bX_i) \} \bI(\bX_i) $, $Z_i^* = Y_i - \nu^*(\bX_i) $, $\widehat{\bV}_i\kk = \{ A_i - e\kk (\bX_i) \} \bI(\bX_i) $, $\widehat{Z}_i\kk =  Y_i - \nu\kk(\bX_i) $, $\xi_i = Z_i^* - \bV_i\sT \bT_\OV^* $, and $\xi_i\kk = \widehat{Z}_i\kk - \widehat{\bV}_i\kT \bT_\OV^* $. Note that the conditions in Lemma \ref{lem-assumption-SSLS} hold.

	We first show the claim in Theorem \ref{coro:GPLM1} by showing $\bT_\OV^* = \bT^*$ if  $\EXP \big[ e^*(\bX_i) \{ 1-e^*(\bX_i) \} \{ \tau^*(\bX_i) - \tau_g^* \} \cond M(\bX_i) = g \big] = 0$. 
	\begin{align*}
			\tau_{\OV,g}^*
			=
			 \frac{\EXP \big[ e^*(\bX_i) \{ 1- e^*(\bX_i) \} \tau^*(\bX_i) \cond M(\bX_i) = g \big] }{\EXP \big[ e^*(\bX_i) \{ 1- e^*(\bX_i) \} \cond M(\bX_i) = g \big] }
			=
			\tau_g^* \frac{\EXP \big[ e^*(\bX_i) \{ 1- e^*(\bX_i) \} \cond M(\bX_i) = g \big] }{\EXP \big[ e^*(\bX_i) \{ 1- e^*(\bX_i) \} \cond M(\bX_i) = g \big] }
			=
			\tau_g^* \ .
	\end{align*}
	Therefore, Theorem \ref{thm:GPLM1} is a special case where $\bT^* = \bT_\OV^*$ of the results below.\\

\noindent \textbf{Step 2}:  For the simplicity, we denote 
	\begin{align*}
		R_{N,1} & : = 
		\frac{1}{N} \sum_{k=1}^2 \sum_{i \in \mathcal{I}_k}\widehat{\bV}_i\kk \widehat{\bV}_{i} \kT
		- \EXP \big( \bV_i^* \bV_i\sT \big) \ , \
		R_{N,2} : = 
		\frac{1}{N} \sum_{k=1}^2 \sum_{i \in \mathcal{I}_k}\widehat{\bV}_i\kk \xi_i\kk
		- \frac{1}{N}  \sum_{i =1}^N \bV_{i}^* \xi_i
		\ .
	\end{align*}
	In following \textbf{Step 3} -- \textbf{Step 6}, we will show that
	\begin{align}					
		& \textbf{Step 3  } \ : \  \hspace*{1cm} \| R_{N,1} \|_2 = o_P(1) \ , \label{proof1-01-01} 
		\\
		& \textbf{Step 4  } \ : \ \hspace*{1cm} \sqrt{N} \|R_{N,2} \|_2 = o_P(1) \ , \label{proof1-01-02} 
		\\
		& \textbf{Step 5  } \ : \ \hspace*{1cm} \frac{1}{\sqrt{N}} \left\|
			 \sum_{i =1}^N \bV_{i}^* \xi_i
		\right\|_2 = O_P(1) \ , \label{proof1-01-03} 
		\\
		& \textbf{Step 6  } \ : \ \hspace*{1cm} \| \widehat{\Sigma} - \Sigma^* \|_2 = o_P(1) \ . \label{proof1-01-04}
	\end{align}

	From Lemma \ref{lem-assumption-SSLS} (c), we observe that the singular values of $\EXP \big\{ \bV_i\kk \bV_i\kT \cond \mathcal{I}_k^C \big\}$ is bounded below by a constant. Therefore, the singular values of $N^{-1} \sum_{k=1}^2 \sum_{i \in \mathcal{I}_k} \widehat{\bV}_i\kk \widehat{\bV}_i\kT$ are positive with probability $1-o(1)$ from the law of large numbers.  This implies $\widehat{\bT}_\SSLS$ is well-defined with probability $1-o(1)$, and in addition, the $\sqrt{N}$-scaled difference between $\widehat{\bT}_\SSLS - \bT_\OV^*$ can be represented as
	\begin{align}
		&
		\nonumber \sqrt{N} \Big( \widehat{\bT}_\SSLS -  \bT_\OV^* \Big) 
		\\
		\nonumber & = \sqrt{N} \bigg\{ \frac{1}{N} \sum_{k=1}^2 \sum_{i \in \mathcal{I}_k}\widehat{\bV}_i\kk \widehat{\bV}_i\kT  \bigg\}^{-1}
		\bigg[
			\frac{1}{N} \sum_{k=1}^2 \sum_{i \in \mathcal{I}_k}\widehat{\bV}_i\kk \widehat{Z}_i\kk
			- \bigg\{ \frac{1}{N} \sum_{k=1}^2 \sum_{i \in \mathcal{I}_k}\widehat{\bV}_i\kk \widehat{\bV}_i\kT \bigg\}   \bT_\OV^*
		\bigg] \\
		\nonumber & = \Big\{ \EXP \big( \bV_i^* \bV_i\sT \big)  + R_{N,1} \Big\}^{-1} 
		\bigg[
			\frac{1}{\sqrt{N}} \sum_{k=1}^2 \sum_{i \in \mathcal{I}_k} \widehat{\bV}_i\kk \big\{ \widehat{Z}_i\kk - \widehat{\bV}_i\kT  \bT_\OV^* \big\}
		\bigg] \\
		& = \Big\{ \EXP \big( \bV_i^* \bV_i\sT \big)  + R_{N,1} \Big\}^{-1} 
		\bigg\{
			\frac{1}{\sqrt{N}} \sum_{i=1}^N {\bV}_{i}^* \big( {Z}_{i}^* - {\bV}_{i}\sT  \bT_\OV^* \big) + \sqrt{N} R_{N,2}
		\bigg\}	\ . \label{proof1-01-05}
	\end{align}
	Note that 	$\big\{ \EXP \big( \bV_i^* \bV_i\sT \big)  + R_{N,1} \big\}^{-1} -  \EXP \big( \bV_i^* \bV_i\sT \big) ^{-1} = -\big\{ \EXP \big( \bV_i^* \bV_i\sT \big)   + R_{N,1} \big\}^{-1} R_{N,1}  \EXP \big(\bV_i^* \bV_i\sT \big)^{-1}$, and, combining \eqref{proof1-01-01} and Assumption \ref{assp:SSLS}, we get
	\begin{align}
		&
	 	\big\| \big\{ \EXP \big( \bV_i^* \bV_i\sT \big)  + R_{N,1} \big\}^{-1} - \EXP \big(\bV_i^* \bV_i\sT \big) ^{-1} \big\|_2 
	 	\nonumber
	 	\\
	 	& \leq \big\| \big\{ \EXP \big(\bV_i^* \bV_i\sT \big)   + R_{N,1} \big\}^{-1} \big\|_2 \big\| R_{N,1}\big\|_2 \big\| \EXP \big(\bV_i^* \bV_i\sT \big)^{-1} \big\|_2 
	 	 = o_P(1) \ .	\label{proof1-01-06}
	 \end{align} 
	 By \eqref{proof1-01-02} and \eqref{proof1-01-03}, the second term in \eqref{proof1-01-05} is
	 \begin{align}
	 	&
	 \bigg\| \frac{1}{\sqrt{N}} \sum_{k=1}^2 \sum_{i \in \mathcal{I}_k} {\bV}_{i}^* \big( {Z}_{i}^* - {\bV}_{i}\sT \bT_\OV^* \big) + \sqrt{N} R_{N,2}  \bigg\|_2  
	 \nonumber
	 \\
	 	& \leq  \bigg\| \frac{1}{\sqrt{N}} \sum_{k=1}^2 \sum_{i \in \mathcal{I}_k} {\bV}_{i}^* \big( {Z}_{i}^* - {\bV}_{i}\sT\bT_\OV^* \big)  \bigg\|_2  + \big\|\sqrt{N} R_{N,2}  \big\|_2  
	 	 = O_P(1) \ . \label{proof1-01-07}
	 \end{align}
	Combining \eqref{proof1-01-06} and \eqref{proof1-01-07} gives 
	\begin{align}
		\bigg\|
			\big[
				 \big\{ \EXP \big(\bV_i^* \bV_i\sT \big)  + R_{N,1} \big\}^{-1} - \EXP \big(\bV_i^* \bV_i\sT \big) ^{-1} 
			\big]
			\bigg\{
				 \frac{1}{\sqrt{N}} \sum_{i=1}^N {\bV}_{i}^* \big( {Z}_{i}^* - {\bV}_{i}\sT \bT_\OV^*  \big) + \sqrt{N} R_{N,2}
			\bigg\}
		\bigg\|_2  = o_P(1) \ .	\label{proof1-01-08}
	\end{align}
	Substituting \eqref{proof1-01-08} into \eqref{proof1-01-05} leads to 
	\begin{align*}
		\sqrt{N} \Big( \widehat{\bT}_\SSLS - \bT_\OV^*  \Big) 
		& = \EXP \big(\bV_i^* \bV_i\sT \big) ^{-1}  \bigg\{
			\frac{1}{\sqrt{N}} \sum_{k=1}^2 \sum_{i \in \mathcal{I}_k} {\bV}_{i}^* \big( {Z}_{i}^* - {\bV}_{i}\sT \bT_\OV^* \big) + \sqrt{N} R_{N,2}
		\bigg\} + o_P(1) \\
		& = 
		\EXP \big(\bV_i^* \bV_i\sT \big) ^{-1}  \bigg\{
			\frac{1}{\sqrt{N}} \sum_{k=1}^2 \sum_{i \in \mathcal{I}_k} {\bV}_{i}^* \big( {Z}_{i}^* - {\bV}_{i}\sT \bT_\OV^* \big) 
		\bigg\} + o_P(1) \ ,
	\end{align*}
	where the second equality used \eqref{proof1-01-02} and Assumption \ref{assp:SSLS}. Therefore, by the Lindberg-Feller central limit theorem and the Cramer-Wold theorem, we obtain the asymptotic normality:
	\begin{align*}
		\sqrt{N} \Big( \widehat{\bT}_\SSLS - \bT_\OV^* \Big) \stackrel{D}{\rightarrow} N \Big( 0 , \Sigma^* \Big) \ ,
	\end{align*}
	where $ \Sigma^* = \EXP \big(\bV_i^* \bV_i\sT \big)^{-1}  \EXP \big( \xi_i^2  \bV_i^*\bV_i\sT \big) \EXP \big(\bV_i^*\bV_i\sT \big)^{-1}$. Note that $\Sigma^*$ is well defined because $ \EXP \big(\bV_i^* \bV_i\sT \big)^{-1}$ is well-defined and $ \EXP \big( \xi_i^2  \bV_i^*\bV_i\sT \big) = \EXP \big \{ \EXP (\xi_i^2 \cond \bV_i^*)   \bV_i^*\bV_i\sT \big\} $ does not diverge. 

	To finish the proof, we need to show \eqref{proof1-01-01} - \eqref{proof1-01-04} in following steps. \\
	
\noindent \textbf{Step 3} : We prove \eqref{proof1-01-01}, which suffices to show that
	\begin{align}				\label{proof1-02-01}
		\bigg\|
			\frac{1}{N/2} \sum_{i \in \mathcal{I}_k}\widehat{\bV}_i\kk \widehat{\bV}_i\kT 
		- \EXP \big(\bV_i^* \bV_i\sT \big)
		\bigg\|_2 = o_P(1) \ 
	\end{align}
	for $k=1,2$. The above value is upper bounded by the sum of two quantities $B_1 + B_2$, where
	\begin{align*}
		B_1 & = \bigg\| \frac{1}{N/2}  \sum_{i \in \mathcal{I}_k} \widehat{\bV}_i\kk \widehat{\bV}_i\kT 
		- \EXP \big\{ \widehat{\bV}_i\kk \widehat{\bV}_i\kT  \cond \mathcal{I}_k^C
		\big\}
		\bigg\|_2  \ , \ 
		B_2 = \Big\|  \EXP \big\{ \widehat{\bV}_i\kk \widehat{\bV}_i\kT  \cond \mathcal{I}_k^C
		\big\}- \EXP \big( \bV_i^* \bV_i\sT \big) 
		\Big\|_2 \ .
	\end{align*}

The conditional expectation of $B_1^2$ conditional on the samples in $\mathcal{I}_k^C$ is upper bounded by a constant from Lemmas \ref{lem1-001} and \ref{lem-assumption-SSLS}:
	\begin{align*}
		\EXP \big(B_1^2 \cond \mathcal{I}_k^C \big) 
		 \leq \frac{\EXP \big\{
			\| \widehat{\bV}_i\kk \widehat{\bV}_i\kT \|_2^2 \cond \mathcal{I}_k^C
		\big\}}{N/2} 
		\leq  \frac{\EXP \big\{
			\| \widehat{\bV}_i\kk \|_2^4 \cond \mathcal{I}_k^C
		\big\}}{N/2} 
		 = \frac{ \| \widehat{\bV}_i\kk \|_{P,4}^4}{N/2} \ .
	\end{align*}
	Therefore, from Lemma \ref{lem1-002}, this implies $B_1 = O_P(N^{-1/2})$, so $o_P(1)$.

	To bound $B_2$, we first observe $\EXP \big( \bV_i^* \bV_i\sT \big) =\EXP \big(\bV_i^* \bV_i\sT \cond \mathcal{I}_k^C \big)$, which leads $B_2= \EXP \big\{
	\widehat{\bV}_i\kk \widehat{\bV}_i\kT  - \bV_i^* \bV_i\sT
		\cond \mathcal{I}_k^C
	\big\} $. We can further find that 
	\begin{align}					
	\hspace*{-0.4cm}
		\widehat{\bV}_i\kk \widehat{\bV}_i\kT  - \bV_i^* \bV_i\sT
		 = \frac{1}{2} \Big[
			\big\{ \widehat{\bV}_i\kk - \bV_i^* \big\} \big\{  \widehat{\bV}_i\kk + \bV_i^*\big\}^\intercal + \big\{  \widehat{\bV}_i\kk + \bV_i^* \big\} \big\{  \widehat{\bV}_i\kk - \bV_i^* \big\}^\intercal
		\Big] \ .		\label{proof1-02-02}
	\end{align}
	Therefore, by applying Lemma \ref{lem1-001} to \eqref{proof1-02-02}, $B_2$ is upper bounded by
	\begin{align*}
		B_2 \leq \norm{  \widehat{\bV}_i\kk - \bV_i^* }_{P,2}\norm{ \widehat{\bV}_i\kk + \bV_i^* }_{P,2} \ .
	\end{align*}
	Lemma \ref{lem-assumption-SSLS} implies that $\norm{  \widehat{\bV}_i\kk + \bV_i^* }_{P,2}$ is bounded by a constant and $\norm{  \widehat{\bV}_i\kk - \bV_i^* }_{P,2}$ vanishes as $N$ increases, so that $B_2$ is $o_P(1)$. Combining the results of $B_1=o_P(1)$ and $B_2=o_P(1)$ gives \eqref{proof1-02-01}. \\
	
	\noindent \textbf{Step 4} : We prove \eqref{proof1-01-02}, which suffices to show that		
	\begin{align}				\label{proof1-03-01}
		\frac{1}{\sqrt{N/2}} \bigg\| \sum_{i \in \mathcal{I}_k}\widehat{\bV}_i\kk \big( \widehat{Z}_i\kk - \widehat{\bV}_i\kT \bT_\OV^* \big)
		- \sum_{i \in \mathcal{I}_k} \bV_{i}^* \big( Z_{i}^* - \bV_{i}\sT \bT_\OV^* \big) \bigg\|_2 = o_P(1) 
	\end{align}
	for $k=1,2$. Above value is upper bounded by the sum of two quantities $B_3+B_4$, where
	\begin{align*}
		B_3 & = \frac{1}{\sqrt{N/2}} \bigg\|
			 \sum_{i \in \mathcal{I}_k} \left[ \widehat{\bV}_i\kk \big\{ \widehat{Z}_i\kk - \widehat{\bV}_i\kT \bT_\OV^* \big\} 
			 - \EXP \big[ \widehat{\bV}_i\kk \big\{ \widehat{Z}_i\kk - \widehat{\bV}_i\kT \bT_\OV^* \big\}  \cond \mathcal{I}_k^C \big]   \right] 
			 \\
			& \hspace*{4cm} - \sum_{i \in \mathcal{I}_k} \left[ \bV_{i}^* \big( Z_{i}^* - \bV_{i}\sT  \bT_\OV^* \big) 
			- \EXP \big\{ \bV^*(Z^*- \bV\sT \bT_\OV^* ) \big\} \right]
		\bigg\|_2 \ , \\
		B_4 & =\sqrt{ \frac{N}{2} } \Big\|
			\EXP \big[ \widehat{\bV}_i\kk \big\{ \widehat{Z}_i\kk - \widehat{\bV}_i\kT \bT_\OV^* \big\} \cond \mathcal{I}_k^C \big] - \EXP \big\{ \bV_i^*(Z_i^*-\bV_i\sT\bT_\OV^*) \big\} 
		\Big\|_2 \ . 
	\end{align*}
	The conditional expectation of $B_3^2$ conditional on the sample in $\mathcal{I}_k^C$ is upper bounded by a constant from Lemma \ref{lem1-001} and \ref{lem-assumption-SSLS}:
	\begin{align*}
		\EXP \big(B_3^2 \cond \mathcal{I}_k^C \big)
		&\leq \EXP \left[
			\left\| \widehat{\bV}_i\kk \big\{ \widehat{Z}_i\kk - \widehat{\bV}_i\kT \bT_\OV^* \big\}  - \bV_i^*(Z_i^*-\bV_i\sT \bT_\OV^*) \right\|_2^2 \, \Big| \, \mathcal{I}_k^C
		\right] \\
		& 
		= \big\| \widehat{\bV}_i\kk \big\{ \widehat{Z}_i\kk - \widehat{\bV}_i\kT \bT_\OV^* \big\}  - \bV_i^*(Z_i^*-\bV_i\sT \bT_\OV^*)  \big\|_{P,2}^2  =o_P(1) \ .
	\end{align*}
	Therefore, from Lemma \ref{lem1-002}, this implies $B_3 = o_P(1)$. Next, we observe $\EXP \big\{ \bV_i^*(Z_i^*- \bV_i\sT \bT_\OV^* ) \big\} = 0$, so $B_4=o_P(1)$ is trivial from Lemma \ref{lem-assumption-SSLS}. As a result, \eqref{proof1-03-01} is obtained. \\
	
	\noindent \textbf{Step 5} : We prove \eqref{proof1-01-03}. From Lemma \ref{lem-assumption-SSLS}, we can find 
	\begin{align*}
		E \bigg\{ \bigg\|
			\frac{1}{\sqrt{N}} \sum_{i=1}^N \bV_i^* \big( Z_i^* - \bV_i\sT \bT_\OV^*  \big)
		\bigg\|_2^2 \, \bigg\}
		= \EXP \big\{ \| \bV_i^* \xi_i \|_2^2 \big\}
		= \big\| \bV_i^* \xi_i \big\|_{P,2}^2
	\end{align*}
	is bounded by a constant, which implies \eqref{proof1-01-03}.  \\
	
	\noindent  \textbf{Step 6} : We prove \eqref{proof1-01-04}. For simpler notations, we denote
	\begin{align*}
		\widehat{G} = \frac{1}{N} \sum_{k=1}^2 \sum_{i \in \mathcal{I}_k} \widehat{\bV}_i\kk \widehat{\bV}_i\kT \quad , \quad
		\widehat{H} = \frac{1}{N} \sum_{k=1}^2 \sum_{i \in \mathcal{I}_k} \{ \widehat{\xi}_i\kk \} ^2 \widehat{\bV}_i\kk \widehat{\bV}_i\kT  \ ,
	\end{align*}
	so that $\widehat{\Sigma} = \widehat{G}^{-1} \widehat{H} \widehat{G}^{-1}$. Then, we find the difference between $\widehat{\Sigma}$ and $\Sigma^*$ is
	\begin{align*}
		\widehat{\Sigma} - \Sigma^*
		& = \widehat{G}^{-1} \widehat{H} \widehat{G}^{-1} - \EXP \big(\bV_i^* \bV_i\sT \big)^{-1} \EXP \big(\xi_i^2 \bV_i^* \bV_i\sT \big)  \EXP \big(\bV_i^* \bV_i\sT \big)^{-1} \\
		& = \widehat{G}^{-1} \Big\{
			\widehat{H} - \EXP \big( \xi_i^2 \bV_i^* \bV_i\sT \big)
		\Big\} \widehat{G}^{-1} + \Big\{ \widehat{G}^{-1} -  \EXP \big(\bV_i^* \bV_i\sT \big)^{-1}  \Big\} \EXP \big(\xi_i^2 \bV_i^* \bV_i\sT \big)  \Big\{ \widehat{G}^{-1} +  \EXP \big( \bV_i^* \bV_i\sT \big)^{-1}  \Big\} \ .
	\end{align*}
	From \eqref{proof1-01-06} and finite $\|\EXP \big(\xi_i^2 \bV_i^* \bV_i\sT \big) \|_2$ induced by Lemma \ref{lem-assumption-SSLS}, we find the second term is $o_P(1)$. Therefore, to prove \eqref{proof1-01-04}, it suffices to show that $\widehat{H} - \EXP \big(\xi_i^2 \bV_i^* \bV_i\sT \big)$ is $o_P(1)$ because $\widehat{G}^{-1}$ is $O_P(1)$ and it is achieved if each component of $\widehat{H} - \EXP \big(\xi_i^2 \bV_i^* \bV_i\sT \big)$ is $o_P(1)$. That is, 
	\begin{align}			\label{proof1-05-01}
		\frac{1}{N/2} \sum_{i \in \mathcal{I}_k} \{ \widehat{\xi}_i\kk \} ^2 \widehat{\bV}_i\kk \widehat{\bV}_i\kT
		- \EXP \big\{ \xi_i^2 \bV_i^* \bV_i\sT \big\}  = o_P(1) \ .
	\end{align}
	The left hand side of \eqref{proof1-05-01} is upper bounded by a sum of quantities $B_5+B_6$, where
	\begin{align*}
		B_5 & = 
			\sum_{i \in \mathcal{I}_k} \{ \widehat{\xi}_i\kk \} ^2 \widehat{\bV}_i\kk \widehat{\bV}_i\kT
			 - \sum_{i \in \mathcal{I}_k} \xi_i^2  \bV_i^* \bV_i\sT 
			  \ , \
		B_6  = \frac{1}{N/2} 
			\sum_{i \in \mathcal{I}_k} \xi_i^2  \bV_i^* \bV_i\sT
			- \EXP \big(  \xi_i^2 \bV_i^* \bV_i\sT  \big)  \ .
	\end{align*}
	Moreover, the summands in $B_5$ are upper bounded by
	\begin{align*}
	&
		\{ \widehat{\xi}_i\kk \} ^2 \widehat{\bV}_i\kk \widehat{\bV}_i\kT
		 -
		\xi_i^2  \bV_i^* \bV_i\sT
		=
		\widehat{ \bm{h} }_i\kk 	\widehat{\bm{h}}_i\kT + \xi_i \widehat{\bm{h}}_i\kk  \bV_i\sT + \xi_i  \bV_i^* \widehat{\bm{h}}_i\kT \ ,
	\end{align*}
	where $\widehat{ \bm{h} }_i\kk = \widehat{\xi}_i\kk  \widehat{\bV}_i\kk - \xi_i \bV_i^* $. As a result, by the H\"older's inequality, we find
	\begin{align}							\label{proof1-05-02}
		\big\| B_5 \big\|_2
		& 
		\leq
		\bigg\|
			\frac{1}{N/2} \sum_{i \in \mathcal{I}_k} \widehat{ \bm{h} }_i\kk 	\widehat{\bm{h}}_i\kT
		\bigg\|_2
		+
		2
		\bigg\|
			\frac{1}{N/2} \sum_{i \in \mathcal{I}_k} \widehat{ \bm{h} }_i\kk  \xi_i \bV_i\sT
		\bigg\|_2
		\nonumber
		\\
		&
		\leq R_{N,3} + 2R_{N,3}^{1/2} \bigg\{ \frac{1}{N/2} \sum_{i \in \mathcal{I}_k} \norm{\bV_i^* \xi_i}_2^2 \bigg\}^{1/2}
	\end{align}
	where $R_{N,3} = (2/N) \sum_{i \in \mathcal{I}_k} \big\|
			\widehat{\bV}_i\kk \widehat{\xi}_i\kk - \bV_i^* \xi_i
		\big\|_2^2$. The summands in $R_{N,3}$ can be decomposed as follows:
	\begin{align*}
		\widehat{\bV}_i\kk \widehat{\xi}_i\kk - \bV_i^* \xi_i
		& = 
		\widehat{\bV}_i\kk(\widehat{Z}_i\kk-\widehat{\bV}_i\kT \widehat{\bT}_\SSLS ) - \bV_i^* ( Z_i^* - \bV_i\sT \bT_\OV^* ) 
		\\
		& = ( \widehat{\bV}_i\kk\widehat{Z}_i\kk - \bV_i^* Z_i^* ) - ( \widehat{\bV}_i\kk\widehat{\bV}_i\kT -\bV_i^* \bV_i\sT ) \bT_\OV^*
		- \widehat{\bV}_i\kk\widehat{\bV}_i\kT (\widehat{\bT}_\SSLS  - \bT_\OV^*) \ .
	\end{align*}
	As a result, 
	\begin{align*}
		 R_{N,3}^{1/2} & = \sqrt{\frac{2}{N}} \bigg\{
			\sum_{i \in \mathcal{I}_k} \left\|
			\widehat{\bV}_i\kk \widehat{\xi}_i\kk - \bV_i^* \xi_i
		\right\|_2^2
		\bigg\}^{1/2} \\
		& \leq
		\sqrt{\frac{2}{N}}  \bigg\{
			\sum_{i \in \mathcal{I}_k} \Big\| \widehat{\bV}_i\kk\widehat{\bV}_i\kT \Big\|_2^2
		\bigg\}^{1/2}
		\Big\| \widehat{\bT}_\SSLS  - \bT_\OV^* \Big\|_2
		\\
		& \hspace*{2cm}
		+ 
		\sqrt{ \frac{2}{N}} \Bigg\{ \sum_{i \in \mathcal{I}_k} \left\| \big\{ \widehat{\bV}_i\kk\widehat{Z}_i\kk - \bV_i^* Z_i^* \big\} - \big\{ \widehat{\bV}_i\kk  \widehat{\bV}_i\kT -\bV_i^* \bV_i\sT \big\}
		 \bT_\OV^* \right\|_2^2 \Bigg\}^{1/2} \ .
	\end{align*}
	Note that the first term in the upper bound is $o_P(1)$ from the main theorem result $\widehat{\bT}_\SSLS \stackrel{P}{\rightarrow} \bT_\OV^*$ and $\widehat{G}= O_P(1)$. The second term is also $o_P(1)$ from \eqref{proof1-03-01} which is already shown in the proof of \eqref{proof1-01-02}. As a result, $R_{N,3} $ is $o_P(1)$. In \eqref{proof1-05-02}, note that $ (2/N) \sum_{i \in \mathcal{I}_k} \norm{\bV_i^* \xi_i}_2^2$ is $O_P(1)$ because $E \big[ \| \bV_i^* \xi_i \|_2^2 \big]$ is bounded from Lemma \ref{lem-assumption-SSLS}, and this leads $B_5=o_P(1)$. The convergence of $B_6$ is straightforward from the law of large numbers. In particular, from the law of large numbers, each component of $B_6$ converges to zero in probability, i.e., $B_{6,j\ell} = o_P(1)$ where $B_{6,j\ell}$ is the$(j,\ell)$th element of $B_6$. Consequently, from the property of the matrix norm, we have 
	\begin{align*}
		\big\| B_6 \big\|_2
		\leq \big\| B_6 \big\|_{F}
		=
		\bigg\{
			\sum_{j,\ell} B_{6,j\ell}^2
		\bigg\}^{1/2}
		=
		o_P(1) \ ,
	\end{align*}
	where $\big\| \cdot \big\|_F$ is the Frobenius norm of a matrix. 	Combining the result of vanishing $B_5$ and $B_6$ shows  \eqref{proof1-05-01}.

Lastly, we show that $\widehat{\bT}_{\SP}$ achieves the semiparametric efficiency bound for $\tau^*$ under model $\model_{\SP}$ at model $\model_{\SP}^\dagger$ (in which the law satisfies either (i) $e^*(\bX_i) = 0.5$ or (ii) the variance is homoskedastic and $\tau^*(\bX_i) = \tau_g^*$ for each group). Let $\model_{\SP}\ETA$ be a collection of laws $P(\bO \con \eta)$ that are parametrized by a one-dimensional parameter $\eta$. Without loss of generality, the true data law is recovered at $\eta^*$. Additionally, $\model_{\SP}\ETA$ satisfies the moment restriction $\EXP\ETA \big[ \big\{ A_i - e(\bX_i \con \eta) \big\}^2 \big\{ \tau(\bX_i \con \eta) - \tau_g(\eta) \big\} \ind \{ M(\bX_i) = g \big\} \big] = 0$ where $\EXP\ETA(\cdot)$ is an expectation operator at law $P(\bO \con \eta)$. 

Let $s_Y(Y_i \cond A_i, \bX_i \con \eta)$, $s_A(A_i \cond \bX_i \con \eta)$, and $s_X(\bX_i \con \eta)$ be the score functions of the densities of $Y_i \cond A_i, \bX_i$, $A_i \cond \bX_i$, and $\bX_i$ at law $P(\bO)$.

It is worth study the derivative of the moment restriction with respect to $\eta$:
\begin{align*}
	&
	\frac{\partial}{\partial \eta}
	\EXP^{(\eta)} \Big[ e(\bX_i \con \eta) \{ 1- e(\bX_i \con \eta) \} \{ \tau(\bX_i \con \eta) - \tau_g(\eta) \} \ind \{ M(\bX_i) = g \} \Big]
		\\
	&
	=
	\EXP\ETA \Big[
		s_X(\bX_i \con \eta)  \{ A- e(\bX_i \con \eta) \}^2 \{ \tau(\bX_i \con \eta) - \tau_g(\eta) \} \ind \{ M(\bX_i) = g \}
	\Big]	
	\\
	&
	\quad
	+
	\EXP\ETA \Big[
		s_A (A=1 \cond \bX_i \con \eta) e(\bX_i \con \eta) \big\{ 1-2e(\bX_i \con \eta) \big\} \{ \tau(\bX_i \con \eta) - \tau_g(\eta) \} \ind \{ M(\bX_i) = g \}
	\Big]	
	\\
	&
	\quad
	+
	\EXP\ETA \Big[
		\{ A- e(\bX_i \con \eta) \}^2 \nabla_\eta \tau(\bX_i \con \eta)
		\ind \{ M(\bX_i) = g \}
	\Big]
	\\
	&
	\quad
	-
	\EXP\ETA \Big[
		\{ A- e(\bX_i \con \eta) \}^2
		\ind \{ M(\bX_i) = g \}
	\Big]	
	\cdot 
	\nabla_\eta \tau_g(\eta)
\end{align*}
where 
\begin{align*}
	\nabla_{\eta} \tau(\bX_i \con \eta)
	& =
	\frac{\partial \tau(\bX_i \con \eta) }{\partial \eta }
	\\
	&
	=
	\EXP\ETA \big\{ Y_i s_Y(Y_i \cond A_i =1,\bX_i\con\eta) \cond A_i=1,\bX_i \big\}
	-
	\EXP\ETA \big\{ Y_i s_Y(Y_i \cond A_i =0,\bX_i\con\eta) \cond A_i=0,\bX_i \big\}
\end{align*}
and
\begin{align*}
	\nabla_\eta \tau_g(\eta)
	& =
	\frac{\partial \tau_g(\eta) }{\partial \eta }
	\\
	&
	= \frac{\partial }{\partial \eta } \frac{\EXP^{(\eta)} \big[ \tau(\bX_i \con \eta) \ind \big\{ M(\bX_i) =g \big\} \big]}{\EXP^{(\eta)} \big[ \ind \big\{ M(\bX_i) =g \big\} \big]}
	\\
	&
	=
	\EXP\ETA \big[ 
		s_X (\bX_i \con \eta) \big\{ \tau(\bX_i \con \eta)  - \tau_g(\eta) \big\}
		+
		\nabla_\eta \tau(\bX_i \con \eta)  \cond M(\bX_i) = g
	\big]
\end{align*}

The tangent space of model $\model_{\SP}$ is therefore a collection of mean-zero, square-integrable functions of $\bO$ that satisfies the restriction induced by the moment restriction, i.e.,
\begin{align}       \label{eq-tangentspace}
\mathcal{T}_{\SP}
& 
=
	\left\{
	S_O(\bO_i)
	\left|	
	\begin{array}{l}
	S_O(\bO_i) = S_Y(Y_i \cond A_i, \bX_i) + S_A(A_i \cond \bX_i) + S_X(\bX_i) 
	\\
	\EXP \big\{ S_Y(Y_i \cond A_i, \bX_i) \cond A_i, \bX_i \big\}
	=
	\EXP \big\{ S_A(A_i \cond \bX_i) \cond \bX_i \big\}
	=
	\EXP \big\{ S_X( \bX_i)  \big\}
	=
	0
	\\
	\EXP \big\{ S_O(\bO_i)^2 \big\} < \infty
	\\
	S_O (\bO_i) \text{ satisfies } \eqref{eq:score} 
	\end{array}
		\right.
	\right\} \ ,
\end{align} 
where \eqref{eq:score}  is
\begin{align}	\label{eq:score}
	&
\EXP \Big[
		S_X(\bX_i)  \{ A_i - e^*(\bX_i) \}^2 \{ \tau^*(\bX_i) - \tau_g^* \}  \Cond M(\bX_i) = g
	\Big]	
		\nonumber
	\\
	&
	\quad
	+
	\EXP \Big[
		S_A(A=1 \cond \bX_i) e^*(\bX_i) \big\{ 1-2e^*(\bX_i) \big\} \{ \tau^*(\bX_i) - \tau_g^* \}
		 \Cond M(\bX_i) = g
	\Big]	
		\nonumber
	\\
	&
	\quad
	+
	\EXP \Bigg[
		\{ A_i - e^*(\bX_i) \}^2
		\Bigg[ 
		\begin{array}{l}
			\EXP \big\{ Y_i S_Y(Y_i \cond A_i =1,\bX_i) \cond A_i=1,\bX_i \big\}
			\\
	-
	\EXP \big\{ Y_i S_Y(Y_i \cond A_i =0,\bX_i) \cond A_i=0,\bX_i \big\}
		\end{array}
	 \Bigg]
	  \, \Bigg| \,  M(\bX_i) = g
	\Bigg]
		\nonumber
	\\
	&
	\quad
	-
	\EXP \Big[
		\{ A_i - e^*(\bX_i) \}^2  \Cond M(\bX_i) = g
	\Big]	
	\nonumber
	\\
	& \quad \quad
	\times
	\EXP \Bigg[
		S_X (\bX_i) \big\{ \tau^*(\bX_i)  - \tau_g^* \big\}
		+
		\Bigg[ 
		\begin{array}{l}
			\EXP \big\{ Y_i S_Y (Y_i \cond A_i =1,\bX_i) \cond A_i=1,\bX_i \big\}
			\\
	-
	\EXP \big\{ Y_i S_Y (Y_i \cond A_i =0,\bX_i) \cond A_i=0,\bX_i \big\}
		\end{array}
	 \Bigg]
	 \, \Bigg| \,  M(\bX_i) = g
	\Bigg]
		\nonumber
	\\
	& = 0 \ .
\end{align}
Let us denote the influence function of $\widehat{\bT}_{\SP}$ as ${\bIF}_\SP = (\bIF_{\SP,1},\ldots,\bIF_{\SP,G})\T$ where
\begin{align*}
	\bIF_{\SP,g}(\bO_i)
	&
	=
	\frac{ \big[ Y_i - \nu^*(\bX_i) - \big\{ A_i - e^*(\bX_i) \big\} \tau_g^*  \big] \big\{ A_i - e^*(\bX_i) \big\}  \ind \big\{ M(\bX_i)= g \big\} }{  \EXP \big[ \big\{ A_i - e^*(\bX_i) \big\}^2 \ind \{ M(\bX_i)=g \} \big]  }
	\\
	&
	=
	\frac{\left[
	\begin{array}{l}
	\big\{ Y_i - \mu^*(A_i, \bX_i) \big\}	\big\{ A_i - e^*(\bX_i) \big\} \ind \big\{ M(\bX_i)= g \big\}
	\\ 
	+
	\big\{ \tau^*(\bX_i) - \tau_g^* \big\}	\big\{ A_i - e^*(\bX_i) \big\} \big\{ 1-2e^*(\bX_i) \big\} \ind \big\{ M(\bX_i)= g \big\}
	\\ 
	+
	\big\{ \tau^*(\bX_i) - \tau_g^* \big\}	\big\{ e^*(\bX_i) \big\} \big\{ 1-e^*(\bX_i) \big\} \ind \big\{ M(\bX_i)= g \big\}
	\end{array}
	\right]}{ \EXP \big[ \big\{ A_i - e^*(\bX_i) \big\}^2 \ind \{ M(\bX_i)=g \} \big]  }
\end{align*}
which is straightforward from $\nu^*(\bX_i) = e^*(\bX_i) \mu^*(1,\bX_i) + \big\{ 1-e^*(\bX_i) \big\} \mu^*(0,\bX_i)$. This influence function satisfies 
\begin{align*}
	&
	\EXP\big\{
		s^*(\bO_i)
		\bIF_{\SP,g} (\bO_i)
	\big\}
	 \EXP \big[ \big\{ A_i - e^*(\bX_i) \big\}^2 \ind \{ M(\bX_i)=g \} \big] 
	\\
	&
	=
	\EXP 
	\Bigg[
		e^*(\bX_i) \big\{ 1-e^*(\bX_i) \big\}
		\bigg[ 
		\begin{array}{l}
		\EXP \big\{ Y_i S_Y (Y_i \cond A_i =1,\bX_i) \cond A_i=1,\bX_i \big\}
		\\
		-
		\EXP \big\{ Y_i S_Y (Y_i \cond A_i =0,\bX_i) \cond A_i=0,\bX_i \big\}
		\end{array}
		\bigg]
		 \ind \big\{ M(\bX_i)= g \big\}
	\Bigg]
	\\
	&
	+ \EXP \Big[
		s_A^*(A_i=1 \cond \bX_i) \big\{ \tau^*(\bX_i) - \tau_g^* \big\} e^*(\bX_i) \big\{ 1 -2 e^*(\bX_i) \big\}  \ind \big\{ M(\bX_i)= g \big\}
	\Big]
	\\
	&
	+
	\EXP \Big[
		s_X^*(\bX_i) \big\{ \tau^*(\bX_i) - \tau_g^* \big\}	\big\{ e^*(\bX_i) \big\} \big\{ 1-e^*(\bX_i) \big\} \ind \big\{ M(\bX_i)= g \big\}
	\Big]
	\\
	&
	=
	P \big\{ M(\bX_i) = g \big\}
	\left[
		\begin{array}{l}
		\EXP \big[
		s_X^*(\bX_i)  \{ A_i - e^*(\bX_i) \}^2 \{ \tau^*(\bX_i) - \tau_g^* \}  \cond M(\bX_i) = g
	\big]	
	\\
	+
	\EXP \big[
		s_A^*(A=1 \cond \bX_i) e^*(\bX_i) \big\{ 1-2e^*(\bX_i) \big\} \{ \tau^*(\bX_i) - \tau_g^* \}
		 \cond M(\bX_i) = g
	\big]	
	\\
	+
	\EXP \bigg[
		\{ A_i - e^*(\bX_i) \}^2
		\bigg[ 
		\begin{array}{l}
			\EXP \big\{ Y_i S_Y(Y_i \cond A_i =1,\bX_i) \cond A_i=1,\bX_i \big\}
			\\
	-
	\EXP \big\{ Y_i S_Y(Y_i \cond A_i =0,\bX_i) \cond A_i=0,\bX_i \big\}
		\end{array}
	 \bigg]
	  \, \bigg| \,  M(\bX_i) = g
	\bigg]
		\end{array}
	\right]
	\\
	&
	\stackrel{(*)}{=}
	P \big\{ M(\bX_i) = g \big\}
	\EXP \Big[
		\{ A_i - e^*(\bX_i) \}^2  \Cond M(\bX_i) = g
	\Big]	
	\nonumber
	\\
	& \quad \quad
	\times
	\EXP \Bigg[
		s_X^* (\bX_i) \big\{ \tau^*(\bX_i)  - \tau_g^* \big\}
		+
		\Bigg[ 
		\begin{array}{l}
			\EXP \big\{ Y_i s_Y^* (Y_i \cond A_i =1,\bX_i) \cond A_i=1,\bX_i \big\}
			\\
	-
	\EXP \big\{ Y_i s_Y^* (Y_i \cond A_i =0,\bX_i) \cond A_i=0,\bX_i \big\}
		\end{array}
	 \Bigg]
	 \, \Bigg| \,  M(\bX_i) = g
	\Bigg]
	\\
	&
	= 
	\EXP \Big[
		\{ A_i - e^*(\bX_i) \}^2  \ind \big\{ M(\bX_i) = g \big\}
	\Big]	
	\EXP \big[ 
		s_X^* (\bX_i ) \big\{ \tau^*(\bX_i)  - \tau_g^* \big\}
		+
		\nabla_\eta \tau(\bX_i \con \eta^*)  \cond M(\bX_i) = g
	\big]
\end{align*}
Here equality $(*)$ holds from \eqref{eq:score}. Therefore, we find that the groupwise treatment effects are differentiable \citep{Newey1990}:
\begin{align*}
	\nabla_\eta \tau_g(\eta) \big|_{\eta=\eta^*}
	=
	\EXP \big[ 
		s_X^* (\bX_i ) \big\{ \tau^*(\bX_i)  - \tau_g^* \big\}
		+
		\nabla_\eta \tau(\bX_i \con \eta^*)  \cond M(\bX_i) = g
	\big]
	=
	\EXP \big\{ s^*(\bO_i)
		\bIF_{\SP,g} (\bO_i) \big\} \ .
\end{align*}

Next, we show that $\bIF_{\SP,g}(\bO_i)$ belongs to the tangent space when (i) the treatment is randomized with probability 0.5 within each group and/or (ii) within each subgroup, the variance is homoscedastic across all $(A_i,\bX_i)$  and $\tau^*(\bX_i)$ is constant across all $\bX_i$. Let us consider a decomposition of $\bIF_{\SP,g}(\bO_i)$ as
\begin{align*}
	\bIF_{\SP,g}(\bO_i)
	=
	\frac{
		\left[
	\begin{array}{l}
	\big\{ Y_i - \mu^*(A_i, \bX_i) \big\}	\big\{ A_i - e^*(\bX_i) \big\} \ind \big\{ M(\bX_i)= g \big\}
	\\ 
	+
	\big\{ \tau^*(\bX_i) - \tau_g^* \big\}	\big\{ A_i - e^*(\bX_i) \big\} \big\{ 1-2e^*(\bX_i) \big\} \ind \big\{ M(\bX_i)= g \big\}
	\\ 
	+
	\big\{ \tau^*(\bX_i) - \tau_g^* \big\}	\big\{ e^*(\bX_i) \big\} \big\{ 1-e^*(\bX_i) \big\} \ind \big\{ M(\bX_i)= g \big\}
	\end{array}
	\right]
	}{\EXP \big[ \big\{ A_i - e^*(\bX_i) \big\}^2 \ind \{ M(\bX_i)=g \} \big]}
	=
	\begin{bmatrix}
	\bIF_{\SP,Y,g}(Y_i \cond A_i, \bX_i)
	\\
	+
	\bIF_{\SP,A,g}(A_i \cond \bX_i)
	\\
	+
	\bIF_{\SP,X,g}(\bX_i)
	\end{bmatrix}
\end{align*}
Note that $\bIF_{\SP,Y,g}$, $\bIF_{\SP,A,g}$, and $\bIF_{\SP,X,g}$ satisfy the (conditional) mean-zero restrictions imposed on the tangent space $\mathcal{T}_{\SP}$. From straightforward algebra, we evaluate the formula in \eqref{eq:score} with respect to $\bIF_{\SP,g}$:
\begin{align*}
	&
\EXP \Big[
		\bIF_{\SP,X,g}(\bX_i)  \{ A_i - e^*(\bX_i) \}^2 \{ \tau^*(\bX_i) - \tau_g^* \} \ind \{ M(\bX_i) = g \}
	\Big]	
		\nonumber
	\\
	&
	\quad
	+
	\EXP \Big[
		\bIF_{\SP,A,g}(A=1 \cond \bX_i) e^*(\bX_i) \big\{ 1-2e^*(\bX_i) \big\} \{ \tau^*(\bX_i) - \tau_g^* \} \ind \{ M(\bX_i) = g \}
	\Big]	
		\nonumber
	\\
	&
	\quad
	+
	\EXP \Bigg[
		\{ A_i - e^*(\bX_i) \}^2
		\Bigg[ 
		\begin{array}{l}
			\EXP \big\{ Y_i \bIF_{\SP,Y,g}(Y_i \cond A_i =1,\bX_i) \cond A_i=1,\bX_i \big\}
			\\
	-
	\EXP \big\{ Y_i \bIF_{\SP,Y,g}(Y_i \cond A_i =0,\bX_i) \cond A_i=0,\bX_i \big\}
		\end{array}
	 \Bigg]
		\ind \{ M(\bX_i) = g \}
	\Bigg]
		\nonumber
	\\
	&
	\quad
	-
	\EXP \Big[
		\{ A_i - e^*(\bX_i) \}^2
		\ind \{ M(\bX_i) = g \}
	\Big]	
	\\
	& \quad \quad
	\times
	\EXP \Bigg[
		\bIF_{\SP,X,g}(\bX_i) \big\{ \tau^*(\bX_i)  - \tau_g^* \big\}
		+
		\Bigg[ 
		\begin{array}{l}
			\EXP \big\{ Y_i \bIF_{\SP,Y,g}(Y_i \cond A_i =1,\bX_i) \cond A_i=1,\bX_i \big\}
			\\
	-
	\EXP \big\{ Y_i \bIF_{\SP,Y,g}(Y_i \cond A_i =0,\bX_i) \cond A_i=0,\bX_i \big\}
		\end{array}
	 \Bigg]
	 \, \Bigg| \,  M(\bX_i) = g
	\Bigg]
		\nonumber
	\\
	 &
	 \propto
	 \EXP \Big[ 
	 \big[ e^*(\bX_i) \big\{ 1 - e^*(\bX_i) \big\} \big]^2
	 \big\{ \tau^*(\bX_i) - \tau_g^* \big\}^2
	 \, \Big| \, M(\bX_i) = g
	 \Big]
	  \\
	 & \quad +
	 \EXP \Big[ 
		e^*(\bX_i) \big\{ 1 - e^*(\bX_i) \big\} \big\{ 1-2e^*(\bX_i) \big\}^2  \big\{ \tau^*(\bX_i) - \tau_g^* \big\}^2 \Cond M(\bX_i) = g
	 \Big]
	 \\
	 & \quad +
	 \EXP \Big[
	 		\big\{ A_i - e^*(\bX_i) \big\}^2
	 		\big[
	 	\sigma_g^2(1,\bX_i) 	\big\{ 1 - e^*(\bX_i) \big\} + \sigma_g^2(0,\bX_i) e^*(\bX_i) 
	 	\big]
	 	\Cond M(\bX_i) = g
	 \Big]
	 \\
	 & \quad -
	 \EXP \big[ \big\{ A_i - e^*(\bX_i) \big\}^2 \cond M(\bX_i)=g \big] 
	  \EXP \Big[
	 \big\{ A_i - e^*(\bX_i) \big\}^2
	 \big\{ \tau^*(\bX_i) - \tau_g^* \big\}^2
	 \, \Big| \, M(\bX_i) = g
	 \Big]
	 \\
	 & \quad -
	 \EXP \big[ \big\{ A_i - e^*(\bX_i) \big\}^2 \cond M(\bX_i)=g \big] 
	  \EXP \Big[ 
	 \sigma_g^2(1,\bX_i) 	\big\{ 1 - e^*(\bX_i) \big\} + \sigma_g^2(0,\bX_i) e^*(\bX_i)
	\, \Big| \, M(\bX_i) = g
	 \Big]
	  \\
	 &
	=
	 \EXP \Bigg[   \big\{ A_i - e^*(\bX_i) \big\}^2
	 \Bigg[ 
	 \begin{array}{l}
	 	 \big\{ A_i - e^*(\bX_i) \big\}^2
	 \big\{ \tau^*(\bX_i) - \tau_g^* \big\}^2
	 \\
	 +
	 	\sigma_g^2(1,\bX_i) 	\big\{ 1 - e^*(\bX_i) \big\} + \sigma_g^2(0,\bX_i) e^*(\bX_i) 
	 \end{array}
	 	\Bigg]
	 \, \Bigg| \, M(\bX_i) = g
	 \Bigg]
	 \\ 
	 & \quad -
	 \EXP \Big[
	  \big\{ A_i - e^*(\bX_i) \big\}^2 \Cond M(\bX_i)=g \Big] 
	 \EXP \Bigg[  
	 \begin{array}{l}
	 	 \big\{ A_i - e^*(\bX_i) \big\}^2
	 \big\{ \tau^*(\bX_i) - \tau_g^* \big\}^2
	 \\
	 +
	 	\sigma_g^2(1,\bX_i) 	\big\{ 1 - e^*(\bX_i) \big\} + \sigma_g^2(0,\bX_i) e^*(\bX_i) 
	 \end{array}
	 \, \Bigg| \, M(\bX_i) = g
	 \Bigg]
\end{align*}
where $\sigma_g^2(A_i, \bX_i)  = \VAR(Y_i \cond A_i, \bX_i)$. If the above quantity is zero, then  ${\bIF}_{\SP}$ belongs to the tangent space and it becomes the efficient influence function for $\bT^*$. This is satisfied when $F_{g,i} := \{ A_i - e^*(\bX_i) \}^2$ and $G_{g,i} := \big\{ A_i - e^*(\bX_i) \big\}^2  \big\{ \tau^*(\bX_i) - \tau_g^* \big\}^2 + \sigma_g^2(1,\bX_i) 	\big\{ 1 - e^*(\bX_i) \big\} + \sigma_g^2(0,\bX_i) e^*(\bX_i) $ have zero-covariance within each subgroup. Some sufficient condition for this are (i) $e^*(\bX_i) =0.5$ for all $\bX_i$ in each subgroup so that $F_{g,i}$ is constant as $F_{g,i}=0.25$ and (ii) $\tau^*(\bX_i) - \tau_g^* = 0$ and $\sigma_g^2 := \sigma_g^2(1,\bX_i)= \sigma_g^2(0,\bX_i)$ for all $\bX_i$ in each subgroup so that $G_{g,i}$ is constant as $G_{g,i} = \sigma_g^2$. That is, when either of the two conditions is satisfied, ${\bIF}_{\SP}$ satisfies the conditions on the efficient influence function for $\bT^*$, implying that $\widehat{\bT}_{\SP}$ achieves the efficiency bound for $\bT^*$ under $\model_{\SP}$.

\subsection{Proof of Theorem \ref{thm:EIF} in the Main Paper}			\label{proof:EIF}

The asymptotic normality of $\widehat{\bT}_{\NP}$ and the consistency of the variance estimator can be shown by following the proof in Section \ref{proof:SSLS} except Lemma \ref{lem-assumption-EIF} is used instead of Lemma \ref{lem-assumption-SSLS}.

Let us denote the influence function of $\widehat{\bT}_{\NP}$ as ${\bIF}_\NP = (\bIF_{\NP,1},\ldots,\bIF_{\NP,G})\T$ where
\begin{align*}
	\bIF_{\NP,g}(\bO_i)
	&
	=
	\frac{\ind \{ M(\bX_i)=g \}}{P \big\{ M(\bX_i) = g \big\}}
	\Bigg[
		\bigg\{ \frac{A_i}{e^*(\bX_i)} - \frac{1-A_i}{1-e^*(\bX_i)} \bigg\}
		\big\{ Y_i - \mu^*(A_i, \bX_i) \big\}
		+
		\big\{ \tau^*(\bX_i) - \tau_g^* \big\}
	\Bigg]
\end{align*}
Following algebra in \citet{Hahn1998}, it is straightforward to check that 
\begin{align*}
	\nabla_\eta \tau_g(\eta) \big|_{\eta=\eta^*}
	=
	\EXP \big[ 
		s_X^* (\bX_i ) \big\{ \tau^*(\bX_i)  - \tau_g^* \big\}
		+
		\nabla_\eta \tau(\bX_i \con \eta^*)  \cond M(\bX_i) = g
	\big]
	=
	\EXP \big\{ s^*(\bO_i)
		\bIF_{\NP,g} (\bO_i) \big\} \ .
\end{align*}
Therefore, we find that the groupwise treatment effects are differentiable \citep{Newey1990}. Additionally, the tangent space under the nonparametric model $\model_{\NP}$ is a collection of entire mean-zero, square-integrable functions of $\bO_i$. Therefore, ${\bIF}_{\NP}$ is the efficient influence function for $\bT^*$ under model $\model_{\NP}$. 

Lastly, we show that ${\bIF}_{\NP}$ does not achieve the efficiency bound for $\bT^*$ under $\model_{\NP}$. Following the approach in Section \ref{proof:SSLS}, ${\bIF}_{\NP}$ belongs to the tangent space $\mathcal{T}_{\SP}$ if ${\bIF}_{\NP,g}$ satisfies the restriction \eqref{eq:score}. From straightforward algebra, we evaluate the formula in \eqref{eq:score} with respect to $\bIF_{\NP,g}$:
\begin{align*}
	&
\EXP \Big[
		\bIF_{\NP,X,g}(\bX_i)  \{ A_i - e^*(\bX_i) \}^2 \{ \tau^*(\bX_i) - \tau_g^* \} \ind \{ M(\bX_i) = g \}
	\Big]	
		\nonumber
	\\
	&
	\quad
	+
	\EXP \Big[
		\bIF_{\NP,A,g}(A=1 \cond \bX_i) e^*(\bX_i) \big\{ 1-2e^*(\bX_i) \big\} \{ \tau^*(\bX_i) - \tau_g^* \} \ind \{ M(\bX_i) = g \}
	\Big]	
		\nonumber
	\\
	&
	\quad
	+
	\EXP \Bigg[
		\{ A_i - e^*(\bX_i) \}^2
		\Bigg[ 
		\begin{array}{l}
			\EXP \big\{ Y_i \bIF_{\NP,Y,g}(Y_i \cond A_i =1,\bX_i) \cond A_i=1,\bX_i \big\}
			\\
	-
	\EXP \big\{ Y_i \bIF_{\NP,Y,g}(Y_i \cond A_i =0,\bX_i) \cond A_i=0,\bX_i \big\}
		\end{array}
	 \Bigg]
		\ind \{ M(\bX_i) = g \}
	\Bigg]
		\nonumber
	\\
	&
	\quad
	-
	\EXP \Big[
		\{ A_i - e^*(\bX_i) \}^2
		\ind \{ M(\bX_i) = g \}
	\Big]	
	\\
	& \quad \quad
	\times
	\EXP \Bigg[
		\bIF_{\NP,X,g}(\bX_i) \big\{ \tau^*(\bX_i)  - \tau_g^* \big\}
		+
		\Bigg[ 
		\begin{array}{l}
			\EXP \big\{ Y_i \bIF_{\NP,Y,g}(Y_i \cond A_i =1,\bX_i) \cond A_i=1,\bX_i \big\}
			\\
	-
	\EXP \big\{ Y_i \bIF_{\NP,Y,g}(Y_i \cond A_i =0,\bX_i) \cond A_i=0,\bX_i \big\}
		\end{array}
	 \Bigg]
	 \, \Bigg| \,  M(\bX_i) = g
	\Bigg]
		\nonumber
	\\
	 & =
	\EXP \Bigg[ 
	 \big\{ A_i - e^*(\bX_i) \big\}^2
	 \bigg[
	 	 \big\{ \tau^*(\bX_i) - \tau_g^* \big\}^2 
	 	 +
	\frac{\sigma_g^2(1,\bX_i)}{e^*(\bX_i) }		
	+
	\frac{\sigma_g^2(0,\bX_i)}{1-e^*(\bX_i) }		
	\bigg]
	\, \Bigg| \, M(\bX_i) = g
	 \Bigg] 
	 \\
	 & \quad 
	 - \EXP \big[  \big\{ A_i - e^*(\bX_i) \big\}^2 \cond M(\bX_i) = g \big]
	 \EXP \Bigg[
	 	\bigg[
	 	 \big\{ \tau^*(\bX_i) - \tau_g^* \big\}^2 
	 	 +
	\frac{\sigma_g^2(1,\bX_i)}{e^*(\bX_i) }		
	+
	\frac{\sigma_g^2(0,\bX_i)}{1-e^*(\bX_i) }		
	\bigg]
	\, \Bigg| \, M(\bX_i) = g
	 \Bigg]
\end{align*}
where $\sigma_g^2(A_i, \bX_i)  = \VAR(Y_i \cond A_i, \bX_i)$. If the above quantity is zero, then  ${\bIF}_{\NP}$ belongs to the tangent space and it becomes the efficient influence function for $\bT^*$. This is satisfied when $F_{g,i} := \{ A_i - e^*(\bX_i) \}^2$ and $G_{g,i} := \big\{ \tau^*(\bX_i) - \tau_g^* \big\}^2 + \sigma_g^2(1,\bX_i)/e^*(\bX_i) + \sigma_g^2(0,\bX_i) /\{1- e^*(\bX_i)\} $ have zero-covariance within each subgroup. In general, this condition is not satisfied for any laws unless some additional conditions are imposed on the nuisance functions. This concludes that $\widehat{\bT}_\NP$ does not achieve the efficiency bound for $\bT^*$ under model $\model_\SP$ in general even though it achieves the efficiency bound for $\bT^*$ under model $\model_\NP$.

\subsection{Proof of Theorem \ref{thm:lin} in the Main Paper} \label{proof:lin}

The asymptotic normality of $(\widehat{\bT}_{\SP}, \widehat{\bT}_{\NP})$ and the consistency of the variance estimator can be established by following the proof in Section \ref{proof:SSLS} except Lemma \ref{lem-assumption-Joint} is used instead of Lemma \ref{lem-assumption-SSLS}.

The variance of the weighted estimator using weight $w$ is 
\begin{align}						\label{eq-proof-304-1}
	&
	w^2\sigma_{\SSLS,g}^2 + 2w(1-w) \sigma_{\SSLS,\EIF,g} + (1-w)^2 \sigma_{\EIF,g}^2
	\nonumber
	\\
	&
	=
	(  {\sigma}_{\SSLS,g}^2 - 2 {\sigma}_{\SSLS,\EIF,g} + {\sigma}_{\EIF,g}^2 ) w^2 
	-2w
	( \sigma_{\EIF,g}^2 - \sigma_{\SSLS,\EIF,g} ) 
	+ \sigma_{\EIF,g}^2 \ . 
\end{align}

Let $\widehat{W} = {\rm diag} (\widehat{w}_1, \ldots, \widehat{w}_G)$ and $W = {\rm diag} (w_1, \ldots, w_G)$.  We first consider the case $ {\sigma}_{\SSLS,g}^2 - 2 {\sigma}_{\SSLS,\EIF,g} + {\sigma}_{\EIF,g}^2 > 0$. Then, $w_g$ is well-defined as 
\begin{align*}
	w_g =	\bigg(
	\frac{{\sigma}_{\EIF,g}^2 - {\sigma}_{\SSLS,\EIF,g}}{ {\sigma}_{\SSLS,g}^2 - 2 {\sigma}_{\SSLS,\EIF,g} + {\sigma}_{\EIF,g}^2} 
	\bigg)_{[0,1]}  \ . 
\end{align*}
From the continuous mapping theorem, we find
\begin{align*}
	\widehat{w}_g
	=
	\bigg(
	\frac{\widehat{\sigma}_{\EIF,g}^2 - \widehat{\sigma}_{\SSLS,\EIF,g}}{ \widehat{\sigma}_{\SSLS,g}^2 - 2 \widehat{\sigma}_{\SSLS,\EIF,g} + \widehat{\sigma}_{\EIF,g}^2} 
	\bigg)_{[0,1]}
	\stackrel{P}{\rightarrow}
		\bigg(
	\frac{{\sigma}_{\EIF,g}^2 - {\sigma}_{\SSLS,\EIF,g}}{ {\sigma}_{\SSLS,g}^2 - 2 {\sigma}_{\SSLS,\EIF,g} + {\sigma}_{\EIF,g}^2} 
	\bigg)_{[0,1]} = w_g \ . 
\end{align*}
Second, we consider the case $\sigma_{\SSLS,g}^2 - 2\sigma_{\SSLS,\EIF,g}^2 + \sigma_{\EIF,g}^2=0$. If $\sigma_{\EIF,g}^2 - \sigma_{\SSLS,\EIF,g}>0$, we find $w_g=1$ and 
\begin{align*}
	\widehat{w}_g
	=
	\bigg(
	\frac{\widehat{\sigma}_{\EIF,g}^2 - \widehat{\sigma}_{\SSLS,\EIF,g}}{ \widehat{\sigma}_{\SSLS,g}^2 - 2 \widehat{\sigma}_{\SSLS,\EIF,g} + \widehat{\sigma}_{\EIF,g}^2} 
	\bigg)_{[0,1]}
	\stackrel{P}{\rightarrow}
	( \infty )_{[0,1]} = 1 = w_g \ . 
\end{align*}
On the other hand,  if $\sigma_{\EIF,g}^2 - \sigma_{\SSLS,\EIF,g}<0$, we find $w_g=0$ and 
\begin{align*}
	\widehat{w}_g
	=
	\bigg(
	\frac{\widehat{\sigma}_{\EIF,g}^2 - \widehat{\sigma}_{\SSLS,\EIF,g}}{ \widehat{\sigma}_{\SSLS,g}^2 - 2 \widehat{\sigma}_{\SSLS,\EIF,g} + \widehat{\sigma}_{\EIF,g}^2} 
	\bigg)_{[0,1]}
	\stackrel{P}{\rightarrow}
	( -\infty )_{[0,1]} = 0 = w_g \ . 
\end{align*}
Under the above cases, using the Slutsky's theorem, we find the asymptotic distribution of $\sqrt{N} ( \widehat{\bT}_W - \bT^* )$ is
\begin{align}							\label{eq-proof-304-2}
	\sqrt{N} ( \widehat{\bT}_W - \bT^* )
	&
	=
	\widehat{W}
	\sqrt{N} ( \widehat{\bT}_{\SSLS} - \bT^* )
	+
	\big( I - \widehat{W} \big)
	\sqrt{N} ( \widehat{\bT}_{\EIF} - \bT^* )
	\nonumber
	\\
	& 
	\stackrel{D}{\rightarrow}
	N \Big( 0, W \Sigma_\SSLS W + 2W \Sigma_{\SSLS,\EIF} (I-W) + (I-W) \Sigma_{\EIF} (I-W) \Big) \ .
\end{align}
This concludes the proof for the above cases.

Lastly, we consider the case $\sigma_{\SSLS,g}^2 - 2\sigma_{\SSLS,\EIF,g}^2 + \sigma_{\EIF,g}^2=0$ and $\sigma_{\EIF,g} ^2 - \sigma_{\SSLS,\EIF,g}=0$, implying $\sigma_{\SSLS,g}^2=\sigma_{\EIF,g}^2 = \sigma_{\SSLS.\EIF,g}$. The asymptotic distribution of $(\widehat{\tau}_{\SSLS,g},\widehat{\tau}_{\EIF,g})\T$ is  degenerate and $\sqrt{N}(\widehat{\tau}_{\SSLS,g} - \widehat{\tau}_{\EIF,g}) \stackrel{D}{\rightarrow} 0$, implying $\sqrt{N}(\widehat{\bT}_{\SSLS} - \widehat{\bT}_{\EIF}) = o_P(1)$. Moreover, any $w_g \in [0,1]$ is a minimizer of the variance in \eqref{eq-proof-304-1}. Therefore, we have
\begin{align*}
	\sqrt{N} ( \widehat{\bT}_W - \bT^* )
	&
	=
	\widehat{W}
	\sqrt{N} ( \widehat{\bT}_{\SSLS} - \bT^* )
	+
	\big( I - \widehat{W} \big)
	\sqrt{N} ( \widehat{\bT}_{\EIF} - \bT^* )
	\\
	&
	=
	\underbrace{
	\widehat{W} }_{O_P(1)}
	\underbrace{
	\sqrt{N} ( \widehat{\bT}_{\SSLS} - \widehat{\bT}_{\EIF} )
	}_{o_P(1)}
	+
	\sqrt{N} ( \widehat{\bT}_{\EIF} - \bT^* )
	\stackrel{D}{\rightarrow}
	N \Big( 0, \Sigma_{\EIF} \Big) \ .
\end{align*}
Here we find $ \Sigma_{\EIF} = D \Sigma_\SSLS D + 2D \Sigma_{\SSLS,\EIF} (I-D) + (I-D) \Sigma_{\EIF} (I-D)$ for any matrix $D$ because of $\Sigma_\SSLS= \Sigma_{\SSLS,\EIF} = \Sigma_\EIF$. Thus, taking $D=W$, we have the same result in \eqref{eq-proof-304-2}.

\newpage

\bibliographystyle{apa}
\bibliography{SSLS_JRSS_A}

\begin{thebibliography}{}

\bibitem[\protect\astroncite{Athey and Imbens}{2016}]{AtheyImbens2016}
Athey, S. and Imbens, G. (2016).
\newblock Recursive partitioning for heterogeneous causal effects.
\newblock {\em Proceedings of the National Academy of Sciences},
  113(27):7353--7360.

\bibitem[\protect\astroncite{Athey et~al.}{2019}]{GRF}
Athey, S., Tibshirani, J., and Wager, S. (2019).
\newblock Generalized random forests.
\newblock {\em The Annals of Statistics}, 47(2):1148--1178.

\bibitem[\protect\astroncite{Benkeser and {van der Laan}}{2016}]{Benkeser2016}
Benkeser, D. and {van der Laan}, M. (2016).
\newblock The highly adaptive lasso estimator.
\newblock In {\em 2016 IEEE International Conference on Data Science and
  Advanced Analytics (DSAA)}, pages 689--696.

\bibitem[\protect\astroncite{Bergmeir and Ben\'itez}{2012}]{RSNNS}
Bergmeir, C. and Ben\'itez, J.~M. (2012).
\newblock Neural networks in {R} using the stuttgart neural network simulator:
  {RSNNS}.
\newblock {\em Journal of Statistical Software}, 46(7):1--26.

\bibitem[\protect\astroncite{Bhattacharya and Zhao}{1997}]{BZ1997}
Bhattacharya, P.~K. and Zhao, P.-L. (1997).
\newblock Semiparametric inference in a partial linear model.
\newblock {\em The Annals of Statistics}, 25(1):244--262.

\bibitem[\protect\astroncite{Bickel et~al.}{1998}]{BKRW1998}
Bickel, P.~J., Klaassen, C.~A., Ritov, Y., and Wellner, J.~A. (1998).
\newblock {\em Efficient and Adaptive Estimation for Semiparametric Models}.
\newblock Springer, New York, 1 edition.

\bibitem[\protect\astroncite{Bickel et~al.}{2009}]{Bickel2009}
Bickel, P.~J., Ritov, Y., and Tsybakov, A.~B. (2009).
\newblock {Simultaneous analysis of Lasso and Dantzig selector}.
\newblock {\em The Annals of Statistics}, 37(4):1705 -- 1732.

\bibitem[\protect\astroncite{Cameron and Miller}{2015}]{Cameron2015}
Cameron, A.~C. and Miller, D.~L. (2015).
\newblock A practitioner’s guide to cluster-robust inference.
\newblock {\em Journal of Human Resources}, 50(2):317--372.

\bibitem[\protect\astroncite{Chamberlain}{1992}]{chamberlain1992}
Chamberlain, G. (1992).
\newblock Efficiency bounds for semiparametric regression.
\newblock {\em Econometrica}, 60(3):567--596.

\bibitem[\protect\astroncite{Chen and Guestrin}{2016}]{xgboost}
Chen, T. and Guestrin, C. (2016).
\newblock Xgboost: A scalable tree boosting system.
\newblock In {\em Proceedings of the 22nd ACM SIGKDD International Conference
  on Knowledge Discovery and Data Mining}, KDD '16, page 785–794.

\bibitem[\protect\astroncite{Chernozhukov et~al.}{2018}]{victor2018}
Chernozhukov, V., Chetverikov, D., Demirer, M., Duflo, E., Hansen, C., Newey,
  W., and Robins, J. (2018).
\newblock Double/debiased machine learning for treatment and structural
  parameters.
\newblock {\em The Econometrics Journal}, 21(1):C1--C68.

\bibitem[\protect\astroncite{{Chernozhukov} et~al.}{2017}]{Victor2017}
{Chernozhukov}, V., {Demirer}, M., {Duflo}, E., and {Fernandez-Val}, I. (2017).
\newblock {Generic machine learning inference on heterogenous treatment effects
  in randomized experiments}.
\newblock {\em Preprint arXiv:1712.04802}.
\newblock Department of Economics, Massachusetts Institute of Technology,
  Cambridge.

\bibitem[\protect\astroncite{Crump et~al.}{2006}]{Crump2006}
Crump, R.~K., Hotz, V.~J., Imbens, G.~W., and Mitnik, O.~A. (2006).
\newblock Moving the goalposts: Addressing limited overlap in the estimation of
  average treatment effects by changing the estimand.
\newblock Working Paper 330, National Bureau of Economic Research.

\bibitem[\protect\astroncite{Crump et~al.}{2009}]{Crump2009}
Crump, R.~K., Hotz, V.~J., Imbens, G.~W., and Mitnik, O.~A. (2009).
\newblock {Dealing with limited overlap in estimation of average treatment
  effects}.
\newblock {\em Biometrika}, 96(1):187--199.

\bibitem[\protect\astroncite{Dorie et~al.}{2019}]{Dorie2019}
Dorie, V., Hill, J., Shalit, U., Scott, M., and Cervone, D. (2019).
\newblock Automated versus do-it-yourself methods for causal inference: Lessons
  learned from a data analysis competition.
\newblock {\em Statistical Science}, 34(1):43--68.

\bibitem[\protect\astroncite{Dunn}{1958}]{dunn1958}
Dunn, O.~J. (1958).
\newblock Estimation of the means of dependent variables.
\newblock {\em The Annals of Mathematical Statistics}, 29(4):1095--1111.

\bibitem[\protect\astroncite{Durbin}{1954}]{Durbin1954}
Durbin, J. (1954).
\newblock Errors in variables.
\newblock {\em Review of the International Statistical Institute}, 22:23--32.

\bibitem[\protect\astroncite{Friedman et~al.}{2010}]{glmnet}
Friedman, J., Hastie, T., and Tibshirani, R. (2010).
\newblock Regularization paths for generalized linear models via coordinate
  descent.
\newblock {\em Journal of Statistical Software}, 33(1):1--22.

\bibitem[\protect\astroncite{Friedman}{1991}]{earth}
Friedman, J.~H. (1991).
\newblock Multivariate adaptive regression splines.
\newblock {\em The Annals of Statistics}, 19(1):1 -- 67.

\bibitem[\protect\astroncite{Friedman}{2001}]{gbm}
Friedman, J.~H. (2001).
\newblock Greedy function approximation: A gradient boosting machine.
\newblock {\em The Annals of Statistics}, 29(5):1189--1232.

\bibitem[\protect\astroncite{Green and Strawderman}{1991}]{JS1991}
Green, E.~J. and Strawderman, W.~E. (1991).
\newblock A {James-Stein} type estimator for combining unbiased and possibly
  biased estimators.
\newblock {\em Journal of the American Statistical Association},
  86(416):1001--1006.

\bibitem[\protect\astroncite{Green et~al.}{2005}]{JS2005}
Green, E.~J., Strawderman, W.~E., Amateis, R.~L., and Reams, G.~A. (2005).
\newblock {Improved Estimation for Multiple Means with Heterogeneous
  Variances}.
\newblock {\em Forest Science}, 51(1):1--6.

\bibitem[\protect\astroncite{Hahn}{1998}]{Hahn1998}
Hahn, J. (1998).
\newblock On the role of the propensity score in efficient semiparametric
  estimation of average treatment effects.
\newblock {\em Econometrica}, 66(2):315--331.

\bibitem[\protect\astroncite{Hahn et~al.}{2020}]{Hahn2020}
Hahn, P.~R., Murray, J.~S., and Carvalho, C.~M. (2020).
\newblock Bayesian regression tree models for causal inference: Regularization,
  confounding, and heterogeneous effects (with discussion).
\newblock {\em Bayesian Analysis}, 15(3):965--1056.

\bibitem[\protect\astroncite{H{\"a}rdle et~al.}{2000}]{Hardle2000}
H{\"a}rdle, W., Liang, H., and Gao, J. (2000).
\newblock {\em Partially Linear Models}.
\newblock Springer Science \& Business Media.

\bibitem[\protect\astroncite{Hastie and Tibshirani}{1986}]{gam}
Hastie, T. and Tibshirani, R. (1986).
\newblock Generalized additive models.
\newblock {\em Statistical Science}, 1(3):297 -- 310.

\bibitem[\protect\astroncite{Hausman}{1978}]{Hausman1978}
Hausman, J.~A. (1978).
\newblock Specification tests in econometrics.
\newblock {\em Econometrica}, 46(6):1251--1271.

\bibitem[\protect\astroncite{Hern\'an and Robins}{2020}]{HR2020}
Hern\'an, M.~A. and Robins, J.~M. (2020).
\newblock {\em Causal Inference: What If}.
\newblock Chapman \& Hall/CRC, Boca Raton.

\bibitem[\protect\astroncite{Hill}{2011}]{Hill2011}
Hill, J.~L. (2011).
\newblock Bayesian nonparametric modeling for causal inference.
\newblock {\em Journal of Computational and Graphical Statistics},
  20(1):217--240.

\bibitem[\protect\astroncite{Imai and Li}{2022}]{ImaiLi2022}
Imai, K. and Li, M.~L. (2022).
\newblock Statistical inference for heterogeneous treatment effects discovered
  by generic machine learning in randomized experiments.
\newblock {\em Preprint arXiv:2203.14511}.

\bibitem[\protect\astroncite{Imai and Ratkovic}{2013}]{Imai2013}
Imai, K. and Ratkovic, M. (2013).
\newblock Estimating treatment effect heterogeneity in randomized program
  evaluation.
\newblock {\em The Annals of Applied Statistics}, 7(1):443--470.

\bibitem[\protect\astroncite{Imbens and Rubin}{2015}]{ImbensRubin2015}
Imbens, G.~W. and Rubin, D.~B. (2015).
\newblock {\em Causal Inference for Statistics, Social, and Biomedical
  Sciences: An Introduction}.
\newblock Cambridge University Press, New York.

\bibitem[\protect\astroncite{Kennedy}{2020}]{Kennedy2020}
Kennedy, E.~H. (2020).
\newblock Towards optimal doubly robust estimation of heterogeneous causal
  effects.
\newblock {\em Preprint arXiv:2004.14497}.

\bibitem[\protect\astroncite{Kooperberg}{2020}]{polspline}
Kooperberg, C. (2020).
\newblock {\em polspline: {Polynomial Spline Routines}}.
\newblock R package version 1.1.19.

\bibitem[\protect\astroncite{K{\"u}nzel et~al.}{2019}]{Kunzel2019}
K{\"u}nzel, S.~R., Sekhon, J.~S., Bickel, P.~J., and Yu, B. (2019).
\newblock Meta-learners for estimating heterogeneous treatment effects using
  machine learning.
\newblock {\em Proceedings of the National Academy of Sciences},
  116(10):4156--4165.

\bibitem[\protect\astroncite{K{\"u}nzel et~al.}{2018}]{Kunzel2018}
K{\"u}nzel, S.~R., Walter, S. J.~S., and Sekhon, J.~S. (2018).
\newblock {Causaltoolbox---Estimator stability for heterogeneous treatment
  effects}.
\newblock {\em Preprint arXiv:1811.02833}.
\newblock Department of Statistics, University of California at Berkeley,
  Berkeley.

\bibitem[\protect\astroncite{Lee et~al.}{2021}]{Lee2021}
Lee, Y., Nguyen, T.~Q., and Stuart, E.~A. (2021).
\newblock Partially pooled propensity score models for average treatment effect
  estimation with multilevel data.
\newblock {\em Journal of the Royal Statistical Society: Series A (Statistics
  in Society)}, 184(4):1578--1598.

\bibitem[\protect\astroncite{Li}{2000}]{Qi2000}
Li, Q. (2000).
\newblock Efficient estimation of additive partially linear models.
\newblock {\em International Economic Review}, 41(4):1073--1092.

\bibitem[\protect\astroncite{Liang and Zeger}{1986}]{GEE}
Liang, K.-Y. and Zeger, S.~L. (1986).
\newblock Longitudinal data analysis using generalized linear models.
\newblock {\em Biometrika}, 73(1):13--22.

\bibitem[\protect\astroncite{Ma et~al.}{2006}]{MCW2006}
Ma, Y., Chiou, J.-M., and Wang, N. (2006).
\newblock {Efficient semiparametric estimator for heteroscedastic partially
  linear models}.
\newblock {\em Biometrika}, 93(1):75--84.

\bibitem[\protect\astroncite{McCoy et~al.}{2016}]{ECLSK1}
McCoy, D.~C., Morris, P.~A., Connors, M.~C., Gomez, C.~J., and Yoshikawa, H.
  (2016).
\newblock Differential effectiveness of head start in urban and rural
  communities.
\newblock {\em Journal of Applied Developmental Psychology}, 43:29--42.

\bibitem[\protect\astroncite{Mittelhammer and Judge}{2005}]{JS2005_2}
Mittelhammer, R.~C. and Judge, G.~G. (2005).
\newblock Combining estimators to improve structural model estimation and
  inference under quadratic loss.
\newblock {\em Journal of Econometrics}, 128(1):1--29.

\bibitem[\protect\astroncite{Nadaraya}{1964}]{Nadaraya1964}
Nadaraya, E.~A. (1964).
\newblock On estimating regression.
\newblock {\em Theory of Probability \& Its Applications}, 9(1):141--142.

\bibitem[\protect\astroncite{Newey}{1990}]{Newey1990}
Newey, W.~K. (1990).
\newblock Semiparametric efficiency bounds.
\newblock {\em Journal of Applied Econometrics}, 5(2):99--135.

\bibitem[\protect\astroncite{Newey}{1994}]{Newey1994}
Newey, W.~K. (1994).
\newblock The asymptotic variance of semiparametric estimators.
\newblock {\em Econometrica}, 62(6):1349--1382.

\bibitem[\protect\astroncite{Nie and Wager}{2020}]{NieWager2020}
Nie, X. and Wager, S. (2020).
\newblock {Quasi-oracle estimation of heterogeneous treatment effects}.
\newblock {\em Biometrika}, 108(2):299--319.

\bibitem[\protect\astroncite{Polley and van~der Laan}{2010}]{Polley2010}
Polley, E.~C. and van~der Laan, M.~J. (2010).
\newblock Super learner in prediction.
\newblock {\em Technical report 200}.
\newblock Division of Biostatistics, Working Paper Series.

\bibitem[\protect\astroncite{Reardon}{2019}]{ECLSK2}
Reardon, S.~F. (2019).
\newblock Educational opportunity in early and middle childhood: Using full
  population administrative data to study variation by place and age.
\newblock {\em RSF: The Russell Sage Foundation Journal of the Social
  Sciences}, 5(2):40--68.

\bibitem[\protect\astroncite{Robins}{1994}]{Robins1994}
Robins, J.~M. (1994).
\newblock Correcting for non-compliance in randomized trials using structural
  nested mean models.
\newblock {\em Communications in Statistics - Theory and Methods},
  23(8):2379--2412.

\bibitem[\protect\astroncite{Robins et~al.}{1992}]{RMN1992}
Robins, J.~M., Mark, S.~D., and Newey, W.~K. (1992).
\newblock Estimating exposure effects by modelling the expectation of exposure
  conditional on confounders.
\newblock {\em Biometrics}, 48(2):479--495.

\bibitem[\protect\astroncite{Robins and Rotnitzky}{2001}]{RR2001}
Robins, J.~M. and Rotnitzky, A. (2001).
\newblock Comment on ``inference for semiparametric models: Some questions and
  an answer,'' by pj bickel and j. kwon.
\newblock {\em Statistica Sinica}, 11:920--936.

\bibitem[\protect\astroncite{Robinson}{1988}]{robinson1988}
Robinson, P.~M. (1988).
\newblock Root-n-consistent semiparametric regression.
\newblock {\em Econometrica}, 56(4):931--954.

\bibitem[\protect\astroncite{Rosenbaum and Rubin}{1983}]{rosenbaum1983}
Rosenbaum, P.~R. and Rubin, D.~B. (1983).
\newblock The central role of the propensity score in observational studies for
  causal effects.
\newblock {\em Biometrika}, 70(1):41--55.

\bibitem[\protect\astroncite{Rosenman and Miratrix}{2022}]{Rosenman2022}
Rosenman, E.~T. and Miratrix, L. (2022).
\newblock Designing experiments toward shrinkage estimation.
\newblock {\em Preprint arXiv:2204.06687}.

\bibitem[\protect\astroncite{Scharfstein et~al.}{1999}]{Scharfstein1999}
Scharfstein, D.~O., Rotnitzky, A., and Robins, J.~M. (1999).
\newblock Adjusting for nonignorable drop-out using semiparametric nonresponse
  models.
\newblock {\em Journal of the American Statistical Association},
  94(448):1096--1120.

\bibitem[\protect\astroncite{Shalit et~al.}{2017}]{Shalit2017}
Shalit, U., Johansson, F.~D., and Sontag, D. (2017).
\newblock Estimating individual treatment effect: Generalization bounds and
  algorithms.
\newblock In {\em Proceedings of the 34th International Conference on Machine
  Learning}, volume~70 of {\em Proceedings of Machine Learning Research}, pages
  3076--3085. JMLR.org.

\bibitem[\protect\astroncite{Sidak}{1967}]{Sidak1967}
Sidak, Z. (1967).
\newblock Rectangular confidence regions for the means of multivariate normal
  distributions.
\newblock {\em Journal of the American Statistical Association},
  62(318):626--633.

\bibitem[\protect\astroncite{Su et~al.}{2009}]{Su2009}
Su, X., Tsai, C.-L., Wang, H., Nickerson, D.~M., and Li, B. (2009).
\newblock Subgroup analysis via recursive partitioning.
\newblock {\em The Journal of Machine Learning Research}, 10:141--158.

\bibitem[\protect\astroncite{Tibshirani et~al.}{2021a}]{grfvig}
Tibshirani, J., Athey, S., Sverdrup, E., and Wager, S. (2021a).
\newblock {Generalized Random Forests: Cluster-Robust Estimation}.

\bibitem[\protect\astroncite{Tibshirani et~al.}{2021b}]{grfpackage}
Tibshirani, J., Athey, S., Sverdrup, E., and Wager, S. (2021b).
\newblock {\em grf: Generalized Random Forests}.
\newblock R package version 2.0.2.

\bibitem[\protect\astroncite{Tourangeau et~al.}{2009}]{ECLSK2009}
Tourangeau, K., Nord, C., L{\^e}, T., Sorongon, A.~G., and Najarian, M. (2009).
\newblock Early childhood longitudinal study, kindergarten class of 1998-99
  ({ECLS-K}): Combined user's manual for the {ECLS-K} eighth-grade and {K-8}
  full sample data files and electronic codebooks. nces 2009-004.
\newblock {\em National Center for Education Statistics}.

\bibitem[\protect\astroncite{van~der Laan et~al.}{2007}]{SL2007}
van~der Laan, M.~J., Polley, E.~C., and Hubbard, A.~E. (2007).
\newblock Super learner.
\newblock {\em Statistical Applications in Genetics and Molecular Biology},
  6(1).

\bibitem[\protect\astroncite{van~der Laan and Robins}{2003}]{vvLaan2003}
van~der Laan, M.~J. and Robins, J.~M. (2003).
\newblock {\em Unified Methods for Censored Longitudinal Data and Causality}.
\newblock Springer, New York.

\bibitem[\protect\astroncite{Wager and Athey}{2018}]{WA2018}
Wager, S. and Athey, S. (2018).
\newblock Estimation and inference of heterogeneous treatment effects using
  random forests.
\newblock {\em Journal of the American Statistical Association},
  113(523):1228--1242.

\bibitem[\protect\astroncite{Wager and Walther}{2016}]{Wager2016}
Wager, S. and Walther, G. (2016).
\newblock Adaptive concentration of regression trees, with application to
  random forests.
\newblock Department of Statistics, Stanford University.

\bibitem[\protect\astroncite{Watson}{1964}]{Watson1964}
Watson, G.~S. (1964).
\newblock Smooth regression analysis.
\newblock {\em Sankhy{\=a}: The Indian Journal of Statistics, Series A}, pages
  359--372.

\bibitem[\protect\astroncite{Westfall and Young}{1993}]{Westfall1993}
Westfall, P.~H. and Young, S.~S. (1993).
\newblock {\em Resampling-based Multiple Testing: Examples and Methods for
  p-value Adjustment}, volume 279.
\newblock John Wiley \& Sons, New York.

\bibitem[\protect\astroncite{Wright and Ziegler}{2017}]{ranger}
Wright, M.~N. and Ziegler, A. (2017).
\newblock {ranger}: A fast implementation of random forests for high
  dimensional data in {C++} and {R}.
\newblock {\em Journal of Statistical Software}, 77(1):1--17.

\bibitem[\protect\astroncite{Wu}{1973}]{Wu1973}
Wu, D.-M. (1973).
\newblock Alternative tests of independence between stochastic regressors and
  disturbances.
\newblock {\em Econometrica}, 41(4):733--750.

\end{thebibliography}

\end{document}